\documentclass[12pt,draftcls, onecolumn, journal]{IEEEtran}
\usepackage{graphicx}
\usepackage{amsmath}
\usepackage{amsfonts}
\usepackage{amssymb}
\usepackage{cite}
\usepackage{color}
\usepackage{multirow}
\usepackage{tabularx}
\newcolumntype{L}[1]{>{\raggedright\arraybackslash}p{#1}}
\newcolumntype{C}[1]{>{\centering\arraybackslash}p{#1}}
\newcolumntype{R}[1]{>{\raggedleft\arraybackslash}p{#1}}

\usepackage{times, epsfig}
\usepackage{subfig}
\usepackage{tikz}
\usepackage{bbm}

\newcommand{\SNR}{\mathsf{SNR}}

\newcommand{\paren}[1]{\left(#1\right)}
\newcommand{\sqparen}[1]{\left[#1\right]}
\newcommand{\brparen}[1]{\left\{#1\right\}}
\newcommand{\parenlo}[1]{\left.\left(#1\right]\right.}
\newcommand{\parenro}[1]{\left.\left[#1\right)\right.}
\newcommand{\abs}[1]{\left| #1\right|}
\newcommand{\norm}[1]{\left\| #1\right\|}
\newcommand{\field}[1]{\ensuremath{\mathbb{#1}}}
\newcommand{\N}{\ensuremath{\field{N}}} 
\newcommand{\R}{\ensuremath{\field{R}}} 
\newcommand{\C}{\ensuremath{\field{C}}} 
\newcommand{\Rp}{\ensuremath{\R_+}} 
\newcommand{\I}[1]{\ensuremath{\mathsf{1}_{\left\{#1\right\}}}} 
\newcommand{\Inb}[1]{\ensuremath{\mathsf{1}_{#1}}} 
\newcommand{\ra}{\ensuremath{\rightarrow}} 
\newcommand{\PR}[1]{\ensuremath{\mathsf{Pr}\left\{#1\right\}}} 
\newcommand{\PRP}[1]{\ensuremath{\mathsf{Pr}\left(#1\right)}} 
\newcommand{\EW}{\ensuremath{\mathsf{E}}} 
\newcommand{\ES}[1]{\ensuremath{\mathsf{E}\left[#1 \right]}} 
\newcommand{\defeq}{\ensuremath{\triangleq}} 
\newcommand{\e}[1]{\ensuremath{{\rm e}^{#1}}} 

\newcommand{\snr}{\ensuremath{{\sf SNR}}}

\renewcommand{\vec}[1]{\ensuremath{\boldsymbol{#1}}} 
\newcommand{\pol}{\ensuremath{\mathcal{P}}}
\newcommand{\TX}{\ensuremath{\vec{x}_{\rm s}}}
\newcommand{\RX}{\ensuremath{\vec{x}_{\rm d}}}
\newcommand{\rateins}{\ensuremath{{\sf r}_{\varphi}}}
\newcommand{\rateave}{\ensuremath{{\sf R}_{\Phi}}}
\newcommand{\Rave}{\ensuremath{{\sf R}_{\rm ave}}}
\newcommand{\xopt}{\ensuremath{\vec{x}_{\rm opt}}}

\newcommand{\Xopt}{\ensuremath{\vec{X}_{\rm opt}}}

\newcommand{\xmid}{\ensuremath{\vec{x}_{\rm mid}}}
\newcommand{\Xmid}{\ensuremath{\vec{X}_{\rm mid}}}
\newcommand{\xs}{\ensuremath{\vec{x}_{\rm s}}}
\newcommand{\xd}{\ensuremath{\vec{x}_{\rm d}}}

\newcommand{\Hsd}{\ensuremath{H_{\rm s, d}}}
\newcommand{\Hsr}{\ensuremath{H_{\rm s, r}}}
\newcommand{\Hrd}{\ensuremath{H_{\rm r, d}}}
\newcommand{\Pout}{\ensuremath{{\sf P}_{\rm out}}}

\newcommand{\relayselect}{\ensuremath{\widehat{s}_{}}}

\newcommand{\relayselectdiffsnr}{\ensuremath{\widehat{s}_{\rm diff}}}

\newcommand{\erf}{\ensuremath{{\rm erf}}}
\newcommand{\erfc}{\ensuremath{{\rm erfc}}}
\newcommand{\diff}{\ensuremath{{\rm d}}}

\DeclareMathOperator{\arcsec}{arcsec}

\DeclareMathOperator{\arccsc}{arccsc}

\IEEEoverridecommandlockouts

\newtheorem{definition}{Definition}

\newtheorem{theorem}{Theorem}

\newtheorem{lemma}{Lemma}

\newlength{\figwidth}
\begin{document}

\setlength{\pdfpagewidth}{8.5in}
\setlength{\pdfpageheight}{11in}

\title{Optimum Location-Based Relay Selection in Wireless Networks}
\author{
\IEEEauthorblockN{ Hazer Inaltekin,~\IEEEmembership{Member, IEEE}, 
Saman Atapattu,~\IEEEmembership{Senior Member, IEEE},\\Jamie S. Evans,~\IEEEmembership{Senior Member, IEEE}\\}
\thanks{H.~Inaltekin is with the School of Engineering, Macquarie University, North Ryde, NSW 2109, Australia.
Email: hazer.inaltekin@mq.edu.au.}
\thanks{S. Atapattu and J. S. Evans are with the Department of Electrical and Electronic Engineering, University of Melbourne, Parkville, VIC 3010, Australia.
Email:\{saman.atapattu, jse\}@unimelb.edu.au.}
\thanks{The material in this paper was presented in part at the IEEE International Conference on Communications, Shanghai, China, May 2019, and in part at the IEEE Global Communications Conference, Waikoloa, HI, USA, December 2019 \cite{Atapattu2019icc, Inaltekin2019gcom}.}
\thanks{This work has been submitted to the IEEE for possible publication. Copyright may be transferred without notice, after which this version may no longer be accessible.}
\vspace{-20mm}
}
\date{}
\maketitle

\begin{abstract}

This paper studies the performance and key structural properties of the optimum location-based relay selection policy for wireless networks consisting of homogeneous Poisson distributed relays. The distribution of the channel quality indicator at the optimum relay location is obtained. A threshold-based distributed selective feedback policy is proposed for the discovery of the optimum relay location with finite average feedback load. It is established that the total number of relays feeding back obeys a Poisson distribution and an analytical expression for the average feedback load is derived. The analytical expressions for the average rate and outage probability with and without selective feedback are also obtained for general path-loss models. It is shown that the optimum location-based relay selection policy outperforms other common relay selection strategies notably. It is also shown that utilizing location information from a small number of relays 
%
%
is enough to achieve almost the same performance with the infinite feedback load case. As generalizations, full-duplex relays, isotropic Poisson point processes, and heterogeneous source-to-relay and relay-to-destination links are also studied.

\end{abstract}

\begin{IEEEkeywords}
Relay selection, cooperative communications, selective feedback, Poisson point processes, stochastic geometry.
\end{IEEEkeywords}

\section{Introduction}
\subsection{Background and Motivation}
In wireless networks with large geographical coverage, direct communication links suffer from high propagation losses with distance \cite{Tse05}. To overcome this performance impediment as well as to ensure connectivity 
%
%
in classical and emerging wireless systems, dual-hop communication or cooperative relaying is considered to be an effective transmission and network deployment strategy from a system design point-of-view \cite{Yanikomeroglu04, Nosratinia04, Poor10, Poor15}. 
%
%
%
At the link level, the classical one-way relay channel supports information flow in one specified direction over potentially shorter transmission distances when compared with direct transmissions.  
%
%
This classical relay channel was originally proposed and studied by van der Meulen in \cite{Meulen71}. 
%
%
Cover and El Gamal derived the capacity theorems for Gaussian and certain discrete memoryless relay channels in \cite{Cover79}. They also obtained an achievable lower bound to the capacity of the general relay channel. Subsequently, these results are extended to networks with many relays, multiple antennas (in the form of approximations with bounded capacity gap) and cooperation architectures \cite{Schein2000isit, Laneman03, Gupta2003it, Sendonaris2003a, Sendonaris2003b, Reznik2004it, Laneman04, Kumar04, Kumar05, Kramer05, Madsen05, Madsen05b, Madsen06, Shamai09, Tse07, Verdu11, Gupta14, Tse11, ElGamal15}. 


The notable benefits of relays in wireless communications are increase in network capacity and transmission diversity \cite{Laneman04, Kumar04, Kumar05, Kramer05}, improved energy efficiency (measured in bits per unit energy) \cite{Tse07, Verdu11, Gupta14} and decrease in network deployment costs \cite{Yanikomeroglu04, Laneman12}. 
%
%
These benefits have already been recognized by the wireless industry, and the possibility of deploying relays for multi-hop communications in emerging wireless systems was included in the latest proposals for LTE-A standards \cite{3GPPRelay2, 3GPPRelay1}. Two apparent major design problems encountered in industry-grade wireless relay network deployments are, besides link-level capacity optimization, {\em relay selection} (due to implementation constraints limiting the number of simultaneous relay connections) and {\em relay placement}. Some recent studies showed that significant system-wide performance gains due to optimization over these design degrees-of-freedom can be achieved \cite{Medard11, Farrell13, Nazaroglu14it, AnuragKumar17}. However, these aspects are usually eclipsed by the mainstream fixed-relay capacity optimization at the link level, e.g., see the recent surveys \cite{Poor15, Leung15} and the references therein.  

\begin{figure}[!t]
\begin{center}
\includegraphics[scale=0.75]{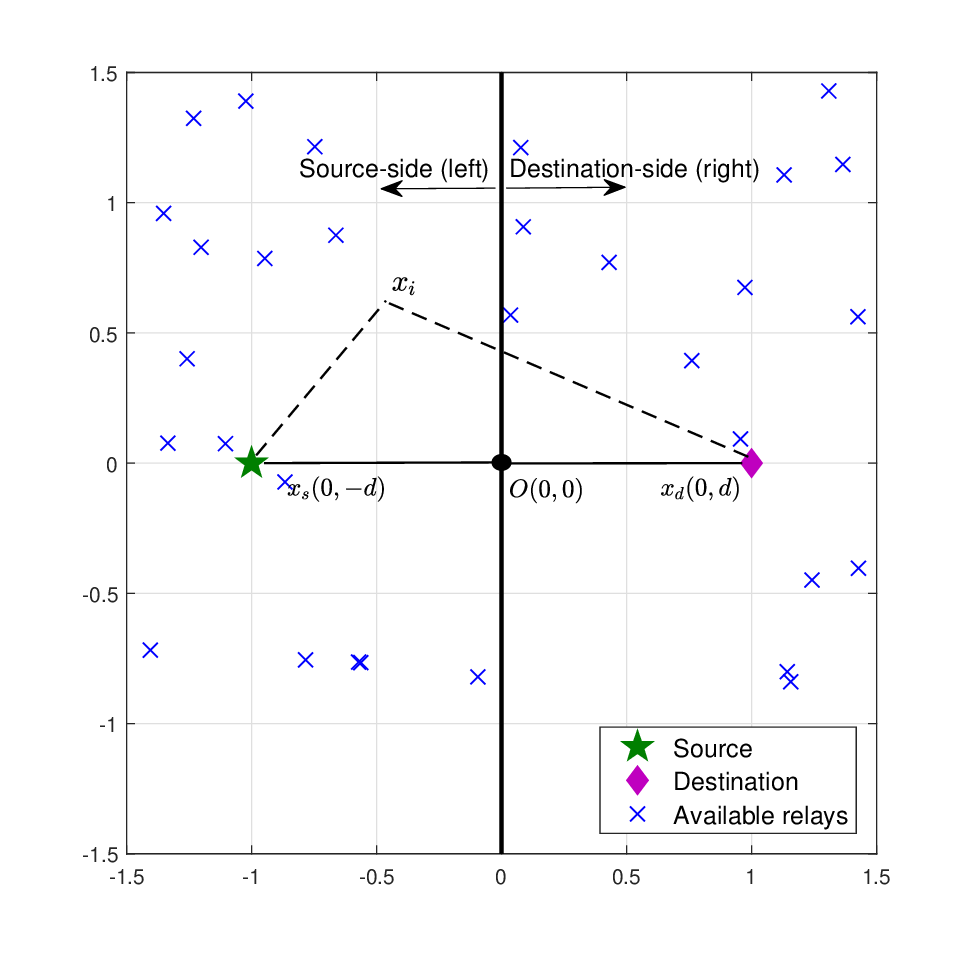}
\end{center}
\vspace{-0.75cm}
\caption{An example configuration for a relay-aided wireless network.} \label{Fig: System Model} 
\end{figure}

In this paper, we focus on the {\em optimum} relay selection problem for randomly deployed single-antenna decode-and-forward (DF) relays connecting source and destination nodes located at arbitrary positions in $\R^2$. An example network configuration is illustrated in Fig. \ref{Fig: System Model}.  In wireless communications, the fading and location processes driving the network capacity and outage events usually vary at different time-scales. For example, the phase of the received electromagnetic waves can change significantly over  millisecond intervals, while the magnitude changes are substantial over  time intervals in the order of seconds or minutes \cite{Tse05, Gallager08}.  By carefully separating the time-scale of changes in fading and location processes, we characterize the key distance balancing and minimum norm (with respect to the mid-point between source and destination nodes) properties for the optimum relay location. We derive the distribution of the channel quality indicator (CQI) at the optimum relay node, which leads to the characterization of the {\em best} achievable average rates and the {\em minimum} outage probability for the resulting class of {\em two-hop} wireless communications paths with optimization over the relay selection dimension. These results hold for general fading distributions and non-increasing path-loss models decaying to zero.   

The time-scale separation approach we adopt for fading and location processes in this paper is motivated by the practical constraints to rule out the prohibitive relay switching rates due to changes in the fading process. In essence, it is similar to the meta-distribution calculations recently introduced for  wireless network performance analysis \cite{Haenggi16, Haenggi18, Haenggi19}. In most parts of the paper, we focus on a homogeneous Poisson point process (HPPP), which we denote by $\Phi$, with intensity $\lambda > 0$ for relay locations. This assumption leads to an insightful analytical structure exposing fundamental dynamics for relay selection. In addition, the uniform relay distribution was also proven to be the optimum configuration for relay placement under some operating regimes such as high signal attenuation as observed in millimeter wave (mmWave) frequency bands \cite{AnuragKumar17}.\footnote{It is well-known that an HPPP can be obtained as the limiting process of a sequence of collection of uniformly distributed points over growing subsets of Euclidean spaces \cite{Durrett96}.} The generalization of the key results to non-homogeneous PPPs is provided in Section \ref{Section: Extensions}, alongside other important extensions such as full-duplex operation (FD) and heterogeneity among the network nodes.      

An important aspect of the optimum relay selection problem studied in this paper is the feedback load required to find the solution. As such, the source node (or, a central entity in this regard) requires the knowledge of {\em all} relay locations to make the optimum relay selection decision, which presents a major design hurdle for practical network deployments with large numbers of relays. To circumvent this feedback bottleneck, we also develop a low-complexity {\em selective} distributed-feedback relay selection policy that requires feedback only from a small number of relays on  average. 

The proposed feedback mechanism is {\em fully distributed} since the relay nodes  utilize only their local information to decide whether they feed back or not. The relay selection is still performed centrally but with significantly reduced feedback. 
We show that the number of relays feeding their CQIs back to the source node for relay selection obeys to a Poisson distribution with a certain mean whose analytical form is completely characterized. We obtain the average rate and outage probability attained by the proposed selective distributed-feedback relay selection policy, and show that it is enough to multiplex only {\em five} relays over the feedback channel to achieve almost the same performance as the {\em all}-feedback scenario. From an implementation and system-design point of view, this finding presents a massive reduction in  feedback load with a negligible loss in  communications performance.   

\subsection{Overview of the Main Results and Contributions in Detail}
The key parameter to obtain the best achievable rates and minimum outage probability with optimized relay selection is the CQI at the optimum relay node, which we denote by $\Gamma_{\rm opt}$. $\Gamma_{\rm opt}$ is determined by minimization of an appropriately defined relay selection function over all relay locations in $\Phi$. An important result of this paper is the derivation of the distribution of $\Gamma_{\rm opt}$. In particular, its cumulative distribution function (cdf) is shown to be    
\begin{equation}
\begin{split}
F_{\Gamma_{\rm opt}}\paren{\gamma} = \left\{
\begin{array}{ll}
0 & \mbox{ if } 
\gamma < d \\
1-\e{-2\lambda d ^2\paren{\paren{\frac{\gamma}{d}}^2 \arcsec\paren{\frac{\gamma}{d}} - \sqrt{\paren{\frac{\gamma}{d}}^2-1}}} & \mbox{ if } 
\gamma \geq d
\end{array} \right.
\end{split}
\end{equation}
with $d$ being the half-distance between source and destination nodes. The network performance limits are simply the appropriate integrals of the $\Phi$-measurable conditional data rates, for which we obtain a single-parameter characterization in the proof of Lemma \ref{Lemma: Rate and Outage Optimality}, with respect to $F_{\Gamma_{\rm opt}}\paren{\gamma}$ on $\Rp$. As explained in Section \ref{Section: Optimum Relay Selection} in detail, the tail distribution of $\Gamma_{\rm opt}$ decays to zero exponentially and all the moments of $\Gamma_{\rm opt}$ are finite. Further, it is also shown that $\Gamma_{\rm opt}$ converges in distribution to $d$ as $\lambda \ra \infty$. 

The previous work on relay selection in random spatial networks is very limited, e.g., see \cite{Cho2011tvt, Mohammadi2012gcom, Behnad2013tcom, Galarza14, Tukmanov2014tcom, Zhou2016twc, Elkotby2015gcom, Elkotby2015twc, Krikidis2014tcom, Belbase2018acc}. These papers mostly focus on {\em sub}-optimum heuristic strategies for selecting relay nodes such as random and closest-to-source relay selection.  The relay nodes selected by the sub-optimum strategies in previous work, as expected, do not possess the fundamental properties of the optimum one that governs the functional form of the distribution of $\Gamma_{\rm opt}$ given above. 
%
%
These are the distance balancing and minimum norm properties of the optimum relay location. It is currently unknown how close a sub-optimum strategy that does not have these properties to the optimum one.  To address this gap in the literature, we obtain a sufficient condition and associated optimality probabilities for a relay selection policy satisfying only one of these fundamental properties to be the optimum selection in Theorems \ref{Lemma: Mid-point Optimality}, \ref{Lemma: Sufficient Probablility} and  \ref{Theorem: Optimality Probability}. These results reveal the analytical dependence of relay selection optimality on the network parameters such as relay node intensity, and provide design guidelines for when a sub-optimum policy can be used with small performance loss.  To the best of our knowledge, our work in this paper is the first study that rigorously and thoroughly investigates the optimum relay selection problem with randomly deployed relay nodes, and derives the fundamental performance limits for multi-hop relay channels with optimized relay selection.

The selective distributed-feedback relay selection policies with autonomous feedback decision computation at relay nodes are studied in Section \ref{Section: Distributed Relay Selection} of the paper. An important result in this part is to show that the total feedback load with threshold-based selection obeys a Poisson distribution with mean $\mu\paren{T}$ given by
\begin{eqnarray}
\mu\paren{T} = \left\{\begin{array}{ll}0 & \mbox{ if } T < d \\\lambda \pi T^2 - 2d\lambda\sqrt{T^2-d^2} - 2T^2\lambda\arctan\paren{\frac{d}{\sqrt{T^2 - d^2}}} & \mbox { if } T \geq d\end{array}\right.,
\end{eqnarray}
where $T$ is the threshold value with which each relay compares its local CQI to decide to feed back or not. It is seen that $\mu(T)$ is a continuous monotone increasing function of $T$ with $\mu(d) = 0$ and $\lim_{T \ra \infty}\mu(T) = \infty$. Hence, any given average feedback load can be met by a threshold value by the intermediate value theorem. Further, based on this statistical characterization of the feedback load, we see that the probability of having at least one relay node feeding its CQI back to the source node is equal to $1-\e{-\mu(T)}$. Hence, we achieve the same rate and outage performance achieved by the all-feedback strategy with probability at least $0.99$ if we use a selective distributed-feedback relay selection policy with $\mu(T) = 5$. The exact characterization of rate and outage with selective feedback is provided in Theorems \ref{Theorem: Average Rate with Feedback} and \ref{Theorem: Outage with Feedback}.

Two significant, as well as surprising, operating regimes for outage probability with selective feedback emerging in Theorem \ref{Theorem: Outage with Feedback} are {\em feedback-limited} and {\em rate-limited} regimes. To the best of our knowledge, this is the first paper that formally characterizes these operating regimes for the relay selection problem.  In particular, the feedback constraints in the feedback-limited operating regime are so tight that we have the same outage probability with or without fading. On the other hand, in the rate-limited operating regime, the target data rate is set so high that we achieve the same outage probability for the all-feedback and selective feedback cases. These operating regimes are discussed in detail in Section \ref{Section: Distributed Relay Selection}, with illustrative numerical examples provided in Section \ref{Section: Numerical Results}.

\subsection{Paper Organization and Notation}
{\it Paper Organization:} The rest of the paper is organized as follows. In Section \ref{Section: Related Work}, we compare and contrast our findings in this paper with relevant previous work in the literature. In Section \ref{s_sys}, we formally present our system model, the notion of relay selection policy and associated performance metrics to evaluate the relay-aided wireless network performance with optimized relay selection. The optimum relay selection problem is introduced in Section \ref{Section: Optimum Relay Selection Problem}. In this section, we also obtain a single-parameter characterization for $\Phi$-measurable conditional data rates, which underpins most of our analysis by exposing the dependence between the CQI at the selected relay node and the resulting network performance. We discuss the fundamental properties that have to be possessed by the optimum relay node in Section \ref{Section: Mid-Point Relay Selection} as well as obtaining a sufficient condition and optimality probabilities for when a relay selection policy possesses only one of these properties. 

The best achievable data rates and the minimum outage probability attained by the optimum relay selection policy are derived in Section \ref{Section: Optimum Relay Selection}. In this section, we mostly focus on the Rayleigh faded wireless medium for obtaining analytical expressions for  network performance, but the same analysis continues to hold for any fading distribution without loss of generality. in Section \ref{Section: Distributed Relay Selection}, we put forward the optimum relay selection problem with selective distributed-feedback, obtain a statistical characterization for the total feedback load in the network and derive the resulting network performance with limitations on the {\em average} total feedback load. An extensive numerical study is presented in Section \ref{Section: Numerical Results} to illustrate the main findings of the paper. We provide the extensions of our analysis to full-duplex relays, non-homogeneous PPPs and more heterogeneous communications scenarios with different signal-to-noise ratio ($\snr$) values at relays and the destination node in Section \ref{Section: Extensions}. Finally, Section \ref{Section: Conclusions} concludes the paper. Most of our proofs are relegated to Appendices \ref{Appendix: Sufficient Probablility Proof}-\ref{Appendix: Different SNR Extension} for the sake of exposition.         

{\it Notation:} The main notation being used throughout the paper is as follows, with some remaining notation introduced later in the paper when needed. We use boldface, upper-case and calligraphic letters to represent vector quantities, random variables and sets, respectively. We use $\R$, $\N$, $\C$ and $\R^2$ to denote the real, natural and complex numbers, and the two-dimensional Euclidean space, respectively. $| x |$ denotes the absolute value of a scalar quantity $x$ (real or complex), whereas $\| \vec{x} \|$ is used to measure the canonical Euclidean norm of a vector quantity $\vec{x}$. 
The expected value of a random variable $X$ is denoted by $\ES{X}$. The probability of an event $\mathcal{E}$ is denoted by $\PRP{\mathcal{E}}$. We use $\I{\cdot}$ to indicate the indicator function of relevant events and mathematical conditions. 


\section{Related Work} \label{Section: Related Work}
Our results and contributions in this paper are related to a large body of work in relay networks. Below, we discuss the most relevant papers under four categories.

\subsection{Relay Networks Without Location Modeling}

The capacity bounds and outage performance of relay networks without explicit modeling of relay locations are well investigated in the literature \cite{Laneman12,Yindi09a,Atapattu13jsac, Lim2011it, Avestimehr2011it, Cardone2014it, Ezzeldin2019it}. In this body of work, the channel states are either fixed and known values, or randomly generated without any modeling of variations in the link signal-to-noise-ratio ($\snr$) as a function of relay locations. The papers such as \cite{Laneman12,Yindi09a,Atapattu13jsac} adopted a communication-theoretic perspective to design single and multiple relay selection strategies, maintaining  the full cooperative {\it diversity}. The analysis is based on the outage probability
%
%
and error rate  
%
%
over fading channels. 
%
%
In particular, it was shown in \cite{Yindi09a}, for the case without a direct link over Rayleigh fading channels, that the best-relay selection strategy achieves full diversity with an outage probability asymptotically decaying to zero according to $\SNR^{-n}$,  
%
%
where $n$ is the number of relays in the network.

Shannon capacity approximations for the {\it Gaussian} $n$-relay diamond networks were obtained in \cite{Lim2011it, Avestimehr2011it, Cardone2014it}. The authors in \cite{Lim2011it} developed a noisy network coding scheme achieving a sum rate within $1.5$ bits/s/Hz of the cut-set upper bound for the single relay case. In \cite{Avestimehr2011it}, Avestimehr et al. used a deterministic approach to approximate the network capacity of single-relay and two-relay Gaussian diamond networks within $1$ bits/s/Hz. The cut-set upper bound was shown to be within $1.96(n + 2)$ bits of network capacity by using noisy network coding arguments for the $n$-relay case in \cite{Cardone2014it}. The papers \cite{Nazaroglu14it, Ezzeldin2019it} developed a network simplification approach by characterising what fraction of total network capacity can be maintained by using only a subset of $k$ relays, out of $n$ available relays in the network. It was shown in \cite{Nazaroglu14it} that every subset of $k$ relays can at most provide approximately a fraction $k/(k + 1)$ of the total capacity with full-duplex relays. The similar results were extended to the half-duplex case in \cite{Ezzeldin2019it}.

Our approach and results differ from this body of work in that we explicitly model the relay locations in the current paper. By doing so, we obtain the best achievable average rates and the  minimum outage probability for the resulting class of two-hop wireless communications paths by optimizing over the relay selection dimension. Our results hold for general non-increasing path-loss models decaying to zero, and for general fading distributions in some cases.

\subsection{Relay Networks With Location Modeling}

The outage and rate performance of relay networks with explicit modeling of relay locations are also investigated in the literature \cite{Cho2011tvt, Behnad2013tcom, Galarza14, Tukmanov2014tcom, Zhou2016twc, Elkotby2015twc, Elkotby2015gcom, Belbase2018acc}, usually utilizing sub-optimum techniques for relay selection to have analytical tractability. In this body of work, the relay locations are modeled as points of a spatial point process. The papers such as \cite{Galarza14, Elkotby2015twc, Elkotby2015gcom} investigated the relay network performance under the heuristic closest-to-source and mid-point relay selection policies. The authors in \cite{Cho2011tvt} considered the relay selection problem from a pre-defined quality-of-service region to minimize the outage probability but they ignored the dependence between source-relay and relay-destination links while calculating the end-to-end network outage performance. In \cite{Behnad2013tcom}, they obtained integral expressions for the outage probability in amplify-and-forward relay networks. The papers \cite{Tukmanov2014tcom,Zhou2016twc, Belbase2018acc} considered the relay selection problem to maximize the relay-destination $\snr$ values. This approach is effectively equivalent to the sub-optimum closest-to-destination relay selection policy.

The papers such as \cite{Elkotby2015twc, Zhou2016twc, Galarza14} considered interference in their relay network model. However, the relay selection policies in these papers do not depend on the network interference level. This is because the network interference introduces a coupling between local relay selection processes at the interfering source-destination links, which makes the optimum multi-user multi-relay selection problem intractable. It also becomes prohibitively hard, even for sub-optimum heuristics taking interference into account for relay selection, to present an accurate statistical characterization of the CQI at the selected relays due to the inter-dependence between local relay selection processes, which does not lead to insightful fundamental expressions for the network performance metrics of interest.

In this paper, different from the previous work on randomly deployed relay networks, we take a more fundamental approach and investigate the performance of the {\em optimum} relay selection policy for general path-loss models when there is no direct link between source and destination nodes. We derive structural properties being possessed by the optimum relay selection policy.  Moreover, we  consider distributed selective feedback relay selection policies, non-homogeneous PPPs and different $\snr$s at source and relay nodes, which were not considered in the previous work described above. We also consider the time-scale separation between fading and relay location processes cautiously, and identify the effect of such separation on the network performance.

\subsection{Relay Communications and Feedback}

Feedback is an important mechanism in relay networks 
%
%
to facilitate effective relay selection. However, there is only a limited body of work investigating the role of feedback in relay communications \cite{Eltayeb15tcom, Elkhalil16tcom, Wang17tvt, Javan17tvt}. 
%
%
The authors in \cite{Eltayeb15tcom, Elkhalil16tcom} proposed a compressed sensing approach for joint identity and $\snr$ estimation to select relays with strong channel conditions. Despite significant feedback load reduction, the rates achieved by the developed algorithms in these papers are always bounded away from the full feedback scenario. In \cite{Wang17tvt}, the author took a different approach to coordinate  asynchronous symbol transmissions from multiple relays by using decision feedback equalizers, shown to achieve full diversity. In \cite{Javan17tvt}, the authors developed numerical techniques to solve the optimum power allocation problem, maximizing data rates for single-relay channels, under quantized feedback.      

In this paper, we focus on solving the feedback problem for relays with random locations by proposing a threshold-based distributed selective feedback relay selection policy to reduce the network feedback load. To the best of our knowledge, there is no prior work addressing this problem when relays are randomly distributed, where availability of relay location and/or channel state information 
%
%
at a central entity or the source node is a common assumption to perform relay selection. This operating assumption requires an excessive feedback load, which does not scale well for large networks. We show that it is enough to multiplex only five relays over the feedback channel to achieve almost the same rate performance as the all-feedback scenario. This is an important reduction in the feedback load from a system-design point of view.

\subsection{Routing in Spatial Networks}

There is a large and growing body of work on routing in random spatial networks \cite{Haenggi05, Haenggi2005it, Sikora2006it, Andrews10, Vaze2011twc, Chen2012it, Stamatiou2014acm, Blaszczyszyn2015twc, Liang2018sipn}, to which our results in this paper are also related. This body of work focuses on the evaluation of network layer performance metrics such as end-to-end delay and throughput to assess the efficiency of a given routing protocol. The analysis is usually carried out for fixed rate transmissions with threshold-based detection and ALOHA- or TDMA-type MAC layer. Some models are very specific, focusing on line networks such as \cite{Sikora2006it, Stamatiou2014acm, Blaszczyszyn2015twc, Liang2018sipn}. In contrast, we take a more fundamental approach in the current paper by considering the information capacity of cooperative relay channels as the performance metric of interest. This change in the approach also triggers a change in the mathematical analysis, different than the above previous work.

The second fundamental difference between our paper and the previous routing work is that we maximize the ``routing" performance of two-hop relay networks by optimizing over the relay selection dimension. The previous routing work fixes the routing protocol such as the $k$-nearest neighbor routing or nearest neighbor routing with minimum bounded progress (i.e., called quasi-equal distance (QED) routing in \cite{Liang2018sipn}). To the best of our knowledge, there is no prior paper in the routing literature optimizing the selection of intermediate nodes (from a random spatial field) to maximize the network performance, as we analyzed thoroughly in the current paper.

\section{Network Model and Performance Metrics}\label{s_sys}


\subsection{Network Model} \label{Subsection: Network Model} 
We consider a relay-aided {\em spatial} wireless network in $\R^2$, as illustrated by Fig. \ref{Fig: System Model}. The network contains a source-destination pair having arbitrary locations $\TX \in \R^2$ (source node) and $\RX \in \R^2$ (destination node). The locations of potential DF relay nodes in the network are given by $\varphi = \brparen{\vec{x}_1, \vec{x}_2, \ldots}$, where $\vec{x}_i \in \R^2$ represents the $i$th DF relay location for $i \in \N$.  We will always assume that $\varphi$ is a locally finite set, i.e., there are only finitely many relays in every bounded subset of $\R^2$. For performance analysis, we will consider the communication scenario in which relay locations are random and determined according to a spatial homogeneous Poisson point process (HPPP) $\Phi = \brparen{\vec{X}_1, \vec{X}_2, \ldots }$. Hence, $\varphi$ should be interpreted as a particular realization of $\Phi$ below, 
which will usually be clear from the context unless otherwise stated. Relay-based cellular networks are an important example of this model \cite{Yuan13, Iwamura10, Yanikomeroglu04}, although our analysis is not restricted to infrastructure-based wireless communications. 
We assume that the primary aim of the relays is to assist data communication between source and destination nodes without generating any additional traffic, as in the latest proposals for LTE-A standards \cite{3GPPRelay2, 3GPPRelay1}. 

For the sake of exposition, we will focus our attention only on the easy-to-implement half-duplex (HD) relaying operation in the remainder of the paper. As explained in Section \ref{Section: Extensions}, this assumption will not limit our results in the full-duplex (FD) case \cite{Kramer05}. 
%
%
The HD operation in our setup follows the classical resource orthogonalization approach \cite{Laneman04}, and we consider that half of the degrees-of-freedom available in the wireless channel is used by the source node, while the other half is allocated for the relay-destination link. This is usually done through time division multiplexing (TDM) in  existing wireless systems when relay and source nodes share the same frequency band \cite{Iwamura10}. 
In a relay-aided wireless network setting, the location of the selected relay node is of critical importance to determine the achievable data rates between source and destination. To this end, we examine the network performance under a given relay selection policy, which is formally defined as follows.      
\begin{definition} \label{Def: Selection Policy}
A relay selection policy $\pol: \Sigma \mapsto \R^2$ is a mapping from the set of all finite subsets of $\R^2$, denoted by $\Sigma$, to $\R^2$ satisfying the condition $\pol\paren{\varphi} \in \varphi$ for all $\varphi \in \Sigma$.    
\end{definition}

Some examples of $\pol$ include the ones choosing the relay closest to the source, the relay closest to the destination and the relay closest to the mid-point between source and destination.  Our focus below is predominantly on the statistical characterization of the performance of a given relay selection policy $\pol$ over a random ensemble of relay locations. For the sake of notational simplicity, we parametrize the relay location selected by $\pol$ as $\vec{x}_\pol$ in the remainder of the paper, with the understanding that $\vec{x}_\pol = \pol\paren{\varphi}$ for any given $\varphi \in \Sigma$. The same meaning is attributed to other variables throughout the paper when we use this parametrization for them as well. The {\em instantaneous} rate of information flow from source to destination (in bits/s/Hz) as a function of the relay selection policy $\pol$ is given as 
\begin{eqnarray}
\rateins\paren{\pol} & = &\frac12 \min\bigl\{\log_2\paren{1 + \snr \abs{\Hsr}^2G\paren{\norm{\TX - \vec{x}_\pol}}}, \nonumber\\
& &\quad\log_2\paren{1 + \snr\paren{\abs{\Hsd}^2G\paren{\norm{\TX - \RX}} + \abs{\Hrd}^2G\paren{\norm{\vec{x}_\pol - \RX}}}} \bigr\} \label{Eqn: DF Rate 1}
\end{eqnarray}
where $G$ is a non-negative non-increasing path-loss function decaying to zero, $H_{a,b} \in \C$ is the random fading coefficient between the source, {\em selected} relay and destination nodes for $a \in \brparen{{\rm s, r}}$ and $b \in \brparen{{\rm r, d}}$, 
and $\snr \defeq \frac{P}{W N_0}$ is the signal-to-noise ratio with $P$, $W$ and $N_0$ being transmission power, communication bandwidth and noise energy per complex dimension, respectively, \cite{Gallager08}.  The rate $\rateins\paren{\pol}$ in \eqref{Eqn: DF Rate 1} is achievable for each fading state $\vec{H} = \paren{\Hsr, \Hrd, \Hsd}^\top$ by using independent Gaussian codebooks at the source and relay nodes and channel state information (CSI) at the receivers \cite{Laneman04}.\footnote{Implicit in this formulation is the assumption of a perfect multiple-access and interference cancellation scheme so that the background noise is the only additive channel distortion at the receivers.}     
%

 
%

One simplifying assumption that we will make for the rate function $\rateins\paren{\pol}$ in order to pose the optimum relay selection problem for general path-loss models is as follows. We will assume that the signal power received over the longer source-destination link is much smaller than the one received over the shorter relay-destination link. This assumption corresponds to the physical circumstances in which 
%
%
either the direct link between source and destination nodes is severely shadowed by an object in the environment (i.e., $\abs{\Hsd} \approx 0$), or the decay of $G$ with distance is very sharp as in the mmWave communications \cite{Iwamura10, Yuan13, Yanikomeroglu04}.  It models the {\em multi-hop} mode of operation for relay channels \cite{Kramer05} as well as the common LTE-A relay deployment scenarios such as dead spot elimination, coverage extension to rural areas and emergency coverage \cite{Iwamura10,Yuan13,Yanikomeroglu04}.  
%

We will also assume that random fading coefficients $\Hsr$ and $\Hrd$ change at a much faster time-scale than the network node locations, which is usually the case in typical wireless communication scenarios \cite{Tse05, Goldsmith05}. In such cases, it is an onerous task, if not practical due to the triggered excessive relay switching rate, for the source node to obtain CSI for all source-to-relay and relay-to-destination channels, and establish a connection to another relay node for handing over the data traffic each time fading coefficients change.  
Hence, our selection criterion will be based only on relay locations for selecting the relay node, which is also embodied in Definition \ref{Def: Selection Policy}.  
This approach is reminiscent of the base station selection strategy in cellular networks.  
This is also the optimum approach when only location information but not the full CSI is available at the source node.\footnote{We continue to assume the availability of full CSI at the receiver side of each link so that the data rates in \eqref{Eqn: DF Rate 1} are achievable.} 

\subsection{Performance Metrics} \label{Subsection: Performance Metrics} 
For determining the performance of a relay selection policy $\pol$, we will use the outage probability and average data rate as our performance metrics \cite{Tse05}. 
The former one is the common metric to measure the system performance for delay-sensitive data traffic requiring a minimum data rate for successful communications, whereas the latter is more appropriate for when the delay requirement is not stringent and transmitters are allowed to adjust their transmission rates based on the observed path-loss values \cite{Wyner94}. 
We assume for both cases that the permissible decoding delay is large enough to average over the fading process. Then, given $\pol$ and the random relay locations $\Phi = \brparen{\vec{X}_1, \vec{X}_2, \ldots}$, the achievable rate over the wireless link connecting the source and selected relay is $\EW\sqparen{\log_2\paren{1 + \snr \abs{\Hsr}^2G\paren{\norm{\TX - \vec{X}_\pol}}} \big| \Phi}$, 
where the expectation is taken only over the randomness due to fading. 
Similarly, the achievable rate between the selected relay and destination is $\EW\sqparen{\log_2\paren{1 + \snr \abs{\Hrd}^2G\paren{\norm{\vec{X}_\pol - \RX}}} \big| \Phi}$.\footnote{For the rate between the selected relay and destination, we ignore the term containing $\abs{\Hsd}$ since we assumed $\abs{\Hsd} \approx 0$. If this is not an appropriate modeling assumption, then the relay-destination rate should be interpreted as the rate obtained through a type of selection combining in which the destination takes only the signals from the relay into account for data decoding. If maximum ratio combining is employed at the destination, similar performance analysis continues to hold as discussed in Section \ref{Section: Numerical Results}. However, we cannot claim the optimality of the proposed relay selection policy for general path-loss models in this case.} 
Hence, we can express the  data rate, averaged over the fading process, from source to destination as
\begin{eqnarray}
\rateave\paren{\pol} = \frac12 \min\bigl\{\EW\sqparen{\log_2\paren{1 + \snr \abs{\Hsr}^2G\paren{\norm{\TX - \vec{X}_\pol}}} \big| \Phi}, \nonumber\\
\EW\sqparen{\log_2\paren{1 + \snr \abs{\Hrd}^2G\paren{\norm{\vec{X}_\pol - \RX}}} \big| \Phi} \bigr\}, \label{Eqn: Rate Given Locations}
\end{eqnarray}
which is a function of $\pol$ and $\Phi$. Using $\rateave\paren{\pol}$ in \eqref{Eqn: Rate Given Locations}, the outage probability for delay-sensitive data traffic and the average rate for elastic data traffic for which the variable rate transmission is permissible (e.g., dynamic adaptive video streaming over HTTP \cite{Inaltekin18b}) 
%
%
as metrics indicative of the relay-assisted network performance are defined as below. 


\begin{definition} \label{Def: Outage Probability}
For a target bit rate $\rho$ and a relay selection policy $\pol$, the outage probability $\Pout\paren{\pol}$ is equal to 
\begin{eqnarray}
\Pout\paren{\pol} = \PR{\rateave\paren{\pol} \leq \rho}. \label{Eqn: Outage Probability}
\end{eqnarray}
\end{definition}

\begin{definition} \label{Def: Average Rate}
For a given relay selection policy $\pol$, the average rate is defined to be 
\begin{eqnarray}
\Rave\paren{\pol} = \ES{\rateave\paren{\pol}}, \label{Eqn: Average Rate}
\end{eqnarray}
where the expectation 
is over the random relay locations. 
\end{definition}

It is important to note that $\Rave\paren{\pol}$ defined in \eqref{Eqn: Average Rate} should be interpreted as the average of the rate in \eqref{Eqn: Rate Given Locations}, which is averaged over the random 
relay locations. It is not the ergodic rate achieved over the long time-horizon for the same source, relay and destination nodes. Rather, the rate in \eqref{Eqn: Rate Given Locations} is attained with a different relay node selected by $\pol$ for each realization of $\Phi$.  
In addition, we observe that we can change the order of minimum and expectation operators while going from \eqref{Eqn: DF Rate 1} to \eqref{Eqn: Rate Given Locations}. This is because the decoding delays permit averaging over the fading process before switching to another relay node. It is more advantageous to minimize expected rates than averaging the minimum of instantaneous rates. This is the {\em waiting-gain} we have 
in \eqref{Eqn: Rate Given Locations}. 

\section{Optimum Relay Selection Problem} \label{Section: Optimum Relay Selection Problem}

Consider the relay selection function $\relayselect\paren{\vec{x}}$, which is defined as
\begin{eqnarray}
\relayselect\paren{\vec{x}} \defeq \max\brparen{\norm{\xs - \vec{x}}, \norm{\vec{x} - \xd}}  \label{Eqn: Relay Selection Function}
\end{eqnarray}
for $\vec{x} \in \R^2$. We say $\relayselect\paren{\vec{x}}$ is the CQI indicator of the relay located at $\vec{x} \in \R^2$. For a given set of relay locations $\varphi \in \Sigma$, we will start with minimization of $\relayselect\paren{\vec{x}}$ over $\varphi$. The solution of this optimization problem is closely related to optimizing the performance metrics $\Pout\paren{\pol}$ and $\Rave\paren{\pol}$, as we will formally establish below. 

We write the optimum relay selection problem as
%
%
\begin{eqnarray}
\begin{array}{ll}
\underset{\vec{x} \in \R^2}{\mbox{minimize}} & \relayselect\paren{\vec{x}} \\
\mbox{subject to} & \vec{x} \in \varphi
\end{array} \label{Eqn: Relay Selection Problem}
\end{eqnarray}
for each $\varphi \in \Sigma$. The solution of \eqref{Eqn: Relay Selection Problem} is well-defined and belongs to $\varphi$ for all $\varphi \in \Sigma$ since $\varphi$ is assumed to be a locally finite set. The optimum relay selection policy, which we denote by $\pol_{\rm opt}$, is the one that solves \eqref{Eqn: Relay Selection Problem} for all $\varphi \in \Sigma$. When we write $\xopt$, we will refer to the location of the relay selected by $\pol_{\rm opt}$. Observing the structure of $\relayselect\paren{\vec{x}}$ and min-max type of optimization in \eqref{Eqn: Relay Selection Problem}, it can be seen that the optimum relay location in $\varphi$ must have two important properties: (i) distance balancing property (with respect to the source and destination locations) and (ii) minimum norm property (with respect to the mid-point between source and destination). 
%
%
We will investigate these properties in detail 
%
in Section \ref{Section: Mid-Point Relay Selection}.  
%

The connection between the solution of the optimization problem in \eqref{Eqn: Relay Selection Problem} and the performance metrics $\Pout\paren{\pol}$ and $\Rave\paren{\pol}$ is not explicit.  In Lemma \ref{Lemma: Rate and Outage Optimality} below, we relate the solution of \eqref{Eqn: Relay Selection Problem} to $\Pout\paren{\pol}$ and $\Rave\paren{\pol}$ by showing that the optimum relay selection policy solving \eqref{Eqn: Relay Selection Problem} for each $\varphi \in \Sigma$ is also the best choice for optimizing $\Pout\paren{\pol}$ and $\Rave\paren{\pol}$. 
\begin{lemma} \label{Lemma: Rate and Outage Optimality}
Let $\Xi$ be the set of all relay selection policies. Then, for identically distributed fading at each wireless link, we have
\begin{eqnarray}
\Pout\paren{\pol_{\rm opt}} = \inf_{\pol \in \Xi} \Pout\paren{\pol} \label{Eqn: Outage Probability Optimality}
\end{eqnarray}
\begin{eqnarray}
\Rave\paren{\pol_{\rm opt}} = \sup_{\pol \in \Xi} \Rave\paren{\pol}. \label{Eqn: Average Rate Optimality}
\end{eqnarray}
\end{lemma}
\begin{IEEEproof}
We first observe the following trivial inequality. If $X_1$ and $X_2$ are two identically distributed non-negative random variables, and $c_1$ and $c_2$ are two non-negative arbitrary constants satisfying $c_1\leq c_2$, then $\ES{\log_2\paren{1 + c_1 X_1}} \leq \ES{\log_2\paren{1 + c_2 X_2}}$. Using this inequality and recalling that for any given relay selection policy $\pol$, $\vec{X}_\pol$ is the location of the relay node selected by $\pol$ when it runs over $\Phi$, we can write
$\rateave\paren{\pol} = \frac12 \EW\sqparen{\log_2\paren{1 + \snr \abs{H}^2G\paren{\relayselect\paren{\vec{X}_\pol}}} \big| \Phi}$ 
where $H$ is the generic fading random variable having the same distribution with $\Hsr$ and $\Hrd$. Since $\pol_{\rm opt}$ solves \eqref{Eqn: Relay Selection Problem} for each $\varphi \in \Sigma$, we also have $\relayselect\paren{\vec{X}_{{\rm opt}}} \leq \relayselect\paren{\vec{X}_\pol}$. Using the above inequality one more time and the property that $G$ is a non-increasing function, we conclude that $\rateave\paren{\pol_{\rm opt}} \geq \rateave\paren{\pol}$ for any $\pol \in \Xi$. This implies $\Rave\paren{\pol_{\rm opt}} = \sup_{\pol \in \Xi} \Rave\paren{\pol}$ and $\Pout\paren{\pol_{\rm opt}} = \inf_{\pol \in \Xi} \Pout\paren{\pol}$.   
\end{IEEEproof}

An important remark about the proof of Lemma \ref{Lemma: Rate and Outage Optimality} is that it also shows $\rateave\paren{\pol}$ can be written as 
\begin{eqnarray}
\rateave\paren{\pol} = \frac12 \EW\sqparen{\log_2\paren{1 + \snr \abs{H}^2G\paren{\relayselect\paren{\vec{X}_\pol}}} \big| \Phi} \label{Eqn: Conditional Rate Expression 1}
\end{eqnarray}  
for any relay selection policy $\pol$. Since $\vec{X}_\pol$ is a $\Phi$-measurable random variable and $H$ is independent of $\Phi$, $\rateave\paren{\pol}$ is as well equal to 
\begin{eqnarray}
\rateave\paren{\pol} = \frac12 \EW\sqparen{\log_2\paren{1 + \snr \abs{H}^2G\paren{\relayselect\paren{\vec{X}_\pol}}} \big| \vec{X}_\pol}. \label{Eqn: Conditional Rate Expression 2}
\end{eqnarray}
Therefore, in order to obtain $\Pout\paren{\pol}$ or $\Rave\paren{\pol}$, we first need to obtain the distribution of $\vec{X}_\pol$ or that of  $\relayselect\paren{\vec{X}_\pol}$, and calculate relevant averages with respect to these distributions. For $\pol_{\rm opt}$ with HPPP distributed relays, the distribution of $\relayselect\paren{\Xopt}$ is not known and it will be derived in Section \ref{Section: Optimum Relay Selection} to obtain $\Pout\paren{\pol_{\rm opt}}$ and $\Rave\paren{\pol_{\rm opt}}$.  




\section{Key Properties of the Optimum Policy} \label{Section: Mid-Point Relay Selection}

As articulated in Section \ref{Section: Optimum Relay Selection Problem}, the optimum relay selection policy $\pol_{\rm opt}$ solving \eqref{Eqn: Relay Selection Problem} must possess two key properties of distance balancing and minimum norm.  In this section, we will characterize these properties in detail before we establish the statistical structure of $\relayselect\paren{\Xopt}$ for Poisson distributed relays over the plane in Section \ref{Section: Optimum Relay Selection}. In order to put our discussion into perspective, we will make use of the heuristic mid-point relay selection policy $\pol_{\rm mid}$, which selects the relay having the closest location $\Xmid$ to the mid-point between source and destination.
%
%
%
$\pol_{\rm mid}$ only possesses one of these key properties, which is the minimum norm property.  We will obtain the probability of $\pol_{\rm mid}$ being optimum for Poisson distributed relays as well as a sufficient condition that guarantees its optimality.  This analysis will reveal the importance of possessing both properties for optimality and the fact that probability of $\pol_{\rm mid}$ being optimum is small for large relay intensity and separation between source and destination nodes. This finding will further motivate our analysis in Section \ref{Section: Optimum Relay Selection}. 
%


\subsection{Minimum Norm and Distance Balancing Properties}
We will start with a set of arbitrary relay locations $\varphi \in \Sigma$ without imposing any statistical structure, and then analyze the case in which 
%
%
relay nodes are randomly distributed over the plane according to an HPPP with intensity $\lambda > 0$. Below, $\vec{w}$ will denote the mid-point between source and destination nodes, i.e., $\vec{w} = \frac{\xs + \xd}{2}$, and $d$ will denote the distance from $\vec{w}$ to $\xs$ or to $\xd$, i.e., $d = \norm{\xs - \vec{w}} = \norm{\xd - \vec{w}}$.  The following simple result formally states the minimum norm property for optimum relay locations. 
%
%
%
%
This result also provides a motivation for the mid-point relay selection rule. 
\begin{lemma} \label{Lemma: Mid-Point Property}
For all $\vec{x} \in \R^2$, $\relayselect\paren{\vec{w}} \leq \relayselect\paren{\vec{x}}$.
\end{lemma}
\begin{IEEEproof}
By triangle inequality, we have $
\norm{\xs - \vec{x}} + \norm{\vec{x} - \xd} \geq 2d
$
for all $\vec{x} \in \R^2$, which implies $\relayselect\paren{\vec{x}} \geq d$ for all $\vec{x} \in \R^2$. This lower bound is achieved with equality when $\vec{x} = \vec{w}$. 
\end{IEEEproof}

Lemma \ref{Lemma: Mid-Point Property} suggests that $\pol_{\rm mid}$ possesses one of the key properties for being a solution for \eqref{Eqn: Relay Selection Problem} since it always chooses the relay node closest to $\vec{w}$, which is globally the best location to place a relay node without direct link signal reception. 
%
%
%
However, $\pol_{\rm mid}$ does not result in the optimum relay selection for all relay configurations $\varphi \in \Sigma$ since, in addition to its distance from the mid-point between $\xs$ and $\xd$, the relay's orientation is also crucial in determining the value of $\relayselect\paren{\vec{x}}$. That is, the closer the relay location $\vec{x}$ to the equidistant hyperplane $\mathcal{H}$ 
between $\xs$ and $\xd$ is, it better balances the distances between source and destination and leads to a smaller value that $\relayselect\paren{\vec{x}}$ takes.  We will utilize the distance balancing property for optimum relay locations to obtain a sufficient condition under which $\pol_{\rm mid}$ is optimum. %
%
We formally state this property in the following lemma. 
%

\begin{lemma} \label{Lemma: Orientation Property}
Let $\mathcal{H} = \brparen{\vec{x} \in \R^2: \paren{\xs - \xd}^\top \vec{x} = \frac12\paren{\norm{\xs}^2 - \norm{\xd}^2}}$ 
be the equidistant hyperplane between $\xs$ and $\xd$. 
For a given $\vec{y} \in \R^2$, let also $\mathcal{C} = \brparen{\vec{x} \in \R^2: \norm{\vec{x} - \vec{w}} = \norm{\vec{y} - \vec{w}}}$, which is the circle around $\vec{w}$ with radius $\norm{\vec{y} - \vec{w}}$. Then, for any $\vec{x} \in \mathcal{H} \cap \mathcal{C}$, we have $\relayselect\paren{\vec{x}} = \sqrt{d^2 + \norm{\vec{y} - \vec{w}}^2}$ and $\relayselect\paren{\vec{x}} \leq \relayselect\paren{\vec{y}}$. 
\end{lemma}
\begin{IEEEproof}
Assume $\vec{y}$ belongs to the half-space closer to $\xd$ and consider the decomposition $\vec{y} = \vec{w} + \vec{z}$ for some $\vec{z} \in \R^2$. Then, we have 
$\paren{\xs - \xd}^\top \vec{z} \leq 0$ and 
\begin{eqnarray}
\relayselect\paren{\vec{y}} &= \norm{\xs - \vec{y}} 
=\sqrt{\norm{\xs - \vec{w}}^2 - \paren{\xs - \xd}^\top \vec{z} + \norm{\vec{z}}^2} 
\geq \sqrt{\norm{\xs - \vec{w}}^2 + \norm{\vec{y} - \vec{w}}^2}. \nonumber 
\end{eqnarray}
Similarly, for any $\vec{x} \in \mathcal{H} \cap \mathcal{C}$, we have $\relayselect\paren{\vec{x}} = \sqrt{\norm{\xs - \vec{w}}^2 + \norm{\vec{y} - \vec{w}}^2}$. Hence, $\relayselect\paren{\vec{x}} \leq \relayselect\paren{\vec{y}}$. The arguments for when $\vec{y}$ belongs to the half-space closer to $\xs$ are the same.  
%
\end{IEEEproof}

\subsection{Optimality Probability for $\pol_{\rm mid}$}

Using Lemma \ref{Lemma: Orientation Property}, we obtain a sufficient condition for the optimality of $\pol_{\rm mid}$ in the next theorem. 
\begin{theorem} \label{Lemma: Mid-point Optimality}
Let 
$\xmid \in \varphi$ be the relay location selected by $\pol_{\rm mid}$ and $\vec{x}_{(2)} \in \varphi$ be the location of the second closest relay to $\vec{w}$. Then, $\pol_{\rm mid}$ solves \eqref{Eqn: Relay Selection Problem} 
%
%
 if $\relayselect\paren{\xmid} \leq \sqrt{d^2 + \norm{\vec{x}_{(2)} - \vec{w}}^2}$.   
\end{theorem}
\begin{IEEEproof}
Let $\vec{x}_{(1)}, \vec{x}_{(2)}, \ldots$ be the ordering of relay locations in $\varphi$ in an ascending manner with respect to their distances to $\vec{w}$, i.e., $\xmid = \vec{x}_{(1)}$ and $\vec{x}_{(i)}$ is the location of the $i$th closest relay to $\vec{w}$. Then, by Lemma \ref{Lemma: Orientation Property}, $\sqrt{d^2 + \norm{\vec{x}_{(i)} - \vec{w}}^2} \leq \relayselect\paren{\vec{x}_{(i)}}$ for all $i \geq 1$. Further, $\sqrt{d^2 + \norm{\vec{x}_{(2)} - \vec{w}}^2} \leq \sqrt{d^2 + \norm{\vec{x}_{(i)} - \vec{w}}^2}$ for all $i \geq 2$. Hence, whenever $\relayselect\paren{\xmid} \leq \sqrt{d^2 + \norm{\vec{x}_{(2)} - \vec{w}}^2}$, we have $\relayselect\paren{\xmid} = \min_{\vec{x} \in \varphi} \relayselect\paren{\vec{x}}$.     
\end{IEEEproof}

For any realization of relay locations, this condition is easy to check since we only need to know the location information of the closest and second closest relay nodes to the mid-point.  Now, by considering random relay locations given according to $\Phi$, which is an HPPP having intensity $\lambda > 0$ (i.e., $\lambda$ is the average number of relays per unit area), we obtain the probability with which the sufficient condition in Theorem \ref{Lemma: Mid-point Optimality} is satisfied. 
%
%

\begin{theorem} \label{Lemma: Sufficient Probablility}
Let $\Phi$ be an HPPP having intensity $\lambda > 0$. Let $\mathcal{E}_{\rm suff}$ be the event for the sufficient condition given in Theorem \ref{Lemma: Mid-point Optimality} for the optimality of $\pol_{\rm mid}$. 
%
%
Then, $\PRP{\mathcal{E}_{\rm suff}}$ is equal to  
\begin{eqnarray}
\PRP{\mathcal{E}_{\rm suff}} = \e{\lambda \pi d^2}\text{erfc}\left(\sqrt{\lambda \pi}\, d\right). \label{Eqn: Sufficiency Probability} 
\end{eqnarray}
\end{theorem}
\begin{IEEEproof}
See Appendix \ref{Appendix: Sufficient Probablility Proof}.
\end{IEEEproof}

In the next theorem, we extend the result in Theorem \ref{Lemma: Sufficient Probablility} and provide an expression for the probability that $\pol_{\rm mid}$ is optimum for Poisson distributed relay locations.
%
%
Even though it will be more complicated than the expression for $\PRP{\mathcal{E}_{\rm suff}}$ given in \eqref{Eqn: Sufficiency Probability}, it is the exact expression for the probability $\PR{\Xmid = \Xopt}$, where $\Xmid \in \Phi$ represents the random location of the relay chosen by $\pol_{\rm mid}$.  
%
\begin{theorem}\label{Theorem: Optimality Probability}
Let $\Phi$ be an HPPP having intensity $\lambda > 0$. For $\vec{x} \in \R^2$ with norm $\psi$ and angle $\theta \in \sqparen{0, \frac{\pi}{2}}$, let %
%
$P\paren{\vec{x}}$ be the function defined as 
\begin{eqnarray}
P\paren{\vec{x}} =  \paren{\paren{\relayselect\paren{\vec{x}}}^2 - \psi^2}\paren{\pi - 2\theta} - d^2\sin\paren{2\theta} - V\paren{\vec{x}, d} + V\paren{\vec{x}, d\sin\paren{\theta}}, \nonumber 
\end{eqnarray}
where $V\paren{\vec{x}, y} = 2y\sqrt{\paren{\relayselect\paren{\vec{x}}}^2 - y^2} + 2\paren{\relayselect\paren{\vec{x}}}^2\arctan\paren{\frac{y}{\sqrt{\paren{\relayselect\paren{\vec{x}}}^2-y^2}}}$. Then, $\PR{\Xmid = \Xopt}$ is equal to
\begin{eqnarray}
\PR{\Xmid = \Xopt} = 4 \ES{\exp\paren{-\lambda P\paren{\Xmid}}\I{0 \leq \Theta_{\rm mid} \leq \frac{\pi}{2}}}, \label{Eqn: Mid Point Optimality}
\end{eqnarray}
where $\I{\cdot}$ is the indicator function and $\Theta_{\rm mid}$ is the angle of $\Xmid$. 
\end{theorem}
\begin{IEEEproof}
See Appendix \ref{Appendix: Probability Xmid equal Xopt Proof}.   
\end{IEEEproof}

We note that it is numerically easy to calculate the expectation in \eqref{Eqn: Mid Point Optimality} by using the probability density function (pdf) of $\Xmid$.  In particular, $\Theta_{\rm mid}$ is uniformly distributed over $\parenro{0, 2\pi}$, $\Psi_{\rm mid} = \norm{\Xmid}$ has the nearest-neighbor pdf $f_{\rm NN}\paren{\psi} = 2\lambda \pi\psi \e{-\lambda\pi\psi^2}\I{\psi\geq0}$, and $\Theta_{\rm mid}$ and $\Psi_{\rm mid}$ are independent random variables. 

\begin{figure*}[!t]
\begin{minipage}[t]{\textwidth}
\begin{center}
\includegraphics[scale=0.25]{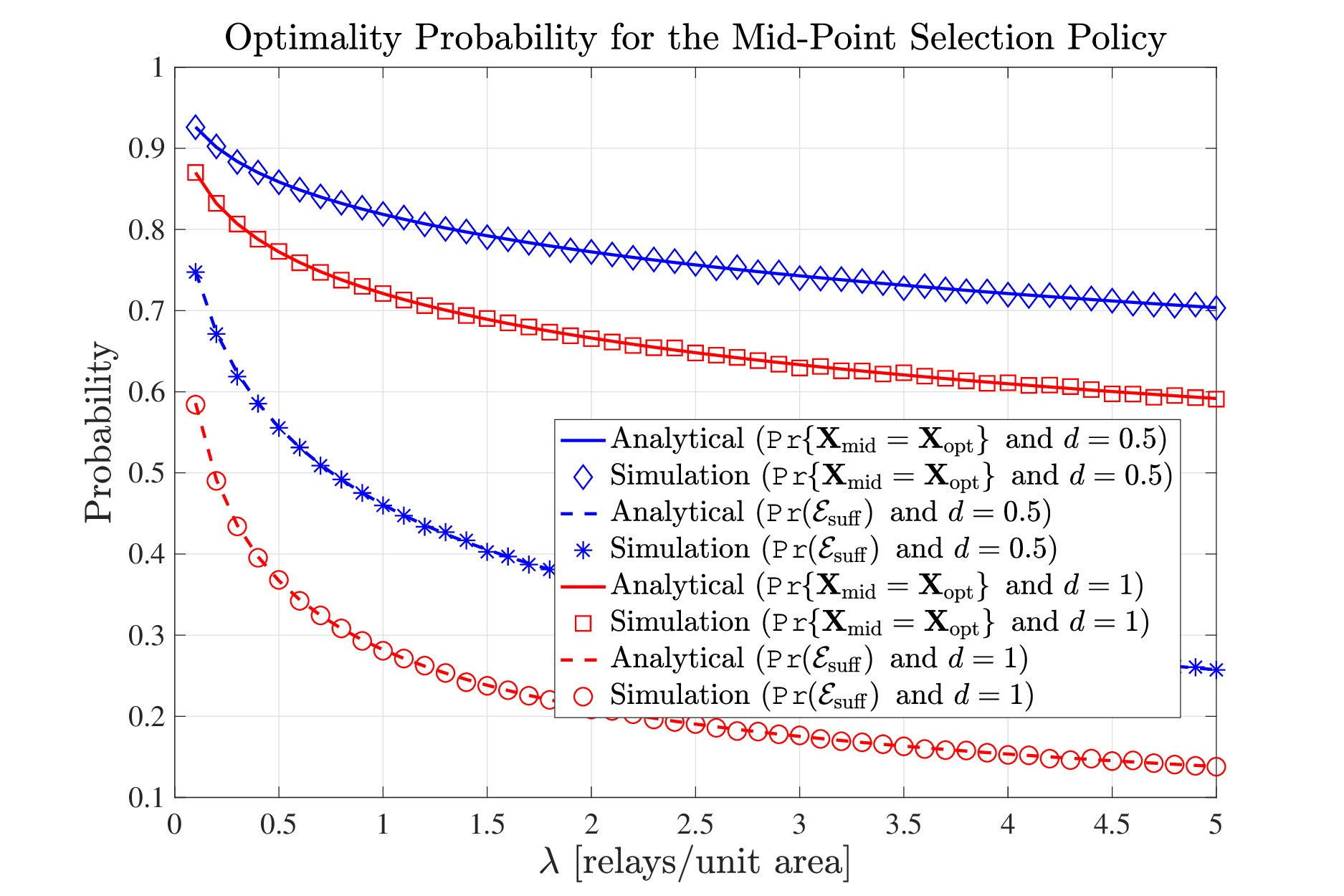}
\hspace{\fill}
\includegraphics[scale=0.25]{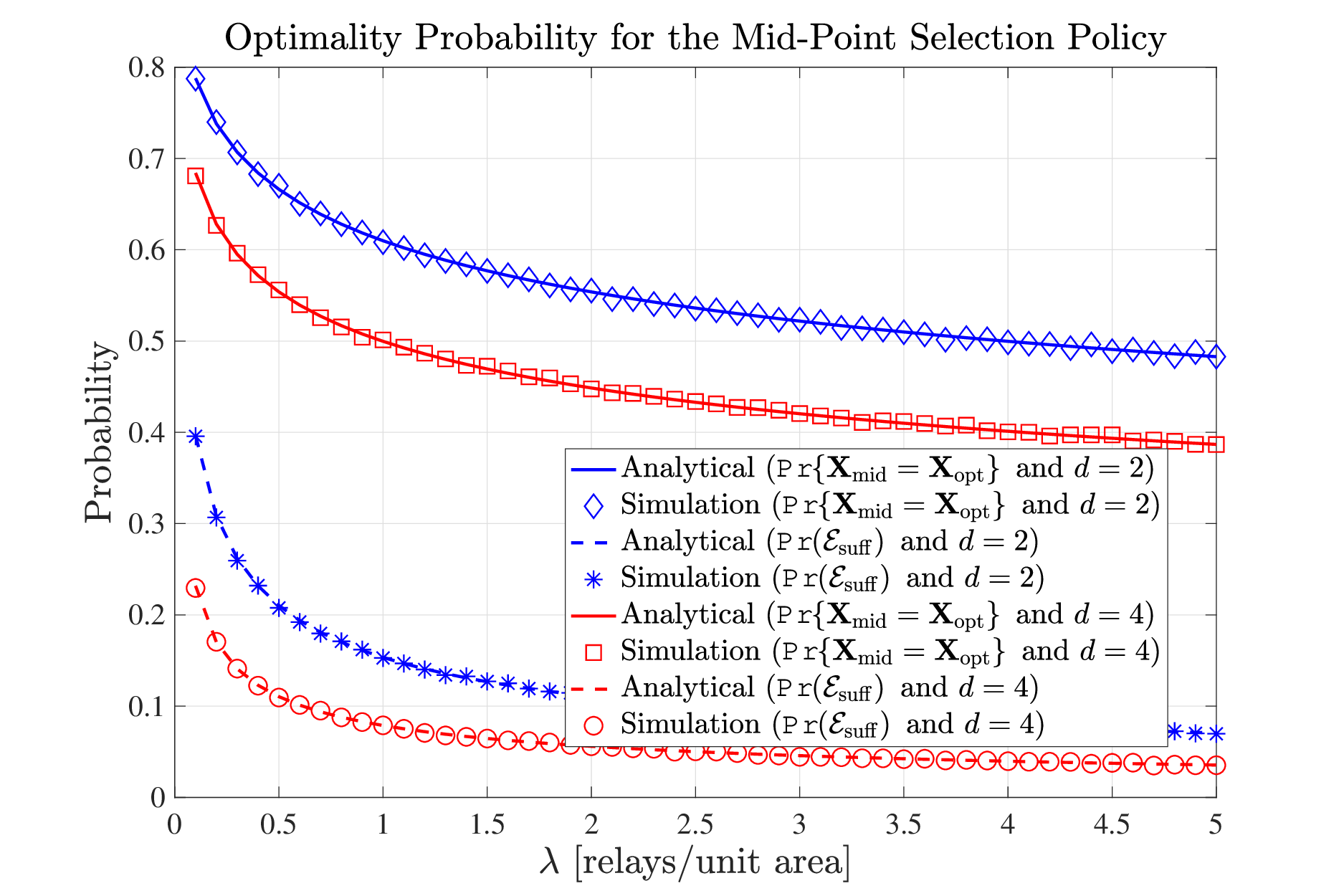}
\end{center}
\end{minipage}
\caption{$\PRP{\mathcal{E}_{\rm suff}}$ and $\PR{\Xmid = \Xopt}$ as a function of $\lambda$ for various values of $d$.}  \label{Fig: Mid-point Optimality}
\end{figure*}

Fig. \ref{Fig: Mid-point Optimality} demonstrates the analytical expressions derived for $\PRP{\mathcal{E}_{\rm suff}}$ and $\PR{\Xmid = \Xopt}$ in Theorem \ref{Lemma: Sufficient Probablility} and Theorem \ref{Theorem: Optimality Probability} as well as the simulated probability values. Our results in these theorems are correct for all source and destination locations, however we will assume $\xs = \paren{-d, 0}$ and $\xd = \paren{d, 0}$ to explain our main observations in Fig. \ref{Fig: Mid-point Optimality} without loss of generality. As can be observed in this figure, $\PR{\Xmid = \Xopt}$ decreases as the relay intensity or the separation between source and destination nodes increases. The diminishing behaviour of $\PR{\Xmid = \Xopt}$ as a function of relay intensity arises from the fact that it becomes more likely to find a relay node $\vec{X} \in \Phi$ with a better distance balancing property, i.e., having angle $\Theta$ close to $\frac{\pi}{2}$ or $\frac{3\pi}{2}$, which is not far away from the mid-point between source and destination, i.e., $\norm{\vec{X}} \approx \norm{\Xmid}$, as the relay intensity increases.   

For $\vec{X} \in \Phi$ having the norm $\Psi$ and angle $\Theta$, we can express the relay selection function $\relayselect\paren{\vec{X}}$ as $\relayselect\paren{\vec{X}} = \sqrt{\Psi^2 + 2d\Psi\abs{\cos\paren{\Theta}} + d^2}$. Hence, as the separation between source and destination nodes increases, the relay locations having angles close to $0$ or $\pi$ are penalized more strongly. Since $\pol_{\rm mid}$ chooses the relay node based only on the distance to mid-point criterion without considering the distance balancing dimension, it becomes less probable for a relay node selected by $\pol_{\rm mid}$ to be optimum when $d$ is large. Correspondingly, the sufficient condition for the optimality of the mid-point relay selection policy is most useful when both the relay intensity and separation between source and destination nodes are small. Overall, despite its analytical convenience for performance evaluation, we observe that $\pol_{\rm mid}$ suffers from its ``distance balancing ignorant operation" notably, which provides a motivation for an in-depth search for the statistical properties of $\pol_{\rm opt}$ in the remainder of the paper.    

{\bf Remark 1:} The performance gap between $\pol_{\rm opt}$ and $\pol_{\rm mid}$ can be significant in terms of average rate and outage probability depending on the network configuration and signal propagation characteristics. This point will be illustrated numerically in Section \ref{Section: Numerical Results}. 


{\bf Remark 2:} Using the results in this section and $\Rave\paren{\pol_{\rm mid}}$, we can readily obtain an upper bound on $\Rave\paren{\pol_{\rm opt}}$. This upper bound is given by 
\begin{eqnarray}
\Rave\paren{\pol_{\rm opt}} \leq \Rave\paren{\pol_{\rm mid}} + \Delta_{\rm ave}, 
\end{eqnarray}
where $\Delta_{\rm ave}$ is equal to 
\begin{eqnarray}
\Delta_{\rm ave} = 2\ES{\paren{1-\e{-\lambda P\paren{\Xmid}}}\log_2\paren{\frac{1 + \snr \cdot G\paren{\sqrt{\Psi_{\rm mid}^2 + d^2}}}{1+\snr \cdot G\paren{\relayselect\paren{\Xmid}}}}\I{0 \leq \Theta_{\rm mid} \leq \frac{\pi}{2}}} \nonumber 
\end{eqnarray}
for the no-fading scenario, and it is equal to 
\begin{eqnarray}
\lefteqn{\Delta_{\rm ave} = \frac{2}{\ln 2}\EW\left[\paren{1-\e{-\lambda P\paren{\Xmid}}}\left(f\paren{\frac{1}{\snr \cdot G\paren{\sqrt{\Psi_{\rm mid}^2 + d^2}}}}\right.\right.} \hspace{14.5cm} \nonumber \\ \lefteqn{\left.\left.-f\paren{\frac{1}{\snr \cdot G\paren{\relayselect\paren{\Xmid}}}}\rule{0cm}{0.9cm} \right)\I{0 \leq \Theta_{\rm mid} \leq \frac{\pi}{2}} \right]} \hspace{6.5cm} \nonumber
\end{eqnarray}
for the Rayleigh fading scenario with the function $P\paren{\vec{x}}$ given as in Theorem \ref{Theorem: Optimality Probability} and the function $f(x)$ defined as $f(x) \defeq \e{x}\text{E}_1(x)$, where $\text{E}_1(x)$ is the exponential integral given by the identity $\text{E}_1(x) = \int_1^\infty \frac{\e{-tx}}{t} \diff t$ for $x>0$. For the Rayleigh fading case, we take $H$ as a circularly symmetric complex Gaussian random variable with unit power, i.e., $H \sim \mathcal{CN}(0,1)$. Hence, $\abs{H}^2$ is an exponential random variable with unit mean, i.e., $f_{\abs{H}^2}(x) = \e{-x}$. The derivations for the upper bounds above are given in Appendix \ref{Appendix: Optimum Rave Upper Bounds}. We note that we can calculate $\Rave\paren{\pol_{\rm mid}}$ by averaging \eqref{Eqn: Conditional Rate Expression 2} with respect to the joint magnitude and angle distribution of $\Xmid$. Similarly, we can calculate $\Delta_{\rm ave}$ by taking above expectations with respect to the joint magnitude and angle distribution of $\Xmid$. In the next section, we will characterize the probability distribution of $\relayselect\paren{\Xopt}$ and obtain an exact expression for $\Rave\paren{\pol_{\rm opt}}$.

\section{Performance Analysis for the Optimum Relay Selection Policy} \label{Section: Optimum Relay Selection}
In this section, we will obtain analytical expressions for $\Rave\paren{\pol_{\rm opt}}$ and $\Pout\paren{\pol_{\rm opt}}$. To this end, we will take random relay location process $\Phi$ as an HPPP having intensity $\lambda>0$. Due to the stationarity and isotropy of HPPPs \cite{Kingman93}, we will assume that $\xs = \paren{-d, 0}^\top$ and $\xd = \paren{d,0}^\top$ without loss of generality. This is the example network configuration illustrated in Fig. \ref{Fig: System Model}.    

\subsection{Distribution of $\relayselect\paren{\Xopt}$} \label{Section: Optimum Relay Distance}
As explained after the proof of Lemma \ref{Lemma: Rate and Outage Optimality}, the conditional data rate, given the relay locations, is equal to 
\begin{eqnarray}\label{Eqn: Conditional Average Rate}
\rateave\paren{\pol} = \frac12 \EW\sqparen{\log_2\paren{1 + \snr \abs{H}^2G\paren{\relayselect\paren{\vec{X}_\pol}}} \big| \Phi}
\end{eqnarray}
for any relay selection policy. Hence, the key step to obtain analytical expressions for $\Rave\paren{\pol_{\rm opt}}$ and $\Pout\paren{\pol_{\rm opt}}$ is to derive the distribution function for $\relayselect\paren{\vec{X}_{{\rm opt}}}$. For notational simplicity, we define $\Gamma_{\rm opt} \defeq \relayselect\paren{\vec{X}_{{\rm opt}}}$. Then, we have
\begin{eqnarray}
\Gamma_{\rm opt} = \min_{\vec{X} \in \Phi} \relayselect\paren{\vec{X}} \label{Eqn: Optimum Relay Distance} 
\end{eqnarray}
by definition of $\pol_{\rm opt}$.  That is, $\Gamma_{\rm opt}$ is the minimum value achieved by the relay selection function over $\Phi$. In the next theorem, we provide the cumulative distribution function (cdf) and the pdf of $\Gamma_{\rm opt}$. 

\begin{theorem} \label{Theorem: Optimal dis distribution}
The cdf $F_{\Gamma_{\rm opt}}\paren{\gamma}$ and the pdf $f_{\Gamma_{\rm opt}}\paren{\gamma}$ of $\Gamma_{\rm opt}$ are given by 
\begin{equation}\label{e_cdfmaxminD}
\begin{split}
F_{\Gamma_{\rm opt}}\paren{\gamma} = \left\{
\begin{array}{ll}
0 & \mbox{ if } 
\gamma < d \\
1-\e{-2\lambda d ^2\paren{\paren{\frac{\gamma}{d}}^2 \arcsec\paren{\frac{\gamma}{d}} - \sqrt{\paren{\frac{\gamma}{d}}^2-1}}} & \mbox{ if } 
\gamma \geq d
\end{array} \right.
\end{split}
\end{equation}
and
\begin{equation}\label{e_pdfmaxminD}
\begin{split}
f_{\Gamma_{\rm opt}}\paren{\gamma} & = 4 \lambda  \gamma \arcsec\left(\frac{\gamma}{d}\right) \e{-2\lambda d^2  \left(\paren{\frac{\gamma}{d}}^2\arcsec\left(\frac{\gamma}{d}\right) - \sqrt{\paren{\frac{\gamma}{d}}^2-1}\right)}\I{\gamma \geq d}
\end{split}.
\end{equation}
\end{theorem}
\begin{IEEEproof}
See Appendix \ref{Appendix: optimal dis distribution Proof}.
\end{IEEEproof}

There are several important remarks worth to mention about the functional forms of $F_{\Gamma_{\rm opt}}$ and $f_{\Gamma_{\rm opt}}$ given in Theorem \ref{Theorem: Optimal dis distribution}. Let $g(x)$ be the function defined as $g(x) = x^2\arcsec(x) - \sqrt{x^2-1}$ for $x \geq 1$. It is easy to see that $g(1) = 0$ and $g^\prime (x) = 2x\arcsec(x)$, which is the first derivative of $g(x)$ with respect to $x$. Since $g^\prime (x) > 0$ for all $x > 1$, we conclude that $g(x)$ is a strictly increasing and positive function of $x$ over $\paren{1, \infty}$. Further, it can also be seen that $\lim_{x \ra \infty} \frac{g(x)}{x^2} = \frac{\pi}{2}$. This shows that the exponents in \eqref{e_cdfmaxminD} and \eqref{e_pdfmaxminD} are non-positive for all values of $\gamma \geq d$ with the tail of $f_{\Gamma_{\rm opt}}$ decaying according to $\lim_{\gamma \ra \infty} - \frac{\ln\paren{f_{\Gamma_{\rm opt}}\paren{\gamma}}}{\gamma^2} = \pi \lambda$. Hence, all the moments of $\Gamma_{\rm opt}$ exist although they may not be calculated in closed form. For example, $\ES{\Gamma_{\rm opt}}$ can be obtained numerically by using \eqref{e_cdfmaxminD} as below
\begin{eqnarray*}
\ES{\Gamma_{\rm opt}} &=& \int_{0}^\infty\PR{\Gamma_{\rm opt} > \gamma} \diff \gamma \\
&=& d\int_1^\infty\e{-2\lambda d^2\paren{\gamma^2\arcsec\paren{\gamma} - \sqrt{\gamma^2-1}}} \diff \gamma, 
\end{eqnarray*}
which cannot be reduced to a closed form. 

Secondly, since $g(x) > 0$ for all $x>1$, it also holds that $\lim_{\lambda \ra \infty} F_{\Gamma_{\rm opt}}\paren{\gamma} = 1$ for $\gamma > d$ and $\lim_{\lambda \ra \infty} F_{\Gamma_{\rm opt}}\paren{\gamma} = 0$ for $\gamma \leq d$. Therefore, $\Gamma_{\rm opt}$ converges in distribution to a deterministic variable having value $d$. This behaviour coincides with our intuition that there is always a relay in any disc with positive radius centered around the origin when the relay intensity becomes large. Indeed, it can be shown that $\Gamma_{\rm opt}$ converges to $d$ almost surely as $\lambda$ grows without a bound by using Borel-Cantelli lemma \cite{Billingsley95}.    

\begin{figure*}[!t]
\begin{minipage}[t]{\textwidth}
\begin{center}
\includegraphics[scale=0.25]{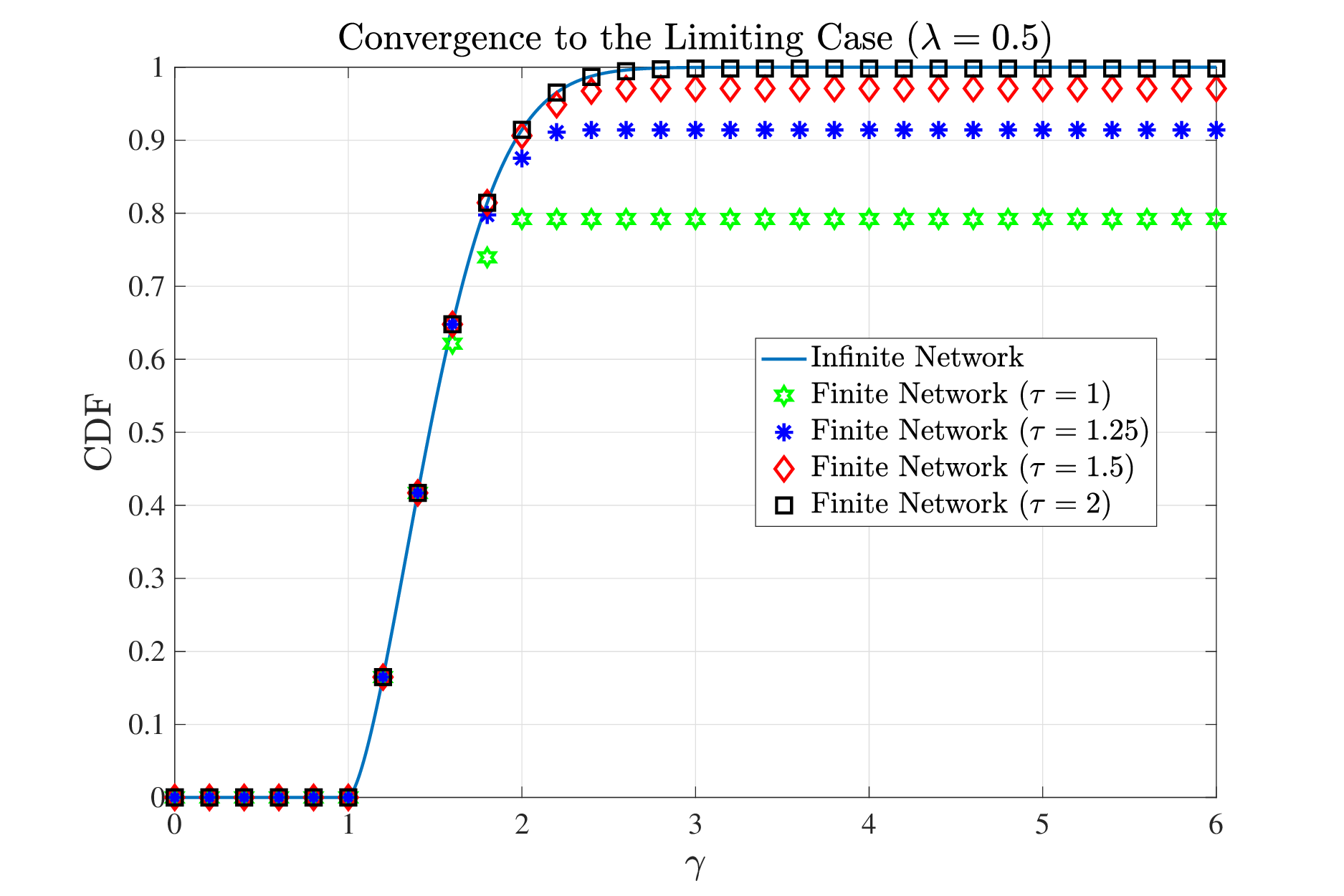}
\hspace{\fill}
\includegraphics[scale=0.25]{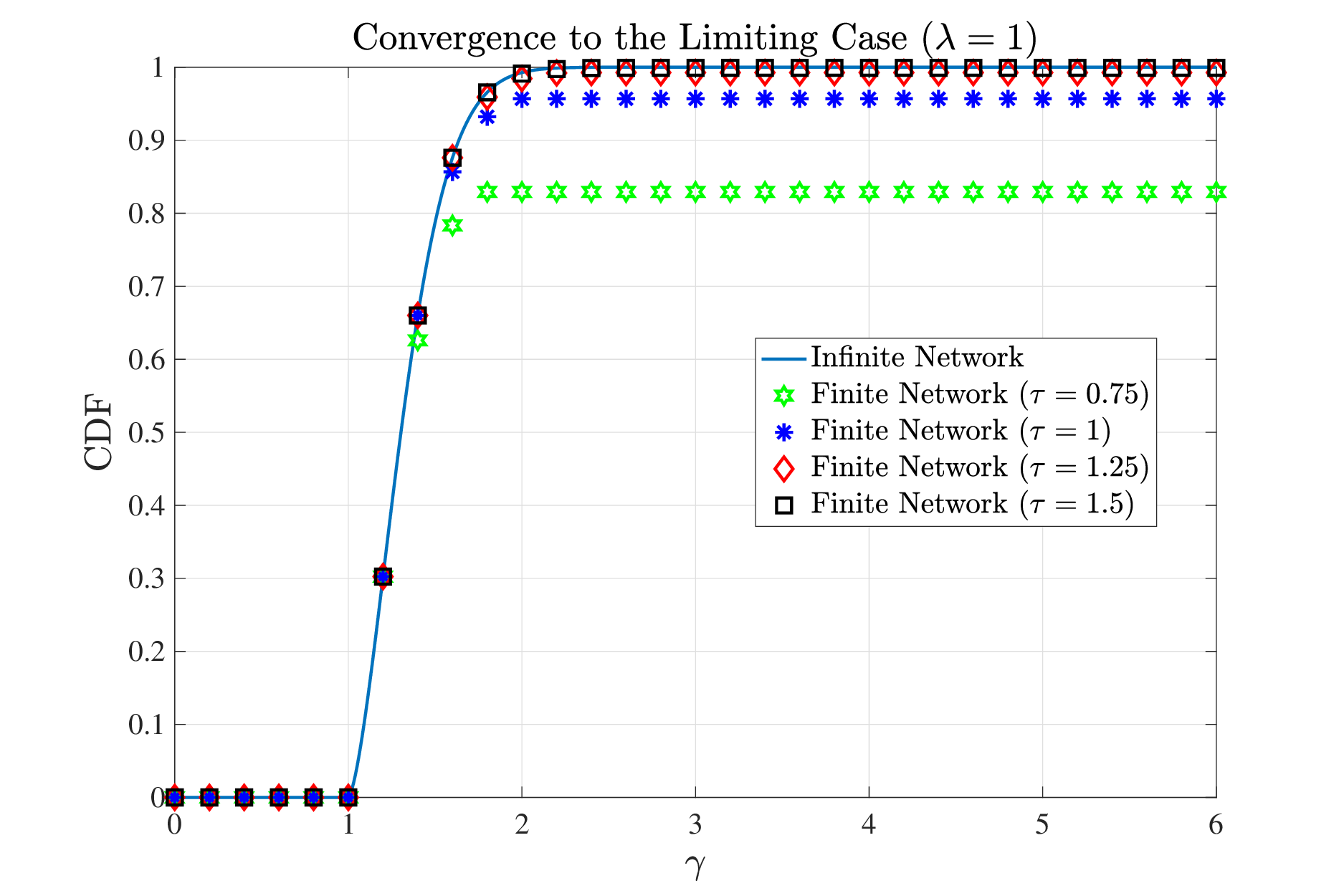}
\end{center}
\end{minipage}
\caption{Convergence to the limiting case.}
\label{Fig: Convergence}
\end{figure*}

Finally, the proof given in Appendix \ref{Appendix: optimal dis distribution Proof} for Theorem \ref{Theorem: Optimal dis distribution}  also reveals the {\em pre-limit} distribution for the minimum of relay selection function in the {\em finite} network case. Let $\Gamma_{{\rm opt}, \tau} = \min_{\vec{X} \in \Phi \cap \mathcal{B}\paren{\vec{0}, \tau}} \relayselect\paren{\vec{X}}$, where $\mathcal{B}\paren{\vec{0}, \tau}$ is the closed disc centered around the origin $\vec{0}$ and having radius $\tau$. Then, the cdf $F_{\Gamma_{{\rm opt}, \tau}}\paren{\gamma}$ can be expressed as 
\begin{eqnarray*}
F_{\Gamma_{{\rm opt}, \tau}}\paren{\gamma} = 1 - \e{-\lambda \pi \tau^2 F_{\Gamma}\paren{\gamma}},
\end{eqnarray*}
where $F_{\Gamma}(\gamma)$ is given by \eqref{Eqn: Distance Distribution} in Appendix D. The convergence of $F_{\Gamma_{{\rm opt}, \tau}}\paren{\gamma}$ to $F_{\Gamma_{\rm opt}}\paren{\gamma}$ as $\tau$ increases is shown in Fig. \ref{Fig: Convergence}, which is quite quick.

The optimum multi-user multi-relay selection problem for random spatial networks with interference is not tractable due to coupling between local relay selection processes at the active interfering links. However, the distribution convergence results shown above hints at an efficient sub-optimum heuristic for multi-user multi-relay selection in the presence of interference. If the layer 2 MAC protocol can achieve a certain minimum separation, sufficiently larger than the typical inter-relay distances, between interfering active transmitter-receiver pairs, then the effect of interference can be mitigated to a large extent in the relay selection process and the local relay selection processes can be decoupled from each other. That is, the differential interference at the optimum relay node (chosen by ignoring interference) and at other relays stays below a certain threshold with high probability to overcome the distance advantage provided by the interference-free optimum relay position when the MAC protocol can sufficiently mitigate interference in the vicinity of active communication links. With the suggested decoupling, the results in our paper can be directly applied to approximate the rate and outage performance of the optimum multi-user multi-relay selection policy by running locally optimum relay selection policies, restricted to a small region around each active link as in the interference-free case. This is specifically a co-design issue of MAC and relay selection algorithms, whose complete analysis is not within the scope of this paper.

As seen in \eqref{Eqn: Conditional Average Rate}, an important determinant of the system performance is the random variable $S_{\rm opt} = \snr \cdot G\paren{\Gamma_{\rm opt}}$. For a non-negative non-increasing path-loss function, the cdf of $S_{\rm opt}$ can be written as
\begin{eqnarray}
F_{S_{\rm opt}}(s) = 1 - F_{\Gamma_{\rm opt}}\paren{G^{-1}\paren{\frac{s}{\snr}}}, \label{Eqn: Sopt cdf}
\end{eqnarray}
where $G^{-1}(s)$ is defined as $G^{-1}(s) = \inf\brparen{x\geq 0: G(x) \leq s}$. We note that $F_{S_{\rm opt}}(s) = 1$ for $s \geq \snr \cdot G(d)$ since $G$ is non-increasing and $\Gamma_{\rm opt} \geq d$. We will make use of $F_{S_{\rm opt}}(s)$ to calculate $\Pout\paren{\pol_{\rm opt}}$ below. Further, if $G$ is monotone decreasing and continuous, the pdf of $S_{\rm opt}$ can be written as     
\begin{eqnarray}
f_{S_{\rm opt}}(s) = \frac{1}{\snr} f_{\Gamma_{\rm opt}}\paren{G^{-1}\paren{\frac{s}{\snr}}}\abs{\frac{1}{G^\prime\paren{G^{-1}\paren{\frac{s}{\snr}}}}}, \label{Eqn: Sopt pdf for monotone G}
\end{eqnarray}
which can used to calculate $\Rave\paren{\pol_{\rm opt}}$ for the power-law decaying path-loss models such as $G(x) = \frac{1}{x^\alpha}$ or $G(x) = \frac{1}{1+x^\alpha}$ \cite{Ak16a, Inaltekin09, Ak16b, Inaltekin12, Inaltekin14, Inaltekin18}.

\subsection{Average Rate}\label{ss_tp}
Now, we provide expressions for the average rate achieved by the optimum relay selection policy $\pol_{\rm opt}$. We will consider both no-fading and Rayleigh fading cases. We recall that we consider a deterministic normalized fading gain with unit power for the no-fading case, and we take $H \sim \mathcal{CN}(0,1)$ for the the Rayleigh fading case. 

%
%
For $\rateave\paren{\pol_{\rm opt}}$, we can write
%
\begin{eqnarray}
\rateave\paren{\pol_{\rm opt}} &=& \frac12 \EW\sqparen{\log_2\paren{1 + \snr \abs{H}^2G\paren{\Gamma_{\rm opt}}} \Big| \Phi} \nonumber \\
&=& \left\{
\begin{array}{ll}
\frac{1}{2}\log_2\left(1+ \snr \cdot G\paren{\Gamma_{\rm opt}}\right) &\text{if no-fading}\\
\frac{1}{2\ln2} \e{\frac{1}{\snr \cdot G\paren{\Gamma_{\rm opt}}}} \text{E}_1\left(\frac{1}{\snr \cdot G\paren{\Gamma_{\rm opt}}}\right) &\text{if Rayleigh fading}
\end{array}, \right. \label{throughput given phi}
\end{eqnarray}
where $\text{E}_1(x) = \int_1^\infty \frac{\e{-tx}}{t} \diff t$, $x>0$, is the exponential integral as defined in Section \ref{Section: Mid-Point Relay Selection}. 
%
%
%
%
%
%
%
%
The expression for $\rateave\paren{\pol}$ for any relay selection policy $\pol$ in the Rayleigh fading case was obtained in Appendix \ref{Appendix: Optimum Rave Upper Bounds} while deriving an upper bound on $\Rave\paren{\pol_{\rm opt}}$, which follows from integration-by-parts and change of variables.       
%
%
From {Definition \ref{Def: Average Rate}}, the average rate, averaged over relay locations, can be calculated as
%
%
%
\begin{equation}\label{e_avgtp1}
\begin{split}
\Rave\paren{\pol_{\rm opt}} 
& = \left\{
\begin{array}{ll}
\frac{1}{2}\int_{d}^{\infty}\log_2\left(1+\snr \cdot G(\gamma) \right)f_{\Gamma_{\rm opt}}(\gamma) \diff \gamma  &\text{if no-fading}\\
\frac{1}{2\ln2}\int_d^{\infty} \e{\frac{1}{\snr \cdot G(\gamma)}} \text{E}_1\left(\frac{1}{\snr \cdot G(\gamma)}\right)f_{\Gamma_{\rm opt}}(\gamma) \diff \gamma &\text{if Rayleigh fading}
\end{array}, \right.
\end{split}
\end{equation}
where $f_{\Gamma_{\rm opt}}$ is as given in \eqref{e_pdfmaxminD}. On the other hand, for a monotone decreasing and continuous path-loss function, $\Rave\paren{\pol_{\rm opt}}$ can be expressed according to   
%
\begin{equation}\label{e_avgtp2}
\begin{split}
\Rave\paren{\pol_{\rm opt}} 
& = \left\{
\begin{array}{ll}
\frac{1}{2}\int_{0}^{\snr \cdot G\paren{d}}\log_2\left(1+s\right)f_{S_{\rm opt}}(s) \diff s  &\text{if no-fading}\\
\frac{1}{2\ln2}\int_0^{\snr \cdot G\paren{d}} \e{\frac{1}{s}} \text{E}_1\left(\frac{1}{s}\right)f_{S_{\rm opt}}(s) \diff s &\text{if Rayleigh fading}
\end{array} \right.
\end{split}
\end{equation}
by using $f_{S_{\rm opt}}(s)$ given in \eqref{Eqn: Sopt pdf for monotone G}.  To the best of our knowledge, it is not possible to reduce the integral expressions in \eqref{e_avgtp1} and \eqref{e_avgtp2} for $\Rave\paren{\pol_{\rm opt}}$ to a closed-form. However, these single integrals can be evaluated very quickly and efficiently by using standard numerical integration techniques. 

\subsection{Outage Probability}\label{ss_out}
Recall that the {outage} probability $\Pout\paren{\pol}$ for a relay selection policy $\pol$ is defined to be the probability that the rate  falls below a certain predetermined threshold $\rho$. From {Definition~\ref{Def: Outage Probability}}, we write
$\Pout\paren{\pol_{\rm opt}} = \PR{\rateave\paren{\pol_{\rm opt}} \leq \rho}$ for the optimum relay selection policy $\pol_{\rm opt}$. 
For the no-fading case, by using \eqref{throughput given phi}, we have
\begin{eqnarray}\label{e_pout_nofad}
\Pout\paren{\pol_{\rm opt}} &=& \PR{\frac{1}{2}\log_2\left(1+\snr \cdot G\paren{\Gamma_{\rm opt}}\right) \leq \rho} \nonumber \\   
 &=& \left\{
\begin{array}{ll}
\hspace{-0mm}F_{S_{\rm opt}}\left( 2^{2\rho} -1 \right)& \hspace{-0mm} \mbox{if } \rho < \frac 1 2 \log_2\left(1+\snr \cdot G(d)\right)\\
\hspace{-0mm} 1& \hspace{-0mm}\text{if } \rho \geq \frac 1 2 \log_2\left(1+\snr \cdot G(d)\right)
\end{array},\right.
\end{eqnarray}
where $F_{S_{\rm opt}}$ is as given in \eqref{Eqn: Sopt cdf}. 
Similarly, for the Rayleigh fading case, we have
\begin{eqnarray}\label{e_pout_ray}
\Pout\paren{\pol_{\rm opt}} &=& \PR{\frac{1}{2\ln2} \e{\frac{1}{\snr\cdot G\paren{\Gamma_{\rm opt}}}} \text{E}_1\left(\frac{1}{\snr\cdot G\paren{\Gamma_{\rm opt}}}\right) \leq \rho} \nonumber \\ 
&=& \left\{
\begin{array}{ll}
F_{S_{\rm opt}}\left( s^\star \right) & \mbox{if } s^\star < \snr \cdot G\paren{d}\\
1 &\text{if } s^\star \geq \snr \cdot G\paren{d}
\end{array},\right.
\end{eqnarray}
where $s^\star$ is the {\em unique} solution for the equation
$\frac{1}{2\ln 2}\e{1/s} \text{E}_1\left(1/s\right) = \rho$.
Uniqueness of $s^\star$ follows from the fact that the function $f(x) = \e{x}\text{E}_1\paren{x}$ defined for $x > 0$ is a continuous and strictly decreasing function of $x$, which attains the values $\lim_{x \ra 0} f(x) = \infty$ (i.e., this can be shown using monotone convergence theorem) and $\lim_{x \ra \infty} f(x) = 0$ (i.e., this can be shown using dominated convergence theorem).  Hence, $s^\star$ can be readily calculated by using numerical methods such as the bisection technique or by using Matlab's \texttt{vpasolve} routine. 
Moreover, by using the inequality $\frac12 \ln\paren{1+\frac{2}{x}} < \e{x}\text{E}_1(x) < \ln\paren{1+\frac{1}{x}}$ (i.e., see  \cite[p.~229,~5.1.20]{Abramowitz1974}), 
we can also bound $\Pout\paren{\pol_{\rm opt}}$ for the Rayleigh fading case according to  
$F_{S_{\rm opt}}\left(2^{2\rho} -1 \right) \leq \Pout\paren{\pol_{\rm opt}} \leq F_{S_{\rm opt}}\left( \frac{2^{4\rho} -1}{2} \right)$. The lower bound on $\Pout\paren{\pol_{\rm opt}}$ coincides with the one that can be obtained by using $\Pout\paren{\pol_{\rm opt}}$ for the no-fading case in \eqref{e_pout_nofad}.      

\section{Relay Selection with Selective Distributed-Feedback} \label{Section: Distributed Relay Selection}
In the previous sections, we have characterized key properties for the optimum relay selection policy, its outage and average rate performance as well as the optimality probability for the mid-point relay selection policy when the relay nodes are randomly distributed over the plane according to an HPPP of any given intensity. This analysis, however, assumes a {\em centralized} operation in which information pertaining to all relay locations is available at the source node (or at a central entity) to solve the optimum relay selection problem in \eqref{Eqn: Relay Selection Problem}. Even though the class of relay selection policies we investigate in this paper only requires relay locations as the CSI, which change at a much slower time-scale than fading, the task of feeding this information back to the source node is still onerous, and hence an impeding factor for physical deployments. One potential means of reducing the feedback load is to adopt a {\em selective distributed-feedback relay selection policy} in which relay nodes, whose locations are given according to $\Phi$, are provided with the autonomy of giving their own feedback decisions independently from other relays in the network based on their local channel quality indicators $\relayselect\paren{\vec{X}}$ for $\vec{X} \in \Phi$.  

\subsection{Threshold Feedback Policies for Relay Selection and Their Statistical Properties} \label{Subsection: Threshold Based Feedback}
We will study simple but practical threshold feedback policies to regulate the feedback load in the network. This class of feedback polices possesses certain optimality properties to maximize data rates \cite{Samarasinghe11a, Inaltekin11, Samarasinghe11b, Samarasinghe11c, Samarasinghe12, Inaltekin13, Samarasinghe13a}. Here, we will utilize them to control the number of relay nodes feeding their channel states back to the source node as a measure of the total feedback load in the network. More explicitly, for any given threshold value $T \geq 0$, we will say that a relay node located at $\vec{X} \in \Phi$ and operating according to a threshold-based selective distributed-feedback relay selection policy with threshold value $T$ will feed its channel quality indicator $\relayselect\paren{\vec{X}}$ back to the source node if and only if $\relayselect\paren{\vec{X}} \leq T$. Hence, the total number of relays feeding back is given by
\begin{eqnarray}
N_{\rm FB} = \sum_{\vec{X} \in \Phi} \I{\relayselect\paren{\vec{X}} \leq T}. \label{Eqn: Random Feedback Load} 
\end{eqnarray}
%
%
The average number of relays feeding back is then equal to 
\begin{eqnarray}
\mu\paren{T} = \ES{N_{\rm FB}}. \label{Eqn: Average Feedback Load 1} 
\end{eqnarray}

The next theorem characterizes the distribution of $N_{\rm FB}$ and the functional form of its average value.
\begin{theorem} \label{Theorem: Feedback}
For any given threshold value $T \geq 0$, $N_{\rm FB}$ is a Poisson distributed random variable whose mean $\mu\paren{T}$ is given according to  
%
\begin{eqnarray} \label{Eqn: Average Feedback Load 2}
\mu\paren{T} = \left\{\begin{array}{ll}0 & \mbox{ if } T < d \\\lambda \pi T^2 - 2d\lambda\sqrt{T^2-d^2} - 2T^2\lambda\arctan\paren{\frac{d}{\sqrt{T^2 - d^2}}} & \mbox { if } T \geq d\end{array}\right..
\end{eqnarray}
\end{theorem}
\begin{IEEEproof}
See Appendix \ref{Appendix: Feedback}. 
\end{IEEEproof}

Using \eqref{Eqn: Average Feedback Load 2}, it can easily be seen that $\mu\paren{d} = 0$ and $\mu\paren{T}$ is a continuous function of $T$. Further, by using \eqref{Eqn: Random Feedback Load} and \eqref{Eqn: Average Feedback Load 1}, it can also be seen that $\mu\paren{T}$ is a monotone increasing function of $T$ with limit $\lim_{T \ra \infty} \mu\paren{T} = \infty$. Hence, for any given feedback load $\mu_0 \geq 0$, we are guaranteed to find a threshold value $T_0$ such that $\mu\paren{T_0} = \mu_0$ by intermediate value theorem. From a network design perspective, this observation shows that the class of threshold-based distributed relay selection policies is rich enough to satisfy any given feedback load constraint on the network by properly allocating a common threshold value to all relay nodes and allowing them to operate autonomously while giving their feedback decisions based on this common threshold value.  

\begin{figure*}[!t]
\begin{minipage}[t]{\textwidth}
\begin{center}
\includegraphics[scale=0.25]{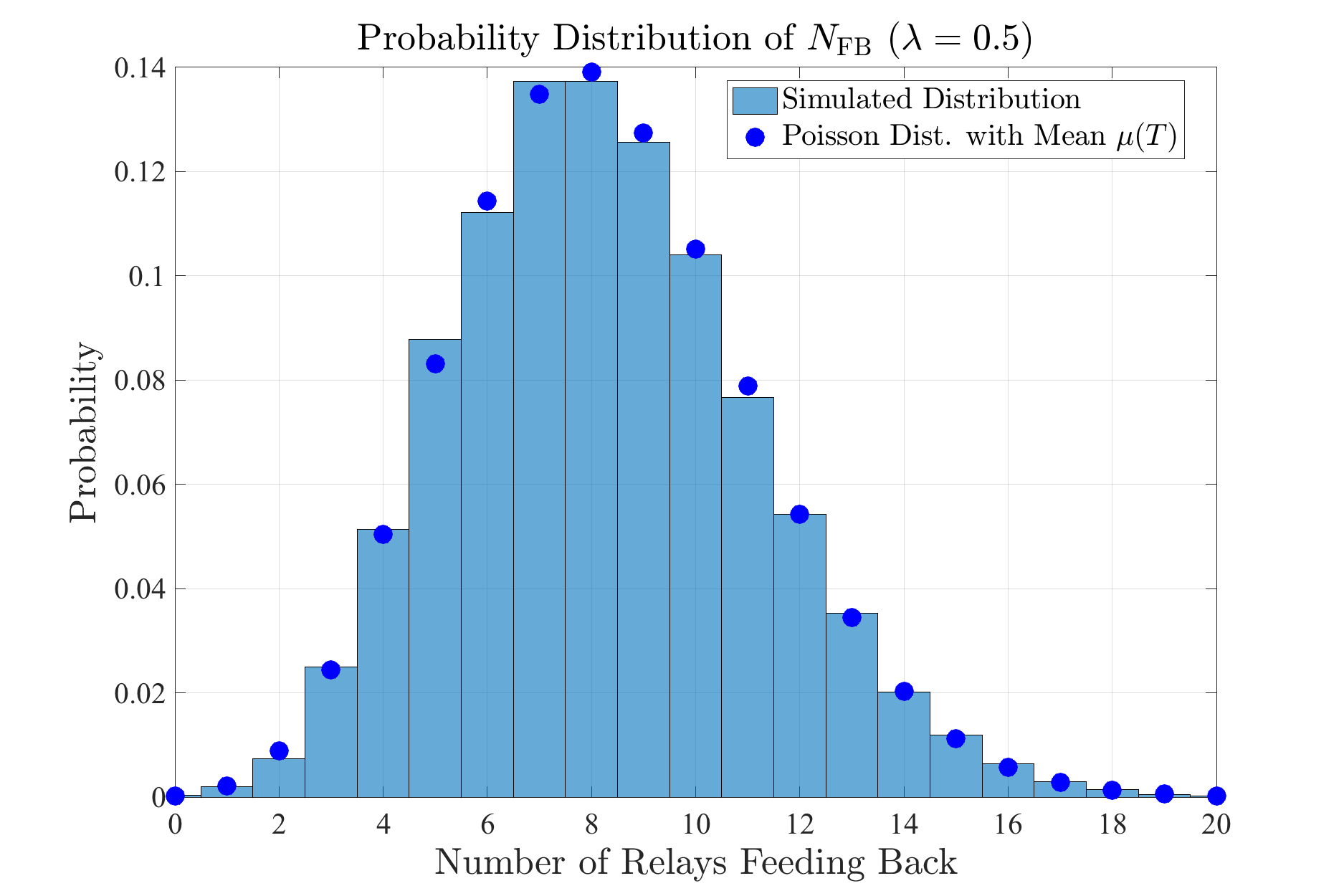}
\hspace{\fill}
\includegraphics[scale=0.25]{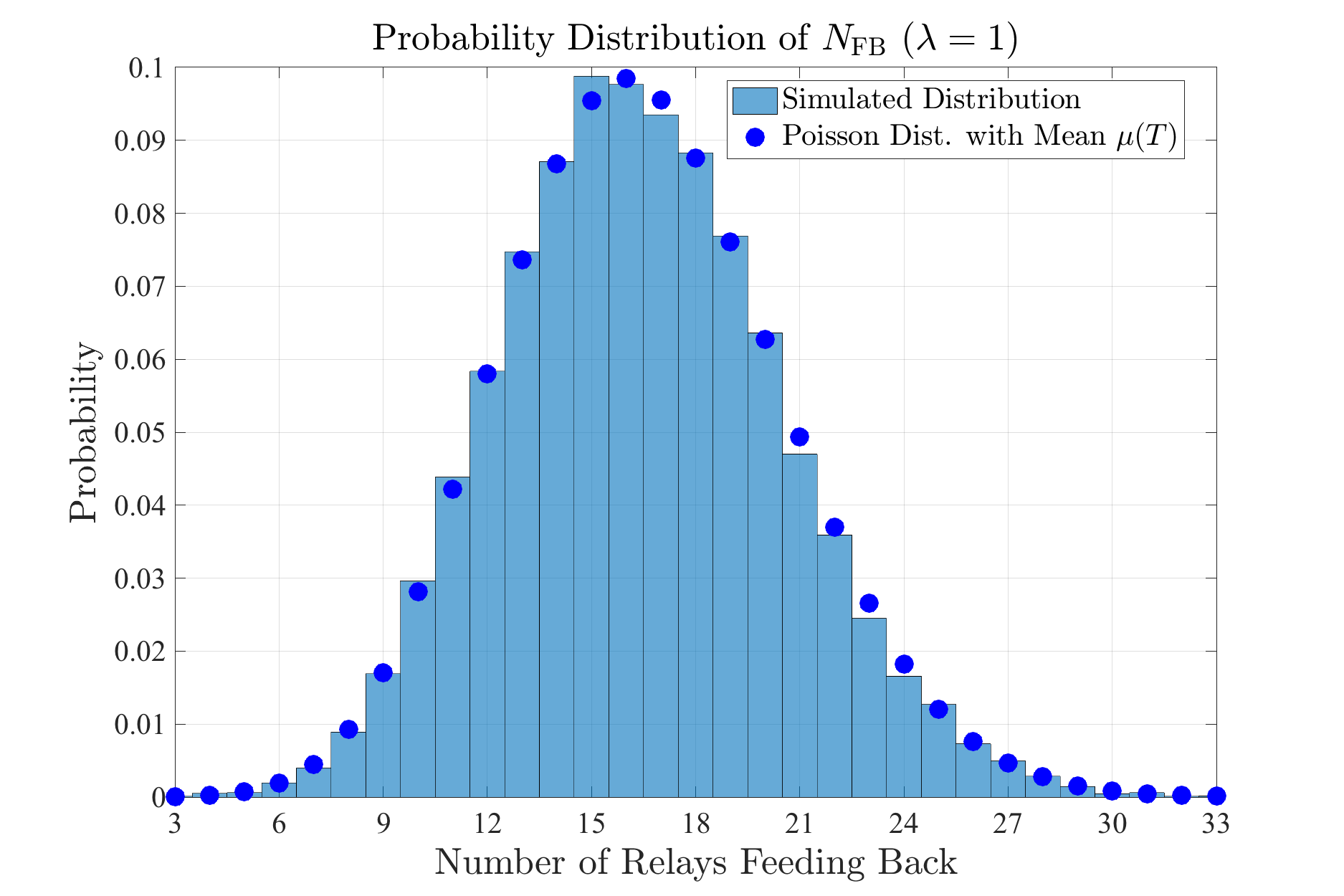}
\end{center}
\end{minipage}
\caption{Probability distribution of the number of relays feeding back for $T=3$ and $d=1$. ($\lambda = 0.5$ for the left-hand side figure and $\lambda = 1$ for the right-hand side figure.)}  \label{Fig: NFB Distribution}
\end{figure*}

\begin{figure*}[!t]
\begin{minipage}[t]{\textwidth}
\begin{center}
\includegraphics[scale=0.25]{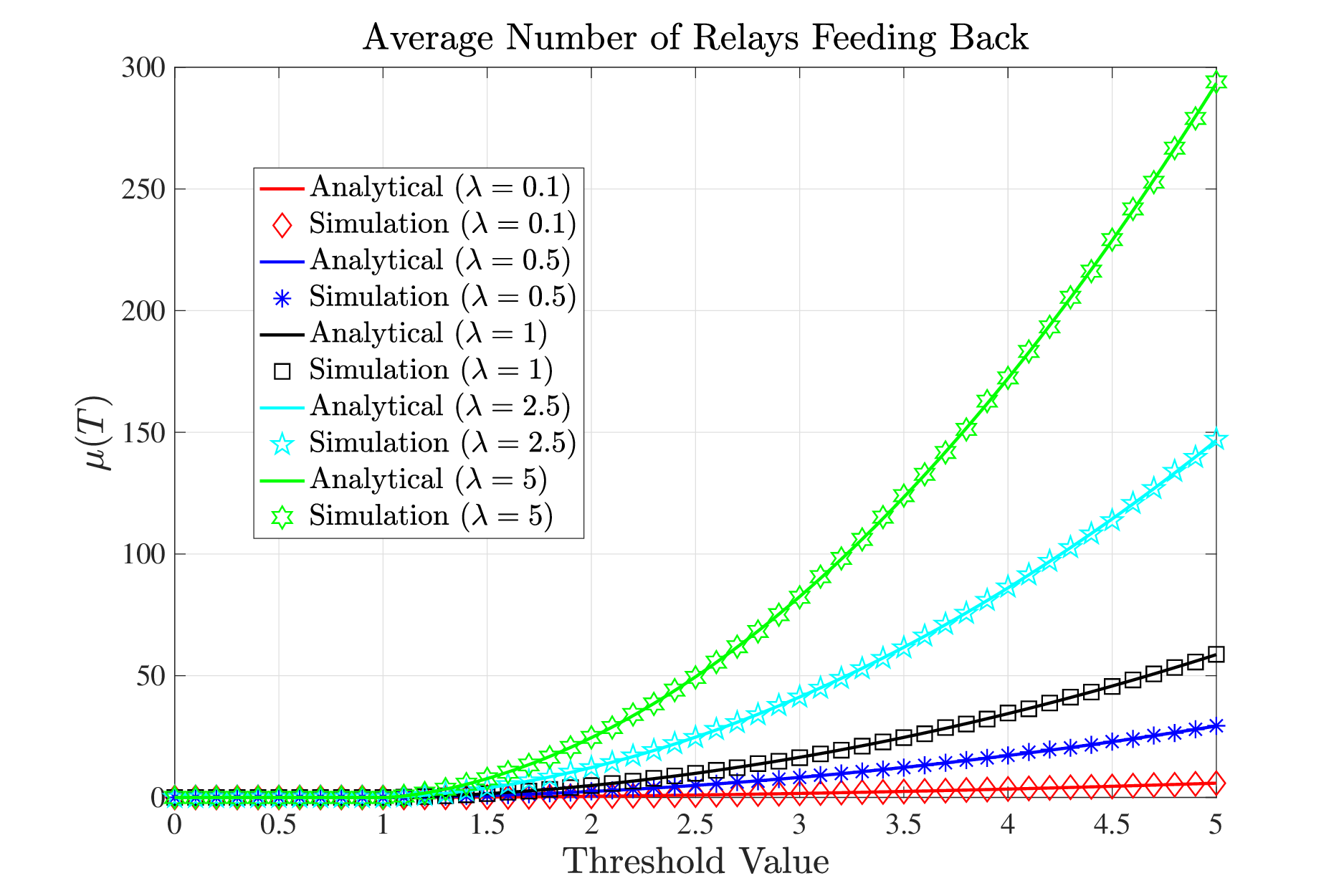}
\hspace{\fill}
\includegraphics[scale=0.25]{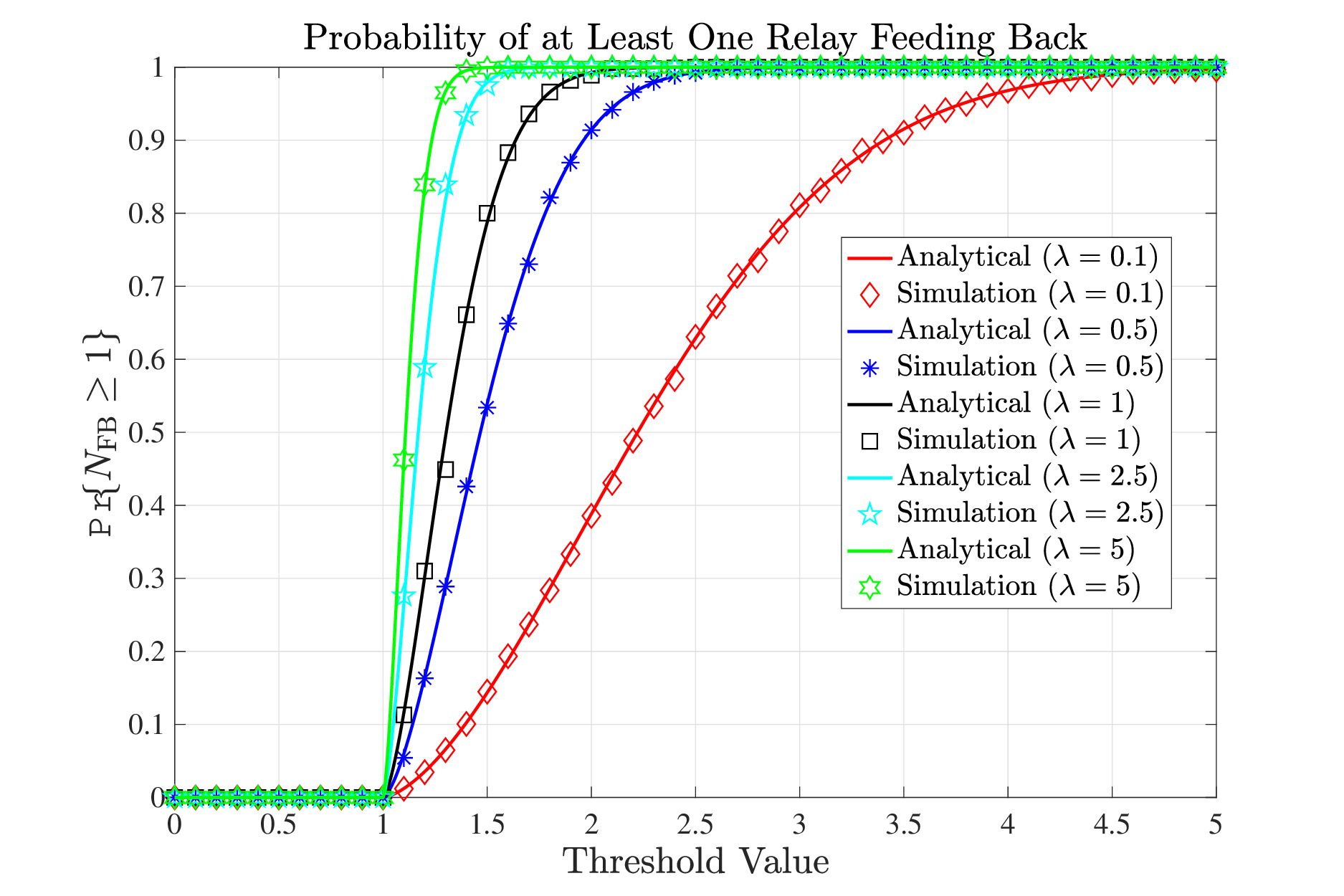}
\end{center}
\end{minipage}
\caption{Average number of relays feeding back and the probability of having at least one relay feeding back for $d=1$ and various values of $\lambda$.}  \label{Fig: muFB}
\end{figure*}

In Fig. \ref{Fig: NFB Distribution}, we plot the simulated distributions of $N_{\rm FB}$ for $\lambda=0.5$ and $\lambda = 1$, and compare them with the Poisson distribution having mean $\mu\paren{T}$. As predicted by Theorem \ref{Theorem: Feedback}, simulated and theoretical distributions match each other perfectly. In Fig. \ref{Fig: muFB}, we plot the average number of relays feeding back $\mu\paren{T}$ and the probability of at least one relay feeding back $\PR{N_{\rm FB} \geq 1}$ as a function of $T$. Again, there is a perfect match between simulated and analytical curves, verifying the predictions in Theorem \ref{Theorem: Feedback}. In particular, $\PR{N_{\rm FB} \geq 1}$ is an important performance indicator for relay-assisted spatial wireless networks with distributed relay selection. The relay with optimum location is always among the relays feeding their channel quality indicators back to the source node if $N_{\rm FB} \geq 1$. Hence, with probability $\PR{N_{\rm FB} \geq 1}$, there is no loss of optimality arising from implementing a threshold-based selective feedback distributed relay selection mechanism. Based on Theorem \ref{Theorem: Feedback}, this probability is equal to $\PR{N_{\rm FB} \geq 1} = 1 - \e{-\mu\paren{T}}$. This shows that the performance loss due to having a threshold-based selective feedback distributed relay selection mechanism diminishes exponentially fast as a function of the feedback load in the network, which we measure in terms of the average number of relays feeding their channel quality indicators back to the source node.  Numerically, we have $\PR{N_{\rm FB} \geq 1} \leq 0.99$ whenever $\mu\paren{T} \geq 5$. As a result, for a given relay intensity and source-destination separation, choosing the threshold value such that $\mu\paren{T} = 5$ implies almost a negligible performance loss, {\em whilst} providing a massive reduction in the total feedback load required to achieve $\Rave\paren{\pol_{\rm opt}}$ and $\Pout\paren{\pol_{\rm opt}}$.  

\subsection{Average Rate and Outage Probability Achieved}
Next, we obtain analytical expressions for the average rate and outage probability achieved by a threshold-based selective feedback distributed relay selection policy $\pol_{\rm FB}$, analogous to those obtained in Section \ref{Section: Optimum Relay Selection}. Different from previous parts, we will denote the average rate and outage probability by $\Rave\paren{\pol_{\rm FB}, T}$ and $\Pout\paren{\pol_{\rm FB}, T}$ with a slight abuse of notation, respectively, to indicate their dependency on the threshold level $T$. Due to the distributed relay selection mechanism, collection of relay nodes whose channel quality indicators are available at the source node is a {\em random} set with size $N_{\rm FB}$. If $N_{\rm FB} \geq 1$, there is one or more relay nodes feeding back to the source node, and the source selects the best relay among them for data transmission. On the other hand, if $N_{\rm FB} = 0$, there is no relay feeding back to the source, and we assume that the source node does not transmit any data due to lack of CSI in this case. Since $N_{\rm FB} = 0$ for $T < d$ (i.e., no relay feeds back for this range of $T$), we will only analyze the case $T \geq d$ below.  Under these operating conditions, the following theorem provides the analytical expressions for the average rate achieved by $\pol_{\rm FB}$. 
\begin{theorem} \label{Theorem: Average Rate with Feedback}
For a given threshold-based selective feedback distributed relay selection policy $\pol_{\rm FB}$ having a threshold value $T \geq d$, the average rate $\Rave\paren{\pol_{\rm FB}, T}$ is equal to
\begin{eqnarray}
\Rave\paren{\pol_{\rm FB}, T} = \left\{\begin{array}{ll}\frac12 \int_{d}^{T}\log_2\paren{1+\snr \cdot G\paren{\gamma}}f_{\Gamma_{\rm opt}}\paren{\gamma} \diff \gamma & \mbox{if no-fading} \\
\frac{1}{2\ln2}\int_{d}^{T} \e{\frac{1}{\snr \cdot G\paren{\gamma}}}\text{E}_1\paren{\frac{1}{\snr \cdot G\paren{\gamma}}} f_{\Gamma_{\rm opt}}\paren{\gamma} \diff \gamma & \mbox{if Rayleigh fading}\end{array}\right., \label{Eqn: Avg Rate with Feedback}
\end{eqnarray}
where $f_{\Gamma_{\rm opt}}(\gamma)$ is given as in Theorem \ref{Theorem: Optimal dis distribution}.   
\end{theorem}
\begin{IEEEproof}
The proof follows from the equivalence of events $\brparen{N_{\rm FB} \geq 1}$ and $\brparen{\Gamma_{\rm opt} \leq T}$, and writing the rate achieved by $\pol_{\rm FB}$, conditioned on the relay locations, according to 
\begin{eqnarray*}
\rateave\paren{\pol_{\rm FB}, T} = \frac12 \log_2\paren{1 + \snr \cdot G\paren{\Gamma_{\rm opt}}} \I{\Gamma_{\rm opt} \leq T}
\end{eqnarray*}
when there is no fading, and according to
\begin{eqnarray*}
\rateave\paren{\pol_{\rm FB}, T} = \frac{1}{2\ln2} \e{\frac{1}{\snr\cdot G\paren{\Gamma_{\rm opt}}}}\text{E}_1\paren{\frac{1}{\snr \cdot G\paren{\Gamma_{\rm opt}}}} \I{\Gamma_{\rm opt} \leq T}
\end{eqnarray*}
when there is Rayleigh fading by using \eqref{throughput given phi}. 
\end{IEEEproof}

We note that \eqref{Eqn: Avg Rate with Feedback} can be written as 
\begin{eqnarray}
\Rave\paren{\pol_{\rm FB}, T} = \left\{\begin{array}{ll}\frac12 \int_{\snr\cdot G\paren{T}}^{\snr \cdot G(d)} \log_2\paren{1+s}f_{S_{\rm opt}}\paren{s} \diff s & \mbox{if no-fading} \\\frac{1}{2\ln2}\int_{\snr \cdot G\paren{T}}^{\snr \cdot G(d)} \e{\frac1s}\text{E}_1\paren{\frac1s} f_{S_{\rm opt}}\paren{s} \diff s & \mbox{if Rayleigh fading}\end{array}\right., \label{Eqn: Avg Rate with Feedback2}
\end{eqnarray}
where $f_{S_{\rm opt}}(s)$ is given as in \eqref{Eqn: Sopt pdf for monotone G} for monotone decreasing and continuous path-loss functions.  In the next theorem, we provide similar expressions for the outage probability $\Pout\paren{\pol_{\rm FB}, T}$ achieved by $\pol_{\rm FB}$.
\begin{theorem} \label{Theorem: Outage with Feedback}
For a given threshold-based selective feedback distributed relay selection policy $\pol_{\rm FB}$ having a threshold value $T \geq d$, the outage probability $\Pout\paren{\pol_{\rm FB}, T}$ is equal to
\begin{eqnarray}
\lefteqn{\Pout\paren{\pol_{\rm FB}, T}} \hspace{15cm} \nonumber \\
\lefteqn{= \left\{\begin{array}{ll}
\e{-\mu\paren{T}} & \mbox{ if } \rho \leq \frac12 \log_2\paren{1 + \snr\cdot G\paren{T}} \\ 
F_{S_{\rm opt}}\paren{2^{2\rho} - 1} & \mbox{ if } \frac12 \log_2\paren{1 + \snr \cdot G\paren{T}} < \rho < \frac12 \log_2\paren{1 + \snr \cdot G\paren{d}} \\
1 & \mbox{ if } \rho \geq \frac12 \log_2\paren{1 + \snr \cdot G\paren{d}}
\end{array}\right.} \hspace{14cm} \label{Eqn: Outage with Feedback No Fading}
\end{eqnarray}
when there is no fading, where $F_{S_{\rm opt}}$ is given as in \eqref{Eqn: Sopt cdf}. On the other hand, for Rayleigh distributed fading, $\Pout\paren{\pol_{\rm FB}, T}$ is equal to  
\begin{eqnarray}
\Pout\paren{\pol_{\rm FB}, T} = \left\{\begin{array}{ll}
\e{-\mu\paren{T}} & \mbox{ if } s^\star \leq \snr \cdot G\paren{T} \\ 
F_{S_{\rm opt}}\paren{s^\star} & \mbox{ if } \snr \cdot G\paren{T} < s^\star < \snr \cdot G\paren{d} \\
1 & \mbox{ if } s^\star \geq \snr \cdot G\paren{d}
\end{array},\right. \label{Eqn: Outage with Feedback Rayleigh Fading}
\end{eqnarray}
where $s^\star$ is the unique solution of the equation $\frac{1}{2\ln2} \e{1/s}\text{E}_1\paren{1/s} = \rho$.  
\end{theorem}
\begin{IEEEproof}
We first consider the no-fading case. In this case, we have
\begin{eqnarray}
\Pout\paren{\pol_{\rm FB}, T} &=& \PR{\rateave\paren{\pol_{\rm FB}, T} \leq \rho} \nonumber \\
&=& \PR{\frac12 \log_2\paren{1 + \snr \cdot G\paren{\Gamma_{\rm opt}}} \I{\Gamma_{\rm opt} \leq T} \leq \rho}. \label{Eqn: Feedback Outage Probability No Fading 1}
\end{eqnarray}
Using \eqref{Eqn: Feedback Outage Probability No Fading 1}, it can be seen that we can write the outage event as the union of two disjoint events according to
\begin{eqnarray}
\brparen{\rateave\paren{\pol_{\rm FB}, T} \leq \rho} = \brparen{\Gamma_{\rm opt} > T} \bigcup \paren{\brparen{\Gamma_{\rm opt} \leq T} \bigcap \brparen{\frac12 \log_2\paren{1 + \snr \cdot G\paren{\Gamma_{\rm opt}}} \leq \rho}}. \hspace{-1.0cm} \nonumber 
\end{eqnarray}
Hence, $\Pout\paren{\pol_{\rm FB}, T}$ is equal to 
\begin{eqnarray}
\Pout\paren{\pol_{\rm FB}, T} &=& \PR{\Gamma_{\rm opt} > T} + \PRP{\brparen{\Gamma_{\rm opt} \leq T} \bigcap \brparen{\frac12 \log_2\paren{1 + \snr \cdot G\paren{\Gamma_{\rm opt}}} \leq \rho}} \nonumber \\
&=& \e{-\mu\paren{T}} + \PRP{\brparen{\Gamma_{\rm opt} \leq T} \bigcap \brparen{\frac12 \log_2\paren{1 + \snr \cdot G\paren{\Gamma_{\rm opt}}} \leq \rho}} \nonumber \\ 
&=& \e{-\mu\paren{T}} + \PR{G^{-1}\paren{\frac{2^{2\rho}-1}{\snr}} \leq \Gamma_{\rm opt} \leq T}.  \label{Eqn: Feedback Outage Probability Rayleigh Fading 1}
\end{eqnarray}

If $G^{-1}\paren{\frac{2^{2\rho}-1}{\snr}} \geq T$, then the second term in \eqref{Eqn: Feedback Outage Probability Rayleigh Fading 1} is equal to zero, and we have $\Pout\paren{\pol_{\rm FB}, T} = \e{-\mu\paren{T}}$. This case corresponds to $\rho \leq \frac12 \log_2\paren{1 + \snr\cdot G\paren{T}}$, which is the first condition in \eqref{Eqn: Outage with Feedback No Fading}.  If $d < G^{-1}\paren{\frac{2^{2\rho}-1}{\snr}} < T$, we have $\Pout\paren{\pol_{\rm FB}, T} = \PR{\Gamma_{\rm opt} \geq G^{-1}\paren{\frac{2^{2\rho}-1}{\snr}}} = F_{S_{\rm opt}}\paren{2^{2\rho} -1}$. This case corresponds to the second condition in \eqref{Eqn: Outage with Feedback No Fading}.
%
%
Finally, if $G^{-1}\paren{\frac{2^{2\rho}-1}{\snr}} \leq d$, we have $\Pout\paren{\pol_{\rm FB}, T} = \PR{\Gamma_{\rm opt} \geq G^{-1}\paren{\frac{2^{2\rho}-1}{\snr}}} = 1$ since $\Gamma_{\rm opt}$ is always greater than $d$. This case corresponds to $\rho \geq \frac12 \log_2\paren{1 + \snr \cdot G\paren{d}}$, which is the the third condition in \eqref{Eqn: Outage with Feedback No Fading}.     

For the Rayleigh fading case, we have
\begin{eqnarray}
\Pout\paren{\pol_{\rm FB}, T} &=& \e{-\mu\paren{T}} + \PRP{\brparen{\Gamma_{\rm opt} \leq T} \bigcap \brparen{\frac{1}{2\ln 2} f\paren{\frac{1}{\snr\cdot G\paren{\Gamma_{\rm opt}}}} \leq \rho}}, 
\end{eqnarray}
where $f(x)$ is as defined in Section \ref{Section: Mid-Point Relay Selection}, i.e., $f(x) \defeq \e{x}\text{E}_1(x)$ for $x > 0$. As explained earlier in the paper, 
%
%
$f(x)$ is a continuous and strictly decreasing function of $x$ with limiting values $\lim_{x \ra 0} f(x) = \infty$ and $\lim_{x \ra \infty} f(x) = 0$. Hence, defining $\gamma^\star$ as the value for the first time $f\paren{\frac{1}{\snr \cdot G\paren{\gamma}}}$ is above $2 \rho \ln2$, i.e., $\gamma^\star \defeq \inf\brparen{\gamma > 0: f\paren{\frac{1}{\snr \cdot G\paren{\gamma}}} \geq 2 \rho \ln2}$, 
%
%
analyzing three cases $\gamma^\star \leq d$, $d < \gamma^\star < T$ and $\gamma^\star \geq T$ separately, and defining $s^\star \defeq \snr \cdot G\paren{\gamma^\star}$, we obtain \eqref{Eqn: Outage with Feedback Rayleigh Fading}.
\end{IEEEproof}

Several important remarks are in order about Theorem \ref{Theorem: Outage with Feedback} characterizing the outage probability  achievable with a threshold-based selective feedback distributed relay selection policy. In particular, we observe two regimes emerging in Theorem \ref{Theorem: Outage with Feedback} when we vary the target rate for both with and without fading. When $\rho \leq \frac12 \log_2\paren{1 + \snr\cdot G\paren{T}}$ without fading or $s^\star \leq \snr \cdot G\paren{T}$ with fading where $s^\star$ is the solution for $\frac{1}{2\ln2} \e{1/s}\text{E}_1\paren{1/s} = \rho$, the outage probability is equal to $\Pout\paren{\pol_{\rm FB}, T} = \e{-\mu\paren{T}}$. This is the {\em feedback-limited} regime in which $\Pout\paren{\pol_{\rm FB}, T}$ is the same for both no-fading and fading cases, depends only on the average number of relays feeding back (which in turn depends on $T$, $\lambda$ and $d$), and is independent of the target rate and fading behaviour. In this regime, the threshold value is set so {\em small} that we are guaranteed to achieve the target rate whenever there is at least one relay feeding its channel quality indicator back to the source node. We recall that the smaller $T$ is, the better CQI relays feeding back have, and achieve higher rates. 

The second regime is the {\em rate-limited} regime that emerges when $\frac12\log_2\paren{1+\snr \cdot G\paren{T}} < \rho < \frac12\log_2\paren{1+ \snr \cdot G\paren{d}}$ for the no-fading case and when $\snr \cdot G\paren{T} < s^\star < \snr \cdot G\paren{d}$ for the fading case. In this regime, the outage probability is equal to $\Pout\paren{\pol_{\rm FB}, T} = F_{S_{\rm opt}}\paren{2^{2\rho}-1}$ for the no-fading case and $\Pout\paren{\pol_{\rm FB}, T} = F_{S_{\rm opt}}\paren{s^\star}$ for the fading case, and hence $\Pout\paren{\pol_{\rm FB}, T}$ is a function of $\rho$, the same for the all-feedback and selective feedback cases, and independent of $T$. $\Pout\paren{\pol_{\rm FB}, T}$ also depends on the fading behaviour since $s^\star$ is not necessarily the same with $2^{2\rho} - 1$. In this regime, the threshold value is set so {\em big} that we are guaranteed to have at least one relay feeding its CQI back to the source node whenever any relay achieves the target rate.

\section{Numerical Results and Discussion} \label{Section: Numerical Results}
In this section, we will present our numerical results in order to illustrate the performance of $\pol_{\rm opt}$ with and without feedback limitations. 
%
All distances are normalized to a {\it unit} distance. 
%
%
The path-loss exponent $\alpha$ is set to $4$, and the path-loss function is taken to be $G(x) = \frac{1}{x^4}$. 
%
%
For the fading scenario, we consider Rayleigh fading coefficients with unit power. To benchmark the performance of the optimum relay selection policy, we also consider the network performance when the relay is chosen in such a way that it is closest to the mid-point and to the destination.\footnote{The performance of the relay selection policy choosing the relay closest to the source is the same with the performance of the one choosing the relay closest to the destination due to symmetry in the problem. Hence, the performance curves pertaining to the former one are not included.} 

The performance curves for the mid-point and closest-to-destination policies are obtained by using the distribution functions given in \eqref{Eqn: Mid Point CDF}-\eqref{Eqn: Closest-to-Destination PDF}, where $\Gamma_{\rm  mid} \defeq \relayselect\paren{\vec{X}_{\rm mid}}$, $\Gamma_{\rm  C2D} \defeq \relayselect\paren{\vec{X}_{\rm C2D}}$, $\vec{X}_{\rm C2D}$ is the relay location closest to the destination, and $F_{\rm NN}\paren{\psi} = \paren{1 - \e{-\lambda\pi\psi^2}}\I{\psi \geq 0}$ is the nearest-neighbor cdf. The derivations are given in Appendix \ref{Appendix: Distributions for Benchmark Strategies}.

\begin{figure*}[!t]
\begin{eqnarray}
F_{\Gamma_{\rm mid}}\paren{\gamma} = \paren{F_{\rm NN}\paren{\sqrt{\gamma^2 - d^2}} - 2\lambda\int_{\paren{\gamma-d}^2}^{\gamma^2 - d^2} \e{-\lambda \pi x}\arccos\paren{\frac{\gamma^2 - x - d^2}{2d\sqrt{x}}}\diff x}\I{\gamma \geq d} \label{Eqn: Mid Point CDF}
\end{eqnarray}
\begin{eqnarray}
f_{\Gamma_{\rm mid}}\paren{\gamma} = \paren{ 4\lambda\gamma\int_{\paren{\gamma-d}^2}^{\gamma^2 - d^2} \frac{\e{-\lambda \pi x}}{\sqrt{4d^2x - \paren{\gamma^2 - x - d^2}^2}}\diff x}\I{\gamma \geq d} \label{Eqn: Mid Point PDF}
\end{eqnarray}
\begin{eqnarray}\label{Eqn: Closest-to-Destination CDF}
F_{\Gamma_{\rm C2D}}\paren{\gamma} = \paren{F_{\rm NN}\paren{\gamma - 2d} + \lambda\int_{\paren{\gamma - 2d}^2}^{\gamma^2} \e{-\lambda\pi x} \arccos\paren{\frac{x + 4d^2 - \gamma^2}{4d\sqrt{x}}}\diff x}\I{\gamma \geq d}
\end{eqnarray}
\begin{eqnarray} \label{Eqn: Closest-to-Destination PDF}
f_{\Gamma_{\rm C2D}}\paren{\gamma} = 2\lambda\gamma\paren{\arccos\paren{\frac{d}{\gamma}}\e{-\lambda\pi\gamma^2} + \int_{\paren{\gamma - 2d}^2}^{\gamma^2} \frac{\e{-\lambda\pi x}}{\sqrt{16d^2x - \paren{x + 4d^2 - \gamma^2}^2}}\diff x}\I{\gamma \geq d}
\end{eqnarray}
\hrulefill
\end{figure*}

We note that the change in scale in how we measure the distances will only scale the $\snr$ values in our analysis below, without changing the emerging fundamental trends in the network performance. Some recent studies indicated that the power-law functions model the path-loss more accurately for distances larger than $350$ meters \cite{Andrews18}. Hence, for the relay densities below, one can consider the unit distance to be on the order of kilometers for higher practical relevance of the presented results.

\begin{figure}
	\centering
	\subfloat[Outage probability vs $\lambda$.]{
		\label{f_tpvslam5}
		\includegraphics[width=0.5\textwidth]{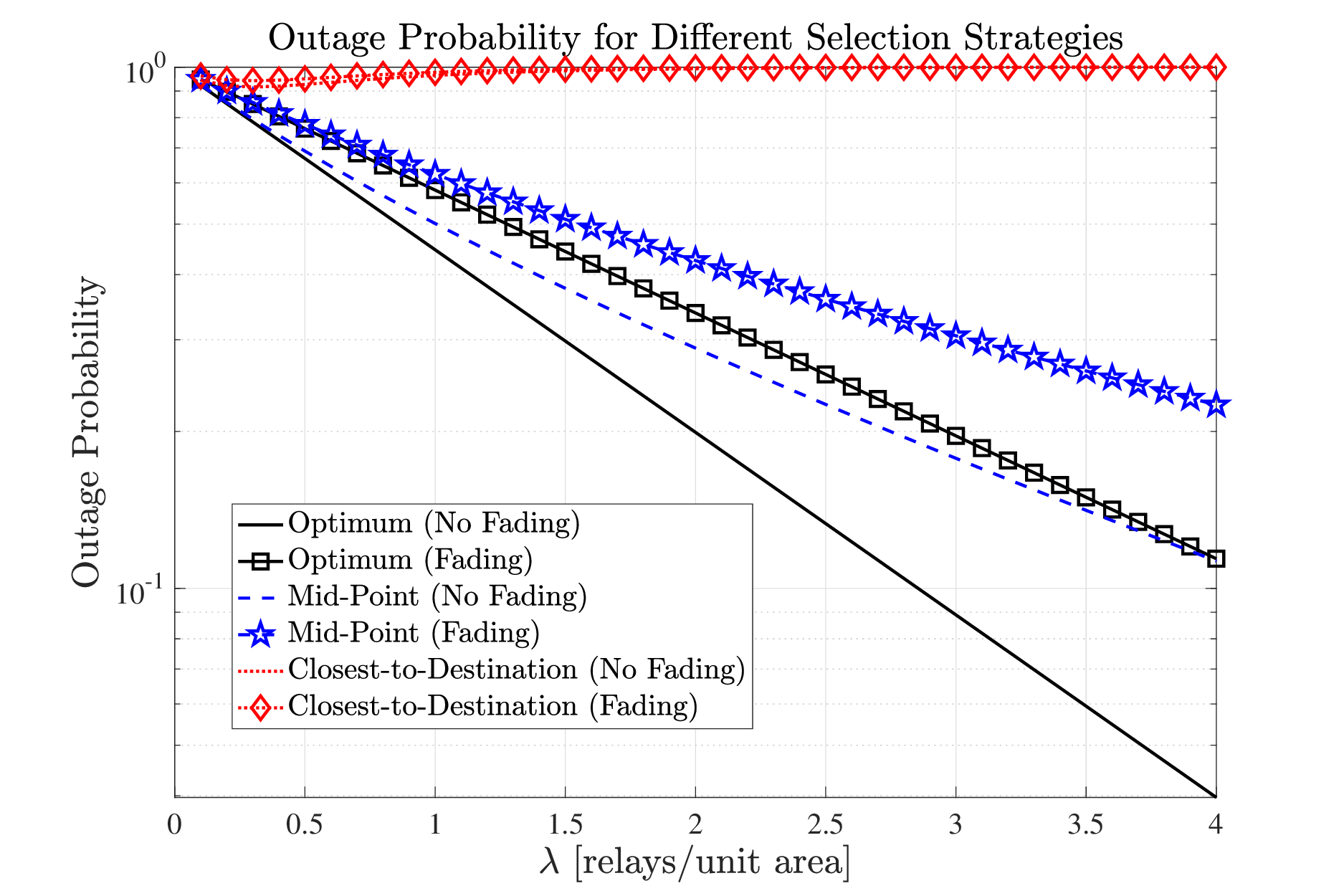}}
			\subfloat[Average rate vs $\lambda$.]{
		\label{f_outvslam5}
		\includegraphics[width=0.5\textwidth]{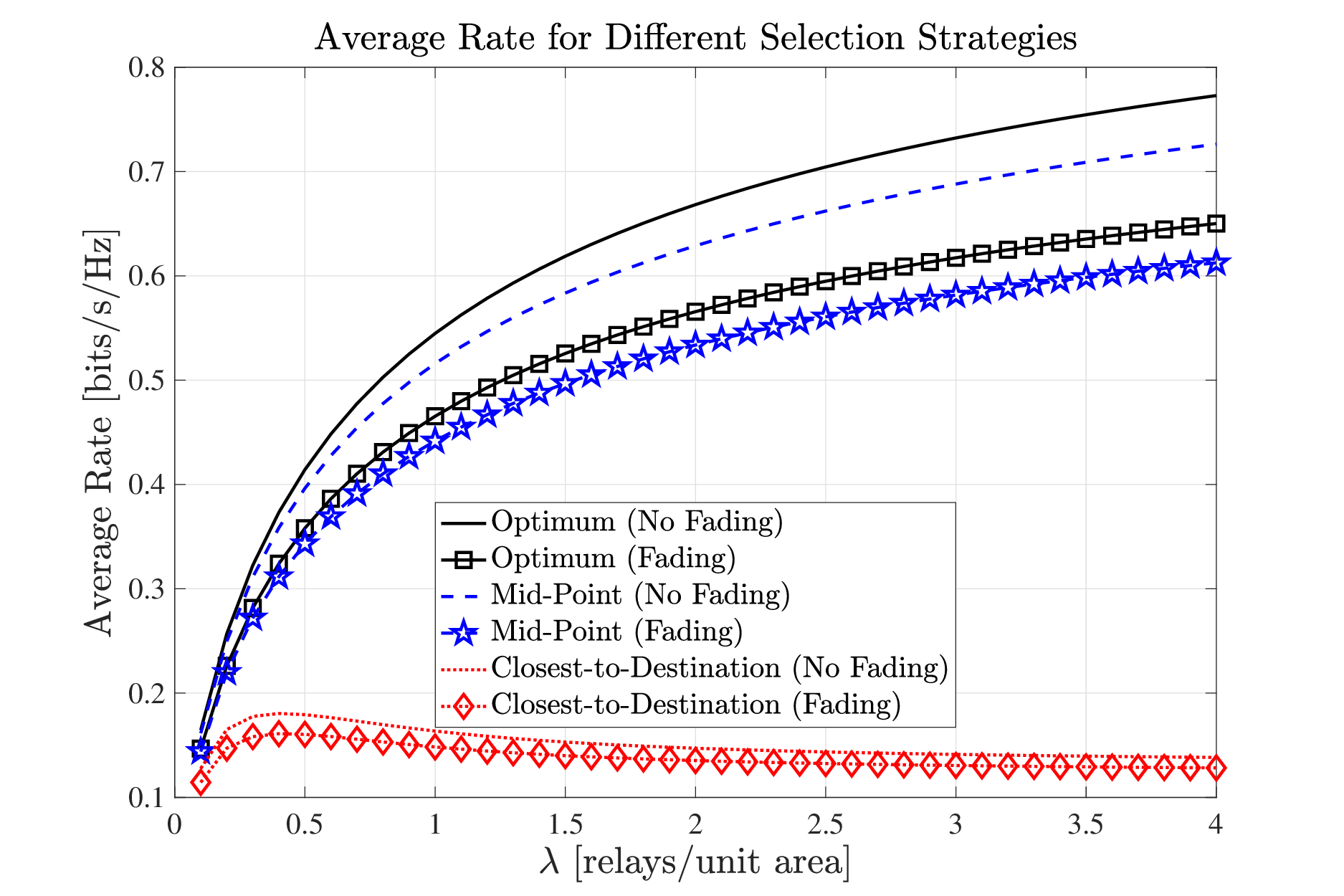}} 
		\caption{Outage probability and average rate achieved by different relay selection strategies as a function of $\lambda$ for $\snr=5$\,dB, $d=1$ and $\alpha=4$.}
	\label{Fig: OutageDiffRS}
	\vspace{-0.75cm}
\end{figure}

Fig.~\ref{Fig: OutageDiffRS} shows the outage probability and average rate curves as a function of $\lambda$ for $\snr=5$ dB. We set  $\rho=0.5$\,[bits/s/Hz] for the outage probability curves and $d=1$. Fig.~\ref{f_tpvslam5} shows that  
%
%
the closest-to-destination scheme has the worst and degrading performance, where the network is in outage with high probability, i.e., $\Pout\paren{\pol}\rightarrow 1$, when $\lambda$ increases. The reason for this phenomenon is the myopic selection of the relay without attempting to balance the relay-to-source and relay-to-destination distances, a problem which is more exacerbated for dense relay deployments. This observation clearly shows the importance of the distance balancing property of the optimum relay selection policy as discussed in Section \ref{Section: Mid-Point Relay Selection}, and how poorly a relay selection policy, which does not respect this property, can perform.   

Secondly, we observe that the optimum relay selection can outperform the mid-point relay selection significantly in terms of outage performance.  For example, at $\lambda=3$, the outage probability improvements of the optimum policy with respect to the mid-point one are around $48$\%  for  the no-fading case and $34$\% for the fading case. The slope of the outage probability in logarithmic scale shows how fast the outage probability decreases with respect to $\lambda$. In particular, logarithm of the outage probability decreases {\em almost} linearly with $\lambda$ for both optimum and mid-point selection policies but with different slopes and a slight decrease in the slope with $\lambda$ for the mid-point policy. For the optimum relay selection policy, the decay rate is around $0.35$ for the no-fading case and $0.24$ for the fading case. On the other hand, for the mid-point relay selection policy, those values are  $0.24$  for the no-fading case, and $0.14$ for the fading case. 

Figs.~\ref{f_outvslam5}  shows the average rate as a function of $\lambda$.
%
%
While we notice similar performance variations, 
we can see that there is a performance gap between optimum and mid-point relay selection. For example, we lose around $0.04$\,[bits/s/Hz] at $\lambda=4$ for both no-fading and Rayleigh fading cases by implementing mid-point relay selection.  
For optimum relay selection, the average rate monotonically increases with $\lambda$ to the limiting values $\frac{1}{2}\log_2\left(1+\frac{\snr }{d^\alpha}\right)$ and $\frac{1}{2\ln2} \e{\frac{d^\alpha}{\snr}} \text{E}_1\left(\frac{d^\alpha}{\snr}\right)$ for the no-fading and Rayleigh fading cases, respectively. 
For closest-to-destination relay selection, the average rate first increases and then decreases with $\lambda$ 
%
%
to the limiting values $\frac{1}{2}\log_2\left(1+\frac{\snr }{2^\alpha d^\alpha}\right)$ and $\frac{1}{2\ln2} \e{\frac{2^\alpha d^\alpha}{\snr}} \text{E}_1\left(\frac{2^\alpha d^\alpha}{\snr}\right)$ for the no-fading and Rayleigh fading cases, respectively. 
The asymptotic values are at $1.03$\,[bits/s/Hz] for the no-fading case and $0.86$\,[bits/s/Hz] for the Rayleigh fading case with optimum relay selection; and they are at $0.13$\,[bits/s/Hz] for the no-fading case and $0.12$\,[bits/s/Hz] for the fading case with closest-to-destination relay selection.

\begin{figure}
	\centering
	\subfloat[Average rate vs $\lambda$ without fading.]{
		\label{Fig: RateFB vs lambda no fading}
		\includegraphics[width=0.5\textwidth]{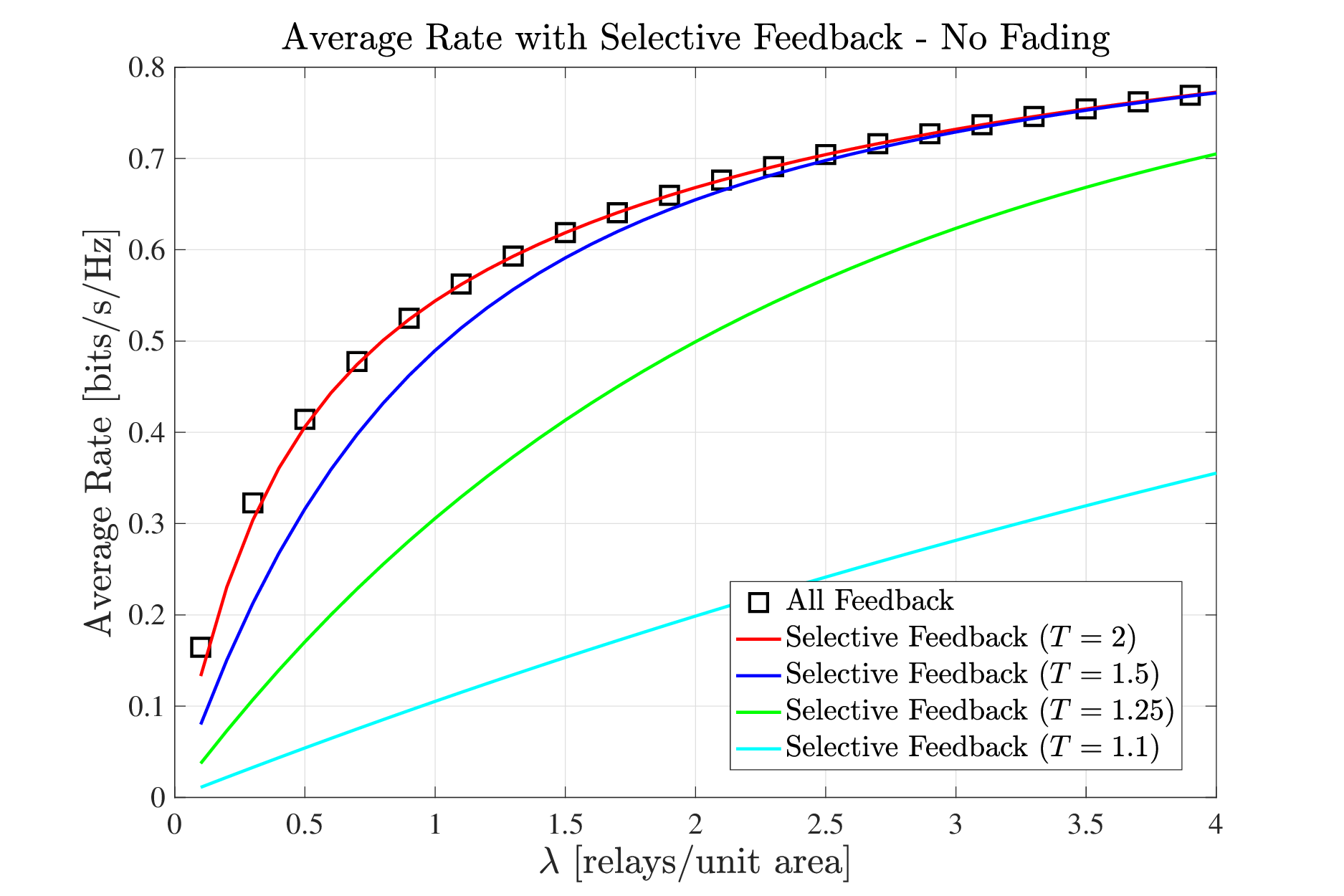}} 
        	\subfloat[Average rate vs $\lambda$ with fading.]{
		\label{Fig: OutageFB vs lambda fading rho=0.3}
		\includegraphics[width=0.5\textwidth]{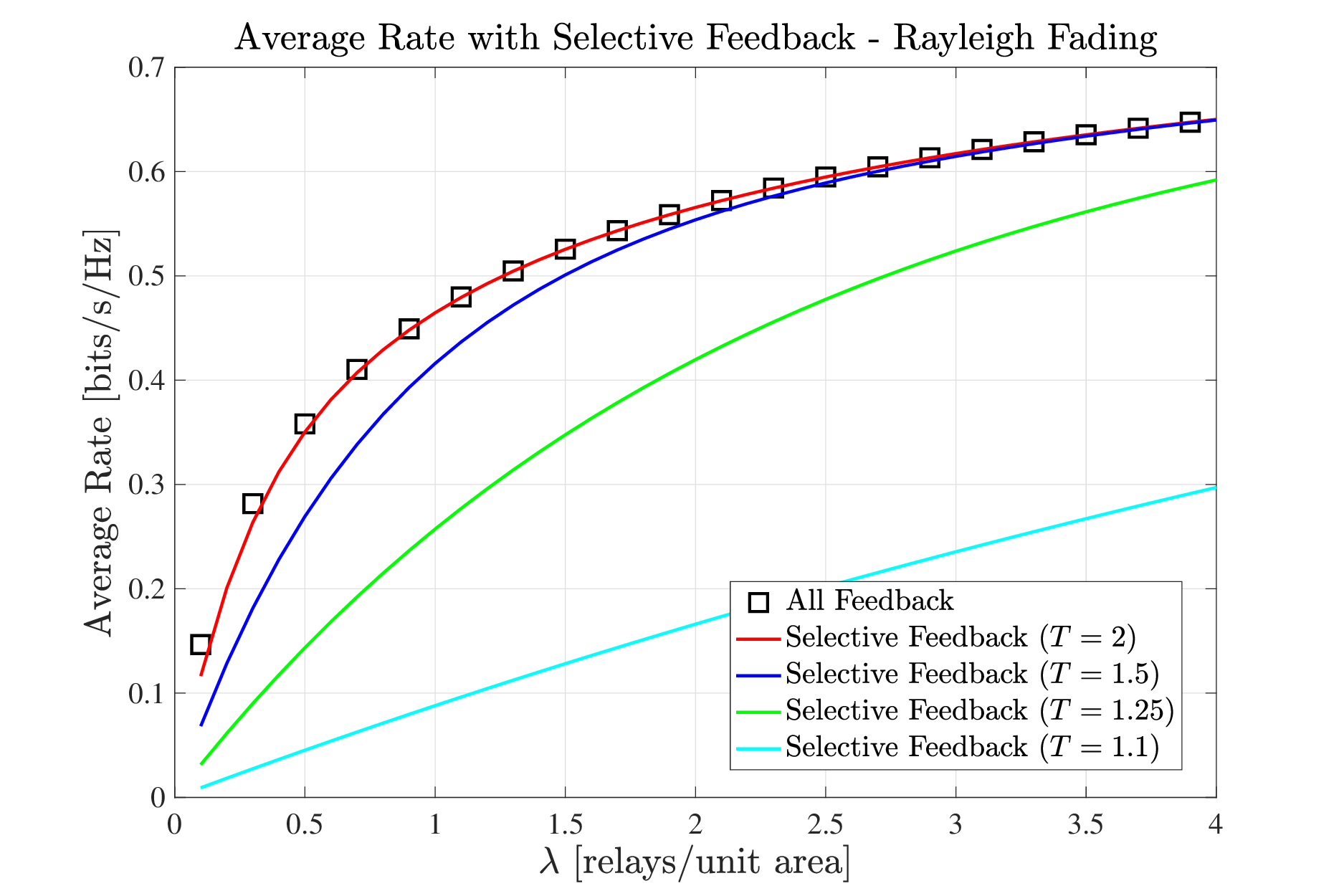}} \\
		\caption{Average rate achieved by the threshold-based selective feedback distributed relay selection policy for various values of the threshold level.} 
	\label{Fig: RateFB}
\end{figure}

%
%

In Fig. \ref{Fig: RateFB}, we plot the average rate $\Rave\paren{\pol_{\rm FB}, T}$ achieved by the threshold-based selective feedback distributed relay selection policy $\pol_{\rm FB}$ as a function of $\lambda$ for various values of $T \geq d$, where we still set $d=1$ and $\snr = 5 \mbox{ dB}$. For small values of $T$, there is a large gap between the average rates achieved by $\pol_{\rm FB}$ and all-feedback policies. This is due to the fact that $\PR{N_{\rm FB} = 0} = \e{-\mu\paren{T}}$ is large for small values of $T$, and hence the source node cannot receive any CQI from relays to choose one with high probability. More specifically, for the considered range of $\lambda \in \sqparen{0.1, 4}$ in Fig. \ref{Fig: RateFB}, $\mu\paren{T}$ ranges from $0.0123$ to $0.4934$ for $T=1.1$, which indicates that the source node is without any CQI more than $60\%$ of time even for the most crowded relay network scenario considered in this figure. 

A similar behaviour with a decreased rate gap between the selective feedback and all-feedback cases continues to hold for $T=1.25$. In this case, the source node cannot access to any CQI around $13\%$ of time for the most crowded relay network scenario. For $T=1.5$, $\mu\paren{T}$ is approximately equal to $3.1$, $4.65$ and $6.2$ for $\lambda=2, 3$ and $4$, respectively. As observed in Fig. \ref{Fig: RateFB}, the rate gap between the selective feedback and all-feedback cases becomes very small after $\lambda \geq 2$ for $T=1.5$, which implies a significant reduction in the feedback load without sacrificing from the achievable data rates. Similar observations but with a smaller rate gap continues to hold for $T=2$. 

\begin{figure}
	\centering
	\subfloat[Outage probability vs $\lambda$ for $\rho=0.3$.]{
		\label{Fig: OutageFB vs lambda no fading rho=0.3}
		\includegraphics[width=0.5\textwidth]{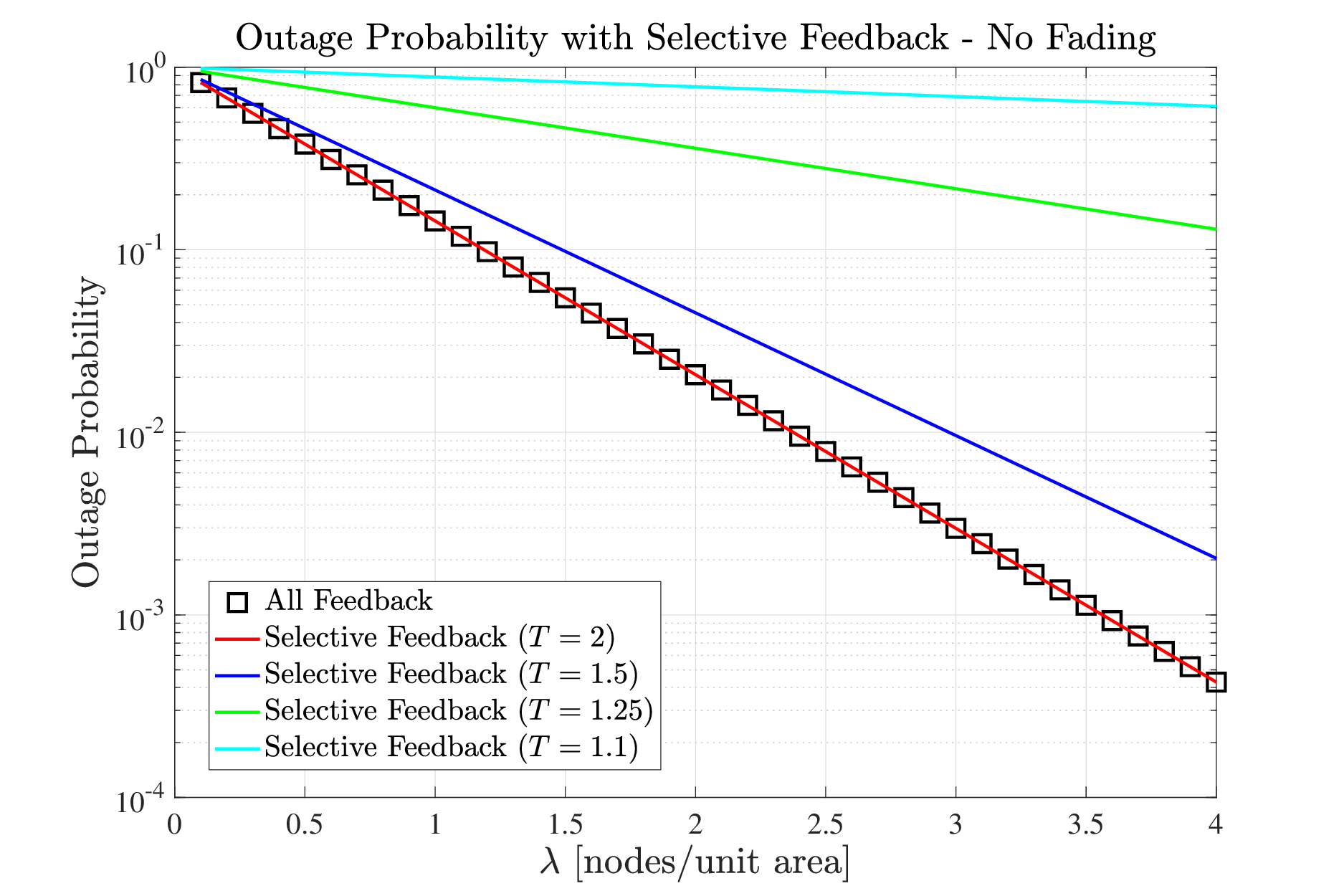}} 
        	\subfloat[Outage probability vs $\lambda$ for $\rho=0.3$.]{
		\label{Fig: OutageFB vs lambda fading rho=0.3 b}
		\includegraphics[width=0.5\textwidth]{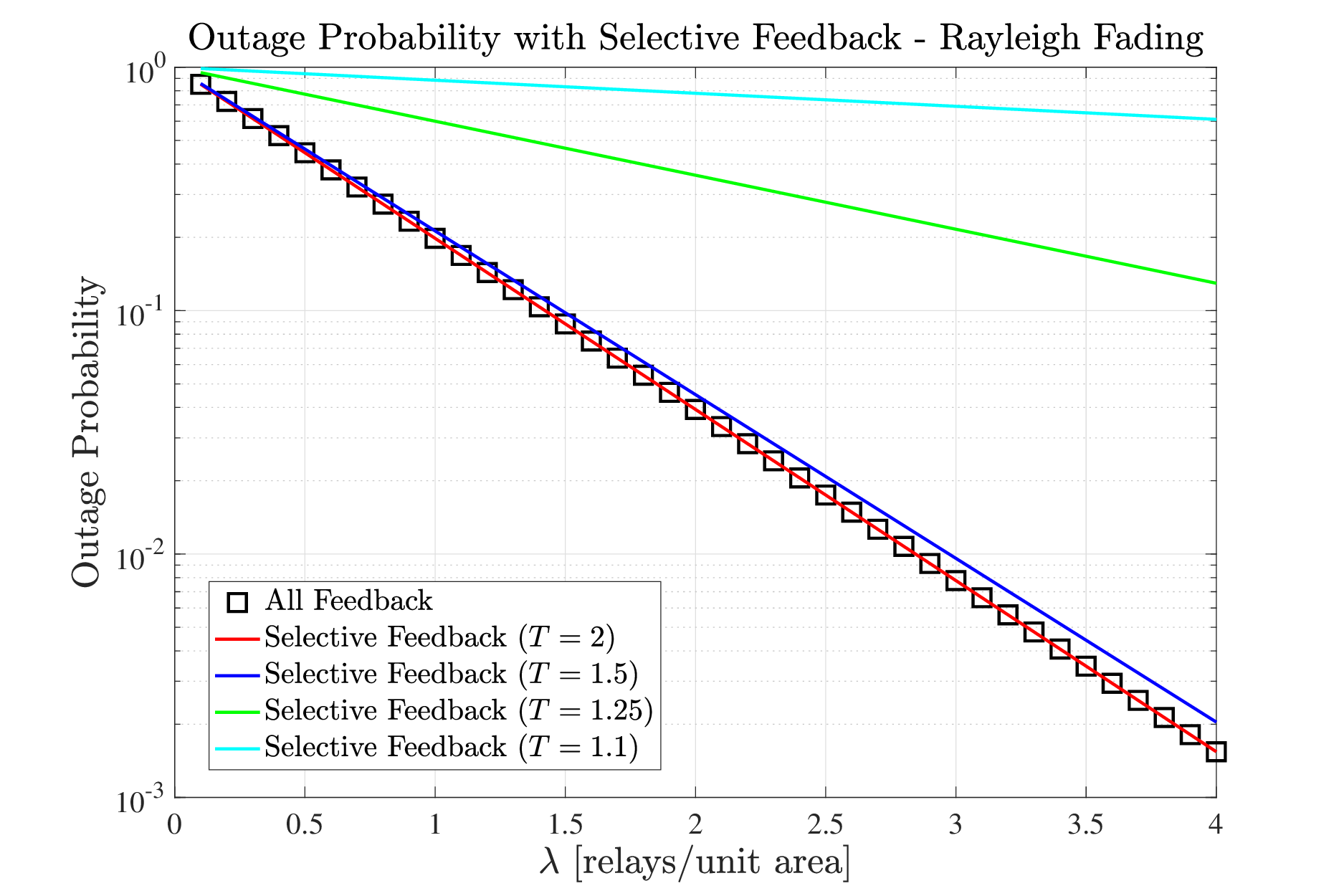}} \\
			\subfloat[Outage probability vs $\rho$ for $\lambda=2$.]{
		\label{Fig: OutageFB vs rhos no fading lambda=2}
		\includegraphics[width=0.5\textwidth]{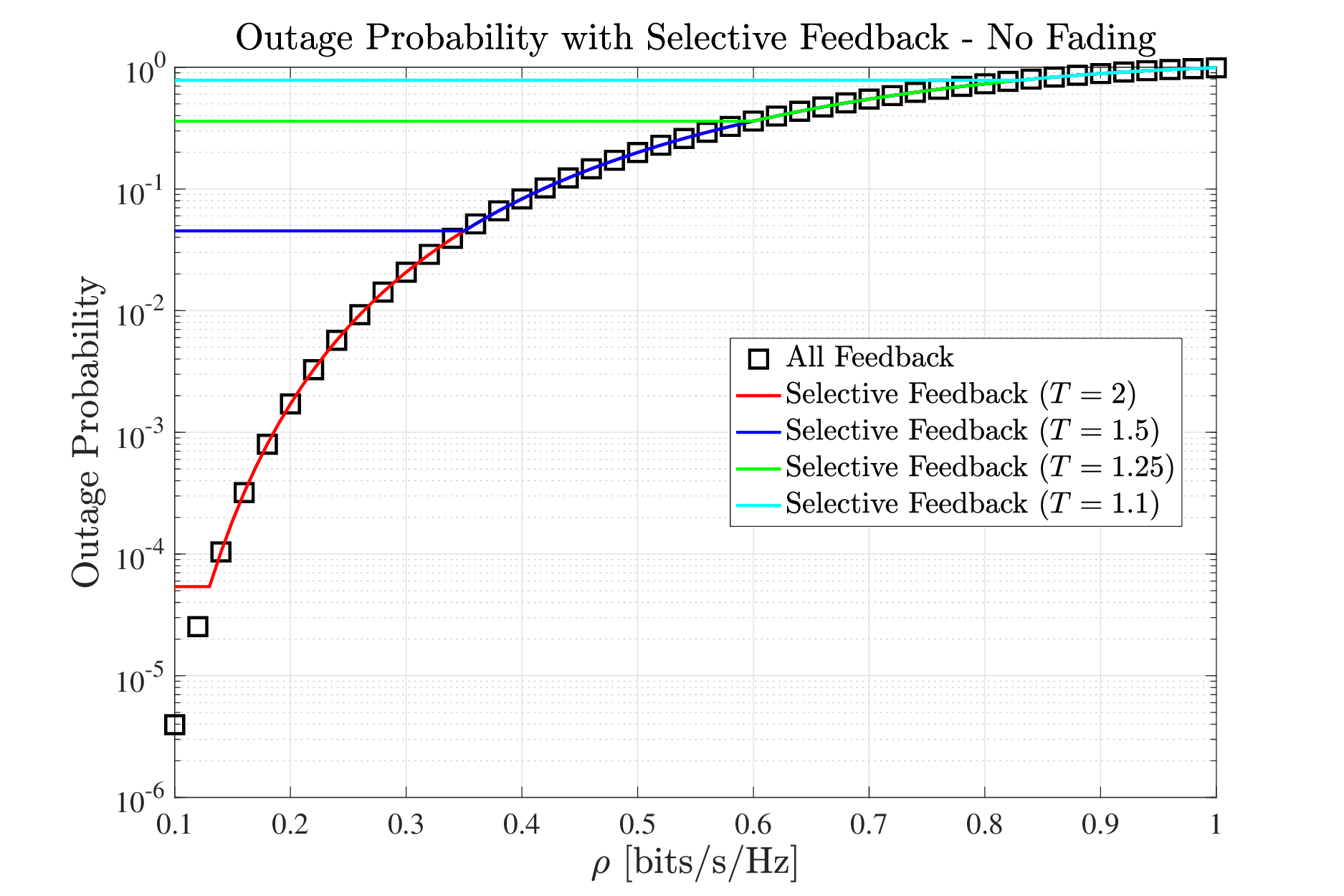}} 
        	\subfloat[Outage probability vs $\rho$ for $\lambda=2$.]{
		\label{Fig: OutageFB vs rhos fading lambda=2}
		\includegraphics[width=0.5\textwidth]{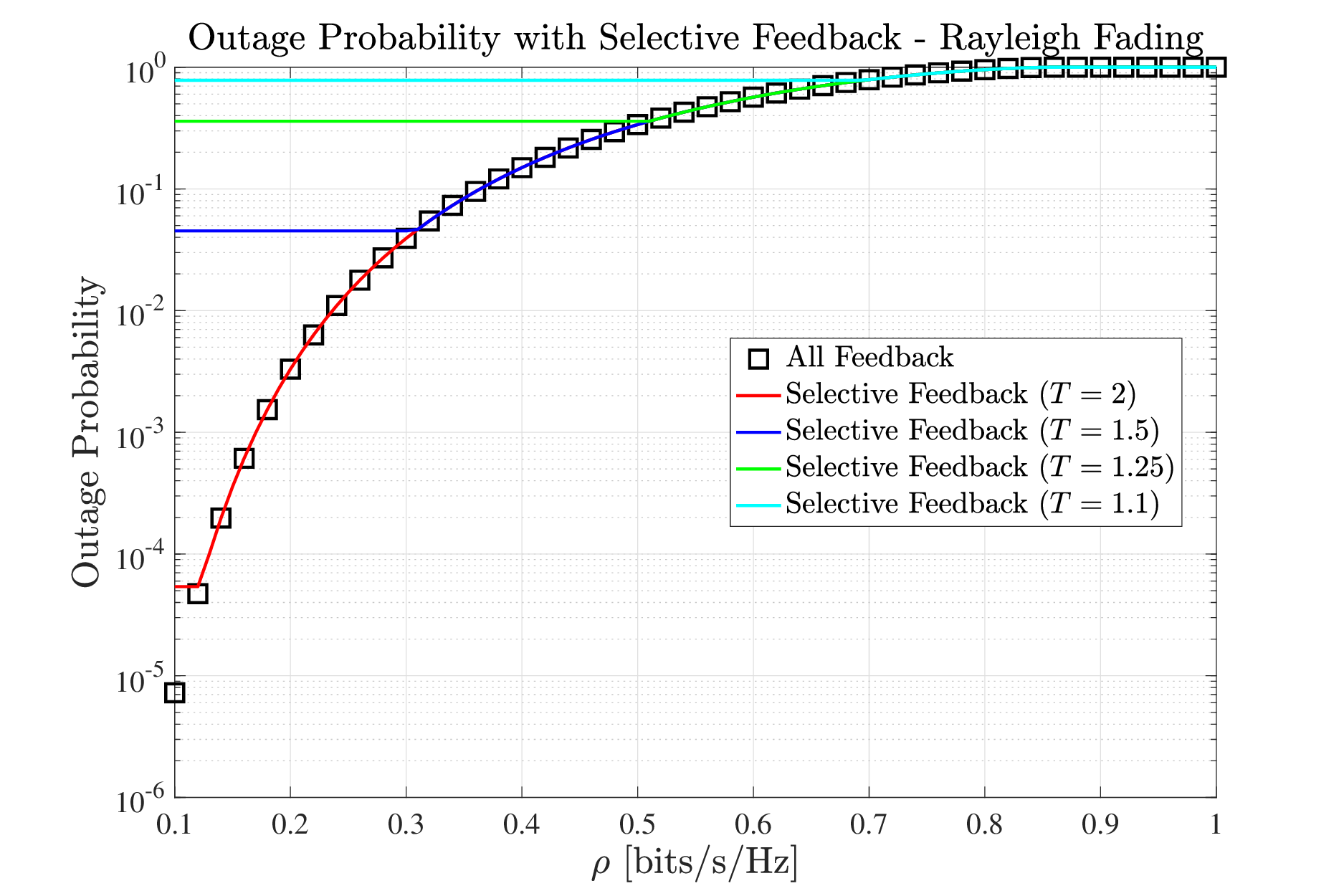}}
		\caption{Outage probability achieved by the threshold-based selective feedback distributed relay selection policy for various values of the threshold level.} 
	\label{Fig: OutageFB}
\end{figure}

%
%
%
%

In Fig. \ref{Fig: OutageFB}, we plot the outage probability curves achieved by $\pol_{\rm FB}$ for both as a function of $\lambda$ and $\rho$. 
%
%
Feedback-limited and rate-limited regimes discussed after Theorem \ref{Theorem: Outage with Feedback} for outage probability are also apparent in this figure. While drawing $\Pout\paren{\pol_{\rm FB}, T}$ in the top figures, we set $\rho = 0.3$. For this value of $\rho$, $s^\star = 0.6022$. Hence, we are in the feedback-limited regime for $T = 1.1, 1.25$ and $1.5$, and we observe exactly the same outage probabilities for both no-fading and fading cases. 

On the other hand, we are in the rate-limited regime for $T=2$, and the outage probabilities become the same for both selective feedback and all-feedback relay selections, i.e., see the red and black curves in the top figures. As a function of $\rho$, the feedback-limited regime is manifested through the flat portion $\Pout\paren{\pol_{\rm FB}, T}$ in the bottom figures. In particular, when drawn as a function of $\rho$, $\Pout\paren{\pol_{\rm FB}, T}$ stays constant until a critical target rate, which is the {\em feedback-limited} regime. In this regime, the outage probability depends only on the average number of relays feeding back, i.e., $\Pout\paren{\pol_{\rm FB}, T} = \e{-\mu\paren{T}}$. On the other hand, after a critical target rate value, outage probability curves coincide with each other and move together as a function of $\rho$ for both selective feedback and all-feedback scenarios. This is the {\em rate-limited} case, and the outage probability is independent of whether we employ a threshold-based selective feedback relay selection policy or not.

\begin{figure}
	\centering
	\subfloat[Average rate vs $\lambda$ for fixed feedback load.]{
		\label{Fig: RateFB vs lambda with mu(T) = 5}
		\includegraphics[width=0.5\textwidth]{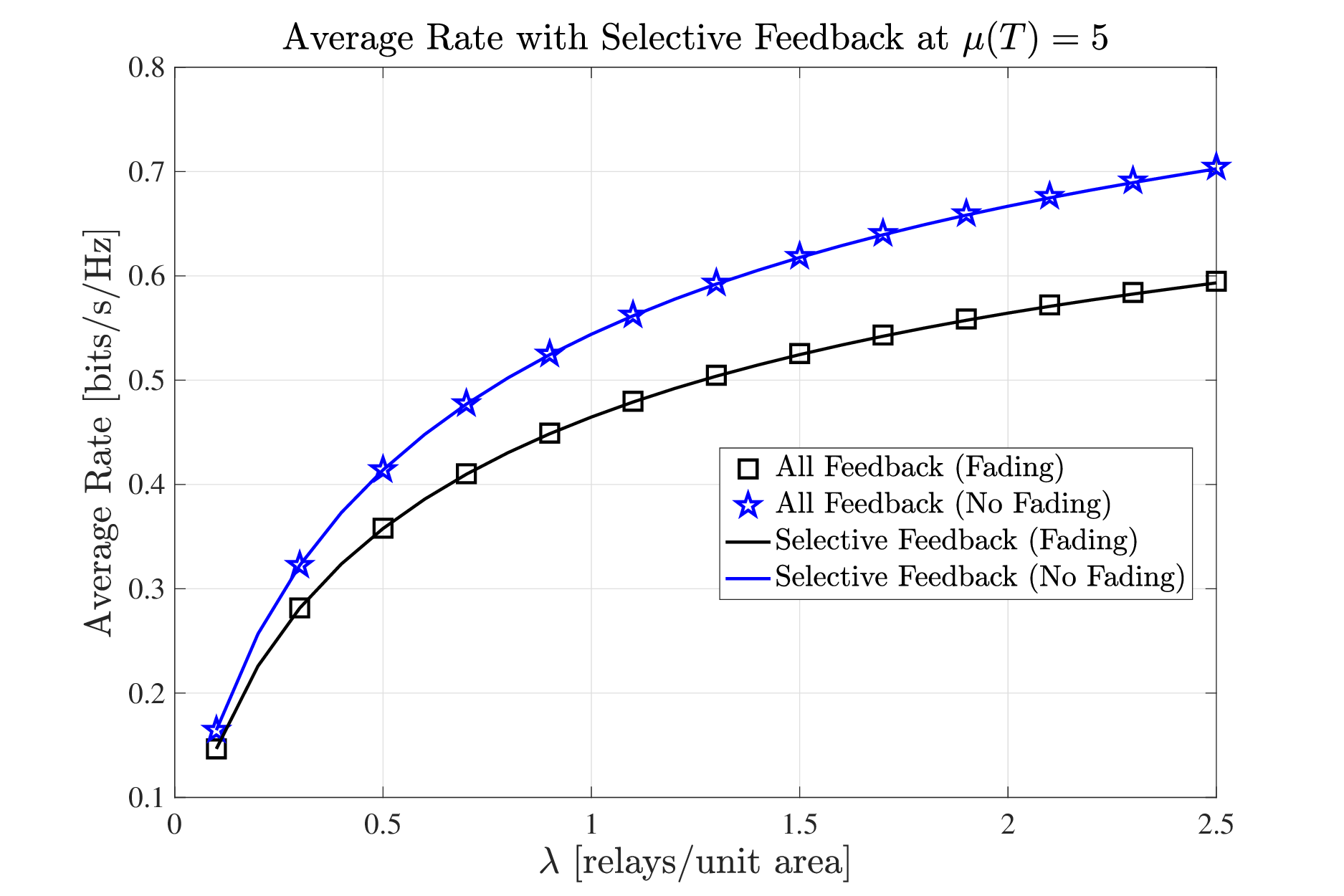}} 
        	\subfloat[Outage probability vs $\lambda$ for fixed feedback load.]{
		\label{Fig: OutageFB vs lambda with mu(T) = 5}
		\includegraphics[width=0.5\textwidth]{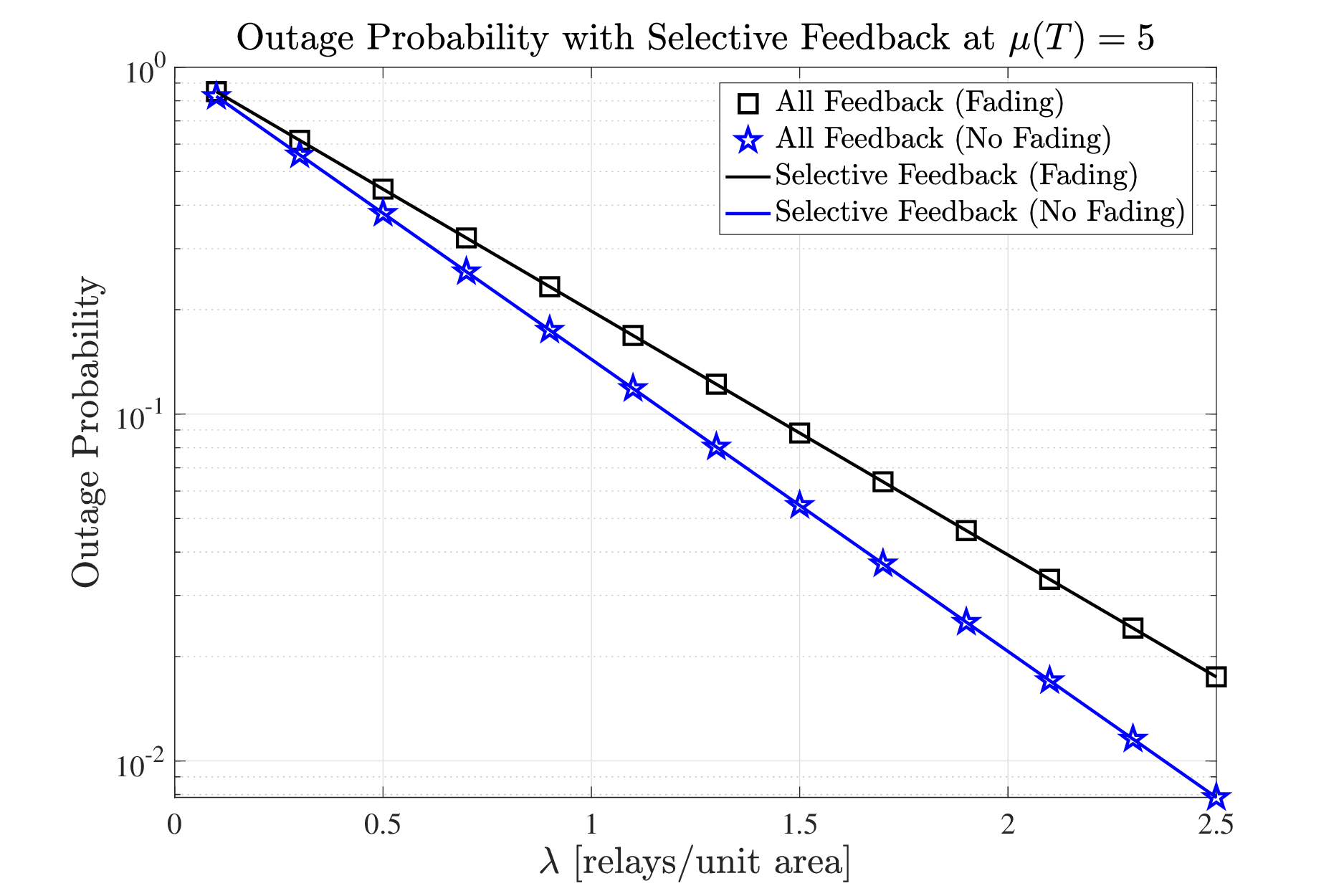}}
		\caption{Average rate and outage probability achieved by the threshold-based selective feedback relay selection policy with a fixed feedback load at $\mu\paren{T} = 5$.}
	\label{Fig: Performance with Fixed Feedback Load}
	\vspace{-0.75cm}
\end{figure}

Based on our observations in Fig. \ref{Fig: RateFB} and earlier explanations after Theorem \ref{Theorem: Feedback}, as a practical network design rule of thumb, we can say that setting $T$ such that $\mu\paren{T} \approx 5$ is enough to achieve the same average rate attained by the all-feedback policy with almost negligible performance loss. This point is specifically illustrated in Fig. \ref{Fig: Performance with Fixed Feedback Load} for both average rate and outage probability, where we observe almost no performance gap between the selective feedback and all-feedback approaches. However, it should be noted that an error floor phenomenon cannot be avoided with selective feedback for the outage probability performance as the network becomes denser due to the functional forms given in \eqref{Eqn: Outage with Feedback No Fading} and \eqref{Eqn: Outage with Feedback Rayleigh Fading}. For dense relay network deployments, the feedback load needs to be traded-off with outage probability to track the all-feedback outage performance.    

\section{Generalizations} \label{Section: Extensions}
\subsection{Relays with FD Capability}

In the FD case, the following modifications are required in \eqref{Eqn: DF Rate 1} to obtain the corresponding data rates.  We need to remove the scaling coefficient $\frac12$ in front of the minimum operator, scale the received signal power at the relay node with $1-\abs{\rho}^2$, add the term in \eqref{Eqn: FD Power Gain} 
\begin{eqnarray}
 \mbox{FD Power Gain} = 2\snr \sqrt{G\paren{\norm{\TX - \RX}}G\paren{\norm{\vec{x}_\pol - \RX}}}\Re\brparen{\rho \Hsd \Hrd^*} \label{Eqn: FD Power Gain}
 \end{eqnarray}
to the received signal power at the destination node and optimize the correlation coefficient $\rho \in \C$ of the source and relay codebooks over the complex unit ball, where $\Re\brparen{x}$ and $x^*$ denote the real part and complex conjugate of $x \in \C$, respectively, \cite{Cover79, Kramer05, Galarza14}. In the particular case of the severely shadowed source-destination links (i.e.,$\abs{\Hsd} \approx 0$) that we analyze in this paper, the FD power gain disappears and the FD rates are maximized when $\rho = 0$, which corresponds to independent codebook designs at the source and relay nodes.  


Hence, for the severely shadowed source-destination links, the data rates achieved by the FD relay deployments are equivalent to those in the HD case given by \eqref{Eqn: DF Rate 1}, up to a scaling coefficient of $\frac12$. This observation implies, for example, that the optimum relay location maximizing the HD rates continues to be the same for the FD operation. 
%
%
%
Therefore, our results obtained for the HD relays can be directly applied to this particular FD relay deployment scenario. 

\subsection{Non-homogeneous PPPs}
Isotropy and complete randomness are two critical properties of HPPPs that enabled the derivation of key results, especially statistical characterization of $\Gamma_{\rm opt}$ in Theorem \ref{Theorem: Optimal dis distribution}, in the previous sections of the paper. In this part, we will derive the cdf and pdf of $\Gamma_{\rm opt}$ for non-homogeneous but isotropic PPPs, which can be in turn used to obtain network performance measures such as $\Rave\paren{\pol_{\rm opt}}$ and $\Pout\paren{\pol_{\rm opt}}$ for more general distributions of relays over $\R^2$. To this end, we consider a non-homogeneous PPP $\Phi$ having a mean measure $\Lambda$, which is defined as 
\begin{eqnarray}
\Lambda\paren{\mathcal{S}} \defeq \ES{\sum_{\vec{X} \in \Phi} \I{\vec{X} \in \mathcal{S}}} \nonumber
\end{eqnarray}
for any Borel subset $\mathcal{S}$ of $\R^2$. We say that $\Phi$ is an {\em isotropic} PPP if $\Lambda$ is invariant under rotations around the origin, i.e., $\Lambda\paren{\mathcal{S}} = \Lambda\paren{\Pi\paren{\mathcal{S}}}$ for all rotations $\Pi$ around the origin and all Borel subsets $\mathcal{S}$ of $\R^2$. The next theorem provides the cdf and pdf of $\Gamma_{\rm opt}$ for isotropic PPPs.  

\begin{theorem} \label{Theorem: Isotropic Extension}
For an isotropic PPP with mean measure $\Lambda$, the cdf of $\Gamma_{\rm opt}$ is given by 
\begin{eqnarray} \label{Eqn: Gamma Opt CDF for Isotropic PPPs}
F_{\Gamma_{\rm opt}}\paren{\gamma} =  \left\{
\begin{array}{ll}
0 & \mbox{ if } 
\gamma < d \\
1-\e{-\frac{2}{\pi}\int_0^\frac{\pi}{2} \Lambda\paren{\mathcal{B}\paren{\vec{0}, \sqrt{\gamma^2 - d^2\sin^2\theta} - d\cos\theta}} \diff \theta} & \mbox{ if } 
\gamma \geq d
\end{array} \right.,
\end{eqnarray}
where $\mathcal{B}\paren{\vec{0}, r}$ is the closed disc centered around the origin with radius $r$. Further, if $\Lambda$ is absolutely continuous with respect to the Lebesgue measure with a {\em continuous} Radon-Nikodym derivative $\lambda\paren{\vec{x}}$, $\vec{x} \in \R^2$, then $\lambda$ is a spherically symmetric function, i.e., $\lambda\paren{\vec{x}} = \lambda\paren{\norm{\vec{x}}}$ for all $\vec{x} \in \R^2$, and the pdf of $\Gamma_{\rm opt}$ is given by 
\begin{eqnarray}
\lefteqn{f_{\Gamma_{\rm opt}}\paren{\gamma} = 4\int_0^\frac{\pi}{2} \frac{\gamma \paren{\sqrt{\gamma^2 - d^2\sin^2\theta} - d\cos\theta}}{\sqrt{\gamma^2 - d^2\sin^2\theta}}\lambda\paren{\sqrt{\gamma^2 - d^2\sin^2\theta} - d\cos\theta} \diff \theta} \hspace{15cm} \nonumber \\
\lefteqn{\cdot \e{-4\int_0^{\frac{\pi}{2}}\int_0^{\sqrt{\gamma^2 - d^2\sin^2\theta} - d\cos\theta}\lambda\paren{\psi}\psi \diff \psi \diff \theta} \cdot \I{\gamma \geq d}.} \hspace{7cm}
\end{eqnarray}
\end{theorem}
\begin{IEEEproof}
See Appendix \ref{Appendix: Isotropic Extension}.
\end{IEEEproof}

In the statement of Theorem \ref{Theorem: Isotropic Extension}, we represented the density $\lambda$, whenever it exists, as a function of a single variable due to its spherically symmetric nature with a slight abuse of notation.  For an HPPP, $\Lambda$ is a scaled version of the Lebesgue measure, i.e., $\Lambda\paren{\mathcal{S}} = \lambda \cdot {\sf area}\paren{\mathcal{S}}$ with constant $\lambda$, and it can be seen that Theorem \ref{Theorem: Isotropic Extension} reduces to Theorem \ref{Theorem: Optimal dis distribution} after some manipulations. Below, we provide some examples, which are of either potential  practical or theoretical interest, to illustrate the applications of Theorem \ref{Theorem: Isotropic Extension}. In all the examples below, we take $\xs = \paren{-d, 0}$, $\xd = \paren{d, 0}$.  

{\bf Example 1:} $\Phi$ is an HPPP with constant intensity $\lambda > 0$ over $\R^2\setminus \mathcal{B}\paren{\vec{0}, r}$ and $\Phi \cap \mathcal{B}\paren{\vec{0}, r} = \emptyset$. This is the situation in which there is a circular exclusion region with radius $r$ around the origin and none of the relays is allowed to lie in this region possibly due to an obstacle. The mean measure in this case is given by 
\begin{eqnarray}
\Lambda\paren{\mathcal{B}\paren{\vec{0}, \tau}} = \left\{
\begin{array}{ll}
0 & \mbox{ if } 
\tau < r \\
\lambda\pi\paren{\tau^2-r^2} & \mbox{ if } 
\tau \geq r
\end{array} \right.. \nonumber
\end{eqnarray}

Using \eqref{Eqn: Gamma Opt CDF for Isotropic PPPs} and calculating the integral $\int_0^{\frac{\pi}{2}} \Lambda\paren{\mathcal{B}\paren{\vec{0}, \sqrt{\gamma^2 - d^2\sin^2\theta} - d\cos\theta}} \diff \theta$ for three different ranges of $\gamma$, we obtain  \begin{eqnarray} \label{Eqn: Gamma Opt CDF with Circular Exclusion 1}
F_{\Gamma_{\rm opt}}\paren{\gamma} =  \left\{
\begin{array}{ll}
0 & \mbox{ if } 
\gamma \leq \sqrt{r^2 + d^2} \\
1-\e{-2\lambda d^2 I_{d, r}\paren{\gamma} + 2\lambda r^2\arcsin\paren{\frac{\gamma^2-d^2-r^2}{2dr}}} & \mbox{ if } 
\sqrt{r^2 + d^2} < \gamma \leq r+d \\
1 - \e{-2\lambda d^2\paren{\paren{\frac{\gamma}{d}}^2\arcsec\paren{\frac{\gamma}{d}} - \sqrt{\paren{\frac{\gamma}{d}}^2-1}} + \lambda\pi r^2} & \mbox{ if } \gamma > r+d
\end{array} \right.,
\end{eqnarray}
where $I_{d, r}\paren{\gamma}$ is given by \eqref{Eqn: Gamma Opt CDF wiht Circular Exclusion 2} and \eqref{Eqn: Gamma Opt CDF wiht Circular Exclusion 3}. 
\begin{figure*}[!t]
\begin{eqnarray}
\lefteqn{I_{d, r}\paren{\gamma} = \paren{\frac{\gamma}{d}}^2\paren{\arcsec\paren{\frac{\gamma}{d}} +\arccsc\paren{\frac{\gamma}{d}\frac{1}{b_{d, r}\paren{\gamma}}} -\arccos\paren{\frac{\gamma^2-d^2-r^2}{2dr}}}} \hspace{15cm} \nonumber \\
\lefteqn{+ b_{d, r}\paren{\gamma}\paren{\sqrt{\paren{\frac{\gamma}{d}}^2 - b_{d, r}^2\paren{\gamma}} - \frac{\gamma^2 - d^2 - r^2}{2dr}} - \sqrt{\paren{\frac{\gamma}{d}}^2 - 1}} \label{Eqn: Gamma Opt CDF wiht Circular Exclusion 2} \hspace{11cm}
\end{eqnarray}
\hrule
\end{figure*}
\begin{figure*}
\begin{eqnarray}
b_{d, r}\paren{\gamma} = \frac{1}{2dr}\sqrt{\paren{d+r-\gamma}\paren{d+r+\gamma}\paren{\gamma + d - r}\paren{\gamma + r - d}} \label{Eqn: Gamma Opt CDF wiht Circular Exclusion 3}
\end{eqnarray}
\hrule
\end{figure*}

We note that $\Phi$ is an HPPP with intensity $\lambda$ over $\R^2$ when $r=0$, and \eqref{Eqn: Gamma Opt CDF with Circular Exclusion 1} reduces to \eqref{e_cdfmaxminD} as expected. Since there is a circular exclusion region around the origin with radius $r$, $\Gamma_{\rm opt}$ always takes values larger than $\sqrt{r^2 + d^2}$, which results in the first case in \eqref{Eqn: Gamma Opt CDF with Circular Exclusion 1}. For $\gamma > r+d$, it can be seen that the function $g\paren{\theta} = \sqrt{\gamma^2 - d^2\sin^2\theta} - d\cos\theta$ is greater than $r$ for all $\theta \in \sqparen{0, \frac{\pi}{2}}$. Hence, $\Lambda\paren{\mathcal{B}\paren{\vec{0}, g\paren{\theta}}} = \lambda \pi \paren{g^2\paren{\theta} - r^2}$ for all $\theta \in \sqparen{0, \frac{\pi}{2}}$, integration of which over $\sqparen{0, \frac{\pi}{2}}$ leads to the third case in \eqref{Eqn: Gamma Opt CDF with Circular Exclusion 1}. The functional form of the cdf of $\Gamma_{\rm opt}$ in this case is identical to the one in \eqref{e_cdfmaxminD}, except an extra $\lambda \pi r^2$ term due to the exclusion region around the origin. The most involved case is the one in which $\sqrt{\gamma^2+d^2} < \gamma \leq r+d$. In this case, $g\paren{\theta} \leq r$ for $\theta \in \sqparen{0, \theta^\star}$ and $g\paren{\theta} > r$ for $\theta \in \parenlo{\theta^\star, \frac{\pi}{2}}$, where $\theta^\star = \arccos\paren{\frac{\gamma^2-d^2-r^2}{2dr}}$. Considering both integration intervals, we arrive at the second case in \eqref{Eqn: Gamma Opt CDF with Circular Exclusion 1}.

In the next two examples, we consider finite number of relays with a Poisson distribution, which will eventually lead to having {\em defective} $\Gamma_{\rm opt}$ distributions.  

{\bf Example 2:} 
In this example, we consider the situation in which the relays form a HPPP $\Phi$ on a circle around the origin, which we represent as the boundary $\partial \mathcal{B}\paren{\vec{0}, r}$ of $\mathcal{B}\paren{\vec{0}, r}$. There is no other relay in the network. Then, the mean measure of  $\Phi$ is given by
\begin{eqnarray}
\Lambda\paren{\mathcal{B}\paren{\vec{0}, \tau}} = \left\{
\begin{array}{ll}
0 & \mbox{ if } 
\tau < r \\
 2 \lambda \pi r & \mbox{ if } 
\tau \geq r
\end{array} \right.. \nonumber
\end{eqnarray}

When compared to the previously studied HPPP cases over $\R^2$ with and without an exclusion region, one important feature of this case is that the total number of relays $N$ contained in $\Phi$ is a Poisson distributed random variable with mean $2\lambda\pi r$, which takes finite values. On the other hand, $\Phi$ contained infinitely many relays in the previous cases.  Using the master equation \eqref{Eqn: Gamma Opt CDF for Isotropic PPPs}, we obtain the cdf of $\Gamma_{\rm opt}$ for Poisson distributed relays on the circle as 
\begin{eqnarray} \label{Eqn: Gamma Opt CDF with Circular Network Region}
F_{\Gamma_{\rm opt}}\paren{\gamma} =  \left\{
\begin{array}{ll}
0 & \mbox{ if } 
\gamma \leq \sqrt{r^2 + d^2} \\
1-\e{-4\lambda r \arcsin\paren{\frac{\gamma^2 - d^2 - r^2}{2dr}}} & \mbox{ if } 
\sqrt{r^2 + d^2} < \gamma \leq r+d \\
1 - \e{-2\lambda \pi r} & \mbox{ if } \gamma > r+d
\end{array} \right.. 
\end{eqnarray}

The first case in \eqref{Eqn: Gamma Opt CDF with Circular Network Region} reflects the fact that $\Gamma_{\rm opt}$ is greater than $\sqrt{r^2 + d^2}$ when relays are located on $\partial \mathcal{B}\paren{\vec{0}, r}$. In the second case, $g(\theta)$, as defined in Example 1, is smaller than $r$ for $\theta \in \sqparen{0, \theta^\star}$ and larger than $r$ for $\theta \in \parenlo{\theta^\star, \frac{\pi}{2}}$, where $\theta^\star = \arccos\paren{\frac{\gamma^2-d^2-r^2}{2dr}}$. For $\gamma > r+d$, we have $g\paren{\theta} > r$ for all $\theta \in \sqparen{0, \frac{\pi}{2}}$ and all the relays available in the network are used to determine the value for $F_{\Gamma_{\rm opt}}\paren{\gamma}$. However, since $\PR{N=0} = \e{-2\lambda\pi r}$, some probability mass for $\Gamma_{\rm opt}$ always escapes to infinity due to lack of a relay in the network that can provide connectivity between source and destination nodes.     

{\bf Example 3:} $\Phi$ is a Gaussian PPP with intensity function $\lambda\paren{\vec{x}} = \frac{n}{2\pi\sigma^2}\e{-\frac{\norm{\vec{x}}^2}{2\sigma^2}}$, where $n$ is the total average number of relays contained in $\Phi$ \cite{Haenggi13}. This is the case in which relays are scattered around the origin according to a bell shaped density with exponentially decaying tails to provide connectivity between source and destination. For Gaussian PPP distribution of relays, the mean measure of  $\Phi$ is equal to $\Lambda\paren{\mathcal{B}\paren{\vec{0}, \tau}} = n\paren{1-\e{-\frac{\tau^2}{2\sigma^2}}}$ and the cdf of $\Gamma_{\rm opt}$ can be obtained as
\begin{eqnarray} \label{Eqn: Gamma Opt CDF with Circular Network Region b}
F_{\Gamma_{\rm opt}}\paren{\gamma} =  \left\{
\begin{array}{ll}
0 & \mbox{ if } 
\gamma < d \\
1-\exp\paren{-n + \frac{2n}{\pi}\e{\frac{-\gamma^2}{2\sigma^2}\int_0^1\frac{\e{\frac{-d}{2\sigma^2}\paren{1-2u^2-2\sqrt{1-u^2}\sqrt{\paren{\frac{\gamma}{d}}^2-u^2}}}}{\sqrt{1-u^2}}\diff u}} & \mbox{ if } \gamma \geq d
\end{array} \right. 
\end{eqnarray}
by using \eqref{Eqn: Gamma Opt CDF for Isotropic PPPs}. 

\subsection{Different $\snr$ Values at Relays and Destination}

As another extension of our baseline model presented in the previous sections, we now consider a more heterogeneous communications scenario in which the $\snr$ values at relays and the destination are different. For example, this is the operating situation of interest when transmission powers and/or RF circuitry at source and relay nodes are distinct. To model this situation, we will denote the $\snr$ value at relays by $\snr_1$ and that at the destination node by $\snr_2$.   
%
%
Conditioned on relay locations $\Phi$, the rate achieved by a relay selection policy $\pol$ is given according to 
\begin{eqnarray}
\rateave\paren{\pol} = \frac12 \min\bigl\{\EW\sqparen{\log_2\paren{1 + \snr_1 \abs{\Hsr}^2\norm{\TX - \vec{X}_\pol}^{-\alpha}} \big| \Phi}, \nonumber\\
\EW\sqparen{\log_2\paren{1 + \snr_2 \abs{\Hrd}^2\norm{\vec{X}_\pol - \RX}^{-\alpha}} \big| \Phi} \bigr\},\label{Eqn: Rate Given Locations Diff SNR}
\end{eqnarray}
where we have taken $G(x) = \frac{1}{x^\alpha}$. Using \eqref{Eqn: Rate Given Locations Diff SNR} and following the arguments similar to those in Section \ref{Section: Optimum Relay Selection Problem}, we can write the optimum relay selection problem as the minimization of the modified relay selection function $\relayselectdiffsnr\paren{\vec{x}}$, which is given by    
\begin{eqnarray}
\relayselectdiffsnr\paren{\vec{x}} =  \max\brparen{\widetilde{\snr}_1 \norm{\xs - \vec{x}}, \widetilde{\snr}_2 \norm{\vec{x} - \xd}}, \label{Eqn: Relay Selection Function Diff SNR}
\end{eqnarray}
where 
%
$\widetilde{\snr}_1 = \snr_1^{-1/\alpha}$ and $\widetilde{\snr}_2 = \snr_2^{-1/\alpha}$. Here, we can interpret $\widetilde{\snr}_1$ and $\widetilde{\snr}_2$ as effective $\snr$s at relays and destination, respectively, for optimum relay selection. The CQI at the optimum relay node in $\Phi$ 
is now given according to
\begin{eqnarray}
\Gamma_{\rm opt, diff} = \min_{\vec{X} \in \Phi} \relayselectdiffsnr\paren{\vec{X}}. \label{Eqn: Optimum Relay Distance Different SNRs} 
\end{eqnarray}
Accordingly, the conditional data rate achieved by the optimum relay selection policy $\pol_{\rm opt}$, given the relay locations, is equal to 
\begin{eqnarray}\label{Eqn: Conditional Average Rate Diff SNR}
\rateave\paren{\pol_{\rm opt}} = \frac12 \EW\sqparen{\log_2\paren{1 +  \abs{H}^2\Gamma_{\rm opt, diff}^{-\alpha}} \big| \Phi}. 
\end{eqnarray}

As in the homogeneous case, the expression in \eqref{Eqn: Conditional Average Rate Diff SNR} shows that it is enough to derive the distribution of $\Gamma_{\rm opt, diff}$ in order to carry out the same analysis for the average rate and outage probability as is done in Section \ref{Section: Optimum Relay Selection}. The cdf of $\Gamma_{\rm opt, diff}$ is obtained in the following theorem. 

\begin{theorem} \label{Theorem: Diff SNR Distribution}
The cdf of $\Gamma_{\rm opt, diff}$ is given by \begin{equation}\label{e_cdfmaxminDdiffSNR}
\begin{split}
F_{\Gamma_{\rm opt, diff}}\paren{\gamma} = \left\{
\begin{array}{ll}
0 & \hspace{-4cm}\mbox{ if } 
\gamma < 2d\left(\frac{\widetilde{\SNR}_1\widetilde{\SNR}_2}{\widetilde{\SNR}_1+\widetilde{\SNR}_2}\right) \\
1-\e{-\lambda \left(\frac{\gamma }{\widetilde{\SNR}_1}\right)^2 \sec ^{-1}\left(\frac{4 d \left(\frac{\gamma }{\widetilde{\SNR}_1}\right)}{\left(\frac{\gamma }{\widetilde{\SNR}_1}\right)^2-\left(\frac{\gamma }{\widetilde{\SNR}_2}\right)^2+ 4d^2}\right)}
\\
\qquad\e{-\lambda \left(\frac{\gamma }{\widetilde{\SNR}_2}\right)^2 \sec ^{-1}\left(\frac{4 d\left(\frac{\gamma }{\widetilde{\SNR}_2}\right)}{\left(\frac{\gamma }{\widetilde{\SNR}_2}\right)^2-\left(\frac{\gamma }{\widetilde{\SNR}_1}\right)^2+ 4d^2}\right)}
\\
\qquad\e{\lambda \sqrt{\left(\frac{\gamma }{\widetilde{\SNR}_1}+\frac{\gamma }{\widetilde{\SNR}_2}-2 d\right) \left(\frac{\gamma }{\widetilde{\SNR}_1}-\frac{\gamma }{\widetilde{\SNR}_2}+2 d\right) \left(-\frac{\gamma }{\widetilde{\SNR}_1}+\frac{\gamma }{\widetilde{\SNR}_2}+2 d\right) \left(\frac{\gamma }{\widetilde{\SNR}_1}+\frac{\gamma }{\widetilde{\SNR}_2}+2 d\right)}} \\
& \hspace{-4cm} \mbox{ if } 
\gamma \geq 2d\left(\frac{\widetilde{\SNR}_1\widetilde{\SNR}_2}{\widetilde{\SNR}_1+\widetilde{\SNR}_2}\right)
\end{array} \right..
\end{split}
\end{equation} 
\end{theorem}
\begin{IEEEproof}
See Appendix \ref{Appendix: Different SNR Extension}.
\end{IEEEproof}

An important remark about the result in Theorem \ref{Theorem: Diff SNR Distribution} is that the optimum relay location does not satisfy the distance balancing property anymore due to different $\snr$ scalings at the source-relay and relay-destination links. In particular, the minimum of $\relayselectdiffsnr\paren{\vec{x}}$ is achieved at a point $\vec{x}^\star$ located on the line segment connecting $\xs$ and $\xd$ that satisfies $\norm{\vec{x}^\star - \xd} = 2d\frac{\widetilde{\SNR}_1}{\widetilde{\SNR}_1+\widetilde{\SNR}_2}$. Hence, the relays in $\Phi$ closer to $\vec{x}^\star$ are likely to be the better candidates to pick to maximize the system performance when the relays and destination are subject to different noise conditions.

\begin{figure}[t]
\centering
\includegraphics[width=0.7\textwidth]{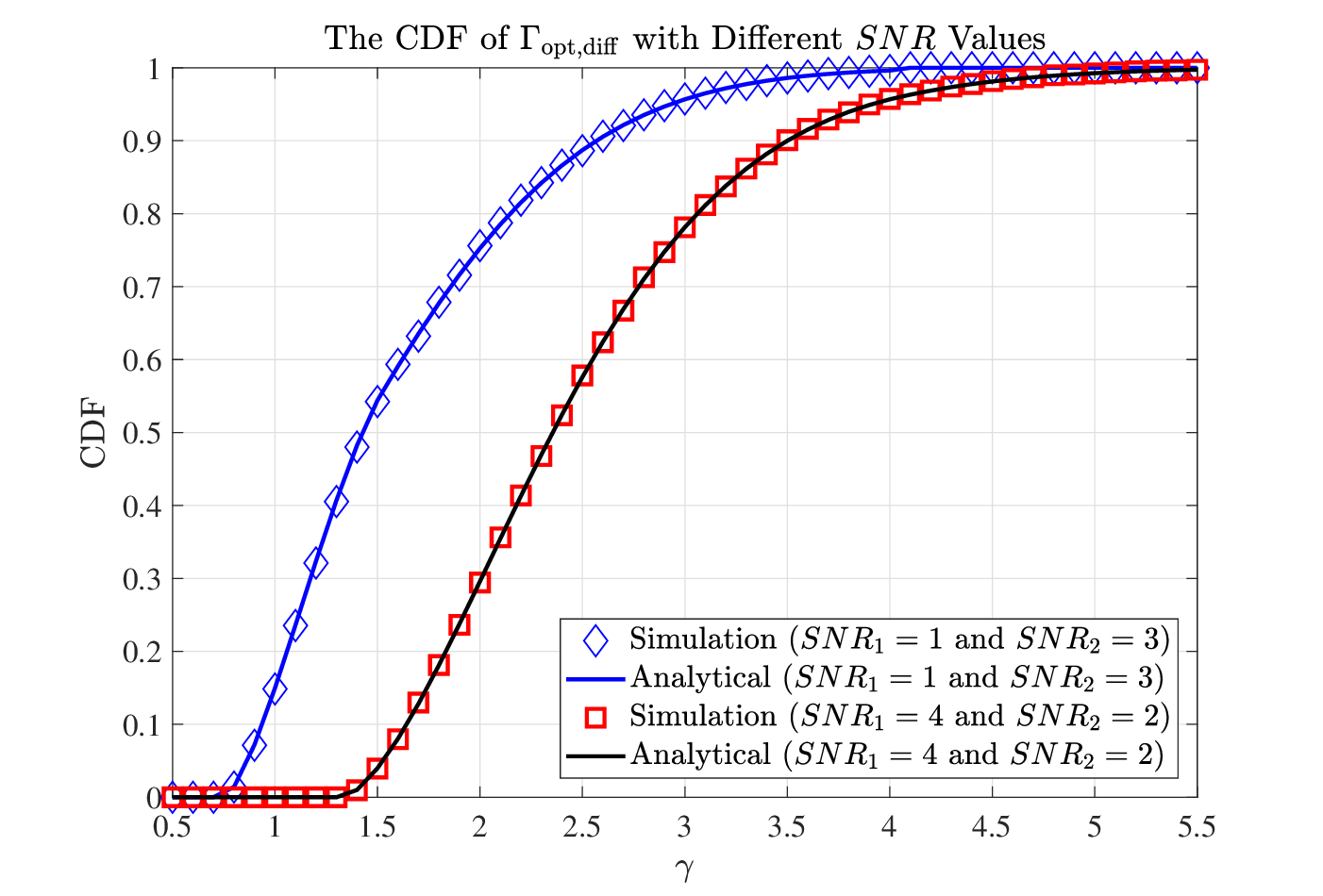}
\caption{The cdf of $\Gamma_{\rm opt, diff}$ for different $\snr$s at relay and destination nodes when $d=0.5$ and $\lambda=1$.}\label{f_cdfdiffsnr}
\end{figure}


\vspace{-0mm}
\section{Conclusions} \label{Section: Conclusions}
In this paper, we have studied a relay-aided wireless network with a single source-destination pair and spatially deployed decode-and-forward relays. We have obtained structural properties for the optimum relay selection policy and the distribution of the channel quality indicator of the relay node selected by the optimum policy. To benchmark the optimum relay selection scheme, we have analyzed the mid-point relay selection policy and obtained a sufficient condition for its optimality. These results hold for general fading distributions and non-increasing path-loss models decaying to zero. Using the derived distribution of the optimum channel quality indicator, we have also characterized best achievable average rates and minimum outage probability for Rayleigh fading channels and for when there is no fading.

An important practical limitation to implement optimum relay selection policy in physical wireless systems is the feedback load required during the relay selection process. To alleviate this limitation, we have proposed a threshold-based distributed relay selection strategy in which each relay node gives an autonomous feedback decision based on its local channel quality indicator. For this class of relay selection policies, we have shown that the total number of relays feeding back is a Poisson distributed random variable. We have characterized the average value for this Poisson distribution analytically, and obtained the analytical expressions for the average rate and outage probability achieved with reduced feedback load for both the no-fading and Rayleigh fading communications scenarios. We have derived useful and practical design rules for selecting relay nodes in a distributed way, which indicates that setting the threshold value to have five relay nodes feeding back on average is enough to achieve almost the same communications performance attained by the all-feedback policy with negligible performance loss. The performance loss becomes insignificant especially when the relay intensity increases.  

Utilizing the developed analytical framework, an important future plan of the authors is the generalization of the derived results to three-dimensional network topologies. A significant application scenario of this generalization is unmanned-aerial-vehicle (UAV) aided wireless communications. The deployment and trajectory optimization of UAVs for coverage and capacity augmentation of terrestrial networks have received increasing interest in recent years \cite{Koyuncu18, Koyuncu19}. This extension will involve a study of the probabilistic path-loss functions for ground-to-air and air-to-ground channels as well as a study of stochastic geometry models on three dimensional manifolds to integrate the UAV specific communications aspects with the developed framework.


\appendices
\section{Proof of Theorem~\ref{Lemma: Sufficient Probablility}}\label{Appendix: Sufficient Probablility Proof}
Without loss of generality, we take $\vec{w} = \vec{0}$ and let $\mathcal{E}_{\rm suff} = \brparen{\relayselect\paren{\Xmid} \leq \sqrt{d^2 + \norm{\vec{X}_{(2)}}^2}}$, where $\Xmid \in \Phi$ and $\vec{X}_{(2)} \in \Phi$ are the closest and second closest points of $\Phi$ to $\vec{0}$. Let $\Psi_{\rm mid} = \norm{\Xmid}$, $\Theta_{\rm mid}$ be the angle of $\Xmid$ and $\Psi_{(2)} = \norm{\vec{X}_{(2)}}$.  Let also $f_{\rm NN}\paren{\psi}$ be the pdf of $\Psi_{\rm mid}$, which is the nearest-neighbor distribution for HPPPs. Due to isotropy of HPPPs, $\Theta_{\rm mid}$ and $\Psi_{(2)}$ are independent random variables and $\Theta_{\rm mid}$ is uniformly distributed over $\parenro{0, 2\pi}$.  Using these observations, $\PRP{\mathcal{E}_{\rm suff}}$ can be expressed as
\begin{eqnarray}
\PRP{\mathcal{E}_{\rm suff}} &=& \frac{2}{\pi}\int_{0}^{\frac{\pi}{2}}\int_0^\infty \PRP{\mathcal{E}_{\rm suff} \ \big| \ \Psi_{\rm mid} = \psi,  \Theta_{\rm mid} = \theta} f_{\rm NN}\paren{\psi} \diff \psi \diff \theta \nonumber \\
&=& \frac{2}{\pi}\int_{0}^{\frac{\pi}{2}}\int_0^\infty \PR{\Psi_{(2)} \geq \sqrt{\psi^2 + 2d\psi\cos\theta} \ \Big| \ \Psi_{\rm mid} = \psi} f_{\rm NN}\paren{\psi} \diff \psi \diff \theta. \label{e_suffCond1} 
\end{eqnarray}
%
%
%

Using spherical contact distributions for HPPPs \cite{Haenggi13}, it can be shown that 
\begin{equation}\label{e_condG2}
\PR{\Psi_{(2)}<\psi^\prime \big| \Psi_{\rm mid} = \psi}
=\left\{\begin{array}{ll}
0 & \mbox{ if } \psi^\prime < \psi \\
1-\e{-\lambda \pi \paren{\paren{\psi^\prime}^2-\psi^2}} & \mbox{ if } \psi^\prime \geq \psi
\end{array}\right..
\end{equation}
Using \eqref{e_suffCond1} and \eqref{e_condG2}, we have
\begin{eqnarray}
\PRP{\mathcal{E}_{\rm suff}}  
& = & \frac{2}{\pi} \int_{0}^{\frac{\pi}{2}} \int_{0}^{\infty} \e{-2\lambda \pi  d \cos(\theta) \psi } f_{\rm NN}(\psi) \diff \psi \diff \theta \nonumber \\
& = & \frac{2}{\pi} \int_{0}^{\frac{\pi}{2}} \mathcal{M}_{\Psi_{\rm mid}}(-2\lambda \pi  d \cos \theta) \diff \theta, \label{e_suffCond2} 
\end{eqnarray}
where $\mathcal{M}_{\Psi_{\rm mid}}(t)=\ES{e^{t \cdot \Psi_{\rm mid}}}$ is the moment generating function of $\Psi_{\rm mid}$.  Since $\Psi_{\rm mid}$ is Rayleigh distributed with pdf $f_{\rm NN}\paren{\psi} = 2\lambda \pi\psi \e{-\lambda\pi\psi^2}$ \cite{Kingman93}, $\mathcal{M}_{\Psi_{\rm mid}}(t)$ is given by 
\begin{eqnarray*}
\mathcal{M}_{\Psi_{\rm mid}}(t) = 1 +  \frac{t}{2\sqrt{\lambda}}\paren{\erf\paren{\frac{t}{2\sqrt{\lambda \pi}}}+1}\e{\frac{t^2}{4\lambda\pi}},
\end{eqnarray*}
where $\erf\paren{\cdot}$ is the Gauss error function. 
%
%
As a result, we can write $\PRP{\mathcal{E}_{\rm suff}}$ in \eqref{e_suffCond2} as
\begin{eqnarray}
\PRP{\mathcal{E}_{\rm suff}} &=& \frac{2}{\pi} \int_{0}^{\frac{\pi}{2}} \left(1 - \sqrt{\lambda}\pi d  \cos (\theta ) \e{\lambda\pi d^2 \cos^2(\theta)} \erfc\left(\sqrt{\lambda \pi} d \cos (\theta)\right)\right)  \diff \theta \nonumber \\
&\stackrel{\rm (a)}{=}& \e{\lambda \pi d^2}\erfc\paren{\sqrt{\lambda\pi}d},
\end{eqnarray}
where $\erfc\paren{x} = 1 - \erf\paren{x}$ is the complementary error function, and (a) follows as we first use the integral identity \cite[eq.~3.2.6]{Ng69}
\[\int_{0}^{\infty}\frac{\e{-a^2t}}{\sqrt{t+\cos^{2}\theta}}\diff t=\frac{\sqrt{\pi}}{a}\e{a^2\cos^{2}\theta}\erfc\left(a\cos\theta\right),\]
 then apply the integral representation \cite[eq.~3.2.7]{Ng69}
\[\erfc\paren{x} = \frac{2}{\pi}\e{-x^2}\int_0^\infty \frac{\e{-x^2 t^2}}{t^2+1} \diff t,\]
and with some further straight forward mathematical manipulations. 
\section{Proof of Lemma \ref{Lemma: Probability Su}}\label{Appendix: Probability Su Proof}

In this appendix, we provide a key lemma that will be used to prove Theorem \ref{Theorem: Optimality Probability} and Theorem \ref{Theorem: Feedback}.  For $\tau \geq \psi \geq 0$, let $\mathcal{D}(\psi,\tau)=\mathcal{B}\paren{\vec{0}, \tau} \setminus \mathcal{B}\paren{\vec{0}, \psi}$ with $\mathcal{B}\paren{\vec{0}, r}$ being the closed disc centered around the origin $\vec{0}$ with radius $r$, and $\vec{U}$ be uniformly distributed over  $\mathcal{D}(\psi,\tau)$. 
%
%
The following lemma provides an expression for $\PR{\relayselect\paren{\vec{U}}>t}$. 
%
%
%

\begin{lemma} \label{Lemma: Probability Su}
Let $\xs = \paren{-d, 0}^\top$ and $\xd = \paren{d, 0}^\top$. 
%
%
For $\tau \geq \sqrt{\psi^2+2d\psi}$, $\PR{\relayselect\paren{\vec{U}}>t}$ is given by 
\begin{equation}\label{Probability Su}
\begin{split}
\PR{\relayselect\paren{\vec{U}}>t}
&=\left\{
\begin{array}{ll}
1 &\text{ if } t<\sqrt{\psi^2+d^2}\\
\frac{\tau^2-t^2}{\tau^2-\psi^2} +  \frac{2a^\star(t^2-\psi^2) + d^2\sin(2a^\star)}{\pi\paren{\tau^2-\psi^2}} \\
\hspace{1.05cm} + p_{\tau, \psi}(t,d) - p_{\tau,\psi}(t, d\sin(a^\star)) &\text{ if }\sqrt{\psi^2+d^2}\leq t < \psi+d \\
\frac{\tau^2-t^2}{\tau^2-\psi^2} + p_{\tau,\psi}(t,d)&\text{ if } \psi+d\leq t < \sqrt{\tau^2+d^2} \\
\frac{2b^\star(\tau^2-t^2) - d^2\sin(2b^\star)}{\pi(\tau^2-\psi^2)} + p_{\tau,\psi}(t, d\sin(b^\star)) &\text{ if }  \sqrt{\tau^2+d^2} \leq t < \tau+d \\
0 & \text{ if } t \geq \tau+d
\end{array}\right., \nonumber
\end{split}
\end{equation} 
where $p_{\tau,\psi}(t, d)\triangleq\frac{2d\sqrt{t^2-d^2}}{\pi(\tau^2-\psi^2)} + \frac{2t^2}{\pi(\tau^2-\psi^2)}\arctan\left(\frac{d}{\sqrt{t^2-d^2}}\right)$,  $a^\star=\arccos\left(\frac{t^2-\psi^2-d^2}{2d\psi}\right)$ and  $b^\star=\arccos\left(\frac{t^2-\tau^2-d^2}{2d\tau}\right)$. 
%
\end{lemma}
\begin{IEEEproof}
The lemma directly follows for $t<\sqrt{\psi^2+d^2}$ and $t \geq \tau+d$ since $\sqrt{\psi^2 + d^2} \leq \relayselect\paren{\vec{x}} \leq \tau + d$ for all $\vec{x} \in \mathcal{D}(\psi,\tau)$. Hence, we will only focus on the case $\sqrt{\psi^2+d^2}\leq t < \tau+d$.  Let $\Psi = \norm{\vec{U}}$ and $\Theta$ be the angle of $\vec{U}$.  $\Psi$ and $\Theta$ are independent random variables due to isotropy, having pdfs $f_{\Psi}(\psi^\prime) = \frac{2\psi^\prime}{\tau^2-\psi^2}$ for $\psi \leq \psi^\prime \leq \tau$ and $f_{\Theta}(\theta) = \frac{1}{2\pi}$ for $\theta \in [0, 2\pi)$, respectively.  By conditioning on $\Theta$, we can express $\PR{\relayselect\paren{\vec{U}}>t}$ as
\begin{eqnarray}
\PR{\relayselect\paren{\vec{U}}>t} &=& \frac{1}{2\pi}\int_{0}^{2\pi}\PR{\Psi^2+2d\abs{\cos\paren{\theta}}\Psi+d^2>t^2} \diff \theta \nonumber \\
&=& \frac{1}{2\pi}\int_{0}^{2\pi} \PR{\Psi > -d\abs{\cos\paren{\theta}} + \sqrt{t^2 - d^2\sin^2\paren{\theta}}} \diff \theta \nonumber \\
&=& \frac{2}{\pi}\int_{0}^{\frac{\pi}{2}} \PR{\Psi > -d\cos\paren{\theta} + \sqrt{t^2 - d^2\sin^2\paren{\theta}}} \diff \theta, \label{Eqn: Uniform Lemma Probability 1} 
\end{eqnarray}
where the last equality follows from the symmetry in the problem.  

Let $g\paren{\theta, t} \defeq \PR{\Psi > -d\abs{\cos\paren{\theta}} + \sqrt{t^2 - d^2\sin^2\paren{\theta}}}$, $t_{\min}\paren{\theta} \defeq \sqrt{\psi^2 + 2d\abs{\cos\paren{\theta}}\psi + d^2}$ and $t_{\max}\paren{\theta} \defeq \sqrt{\tau^2 + 2d\abs{\cos\paren{\theta}}\tau + d^2}$. It can be seen that $g\paren{\theta, t} = 1$ if $t < t_{\min}\paren{\theta}$ and $g\paren{\theta, t} = 0$ if $t > t_{\max}\paren{\theta}$. Hence, $g\paren{\theta, t}$ can be written as 
\begin{eqnarray}
g\paren{\theta, t} = \I{t < t_{\min}\paren{\theta}} + \frac{\tau^2 - \paren{-d\abs{\cos\paren{\theta}} + \sqrt{t^2 - d^2\sin^2\paren{\theta}}}^2}{\tau^2-\psi^2} \I{t_{\min}\paren{\theta} \leq t \leq t_{\max}\paren{\theta}}. \label{Eqn: Uniform Lemma Probability 2} 
\end{eqnarray}

We will consider three disjoint intervals for $t$ to calculate $\PR{\relayselect\paren{\vec{U}}>t} = \frac{2}{\pi}\int_{0}^{\frac{\pi}{2}} g\paren{\theta, t} d\theta$ by using \eqref{Eqn: Uniform Lemma Probability 2}. We first consider $\sqrt{\psi^2 + d^2} \leq t < \psi + d$. We note that $t_{\min}\paren{\theta}$ is a continuous and strictly decreasing function of $\theta$ for $\theta \in \sqparen{0, \frac{\pi}{2}}$. Further, $t_{\min}\paren{0} = \psi + d$ and $t_{\min}\paren{\frac{\pi}{2}} = \sqrt{\psi^2 + d^2}$. Hence, $t < t_{\min}\paren{\theta}$ for $\theta \in \parenro{0, a^\star}$, where $a^\star = \arccos\paren{\frac{t^2-\psi^2-d^2}{2d\psi}}$.  For $\theta \in \sqparen{a^\star, \frac{\pi}{2}}$, we have $t_{\min}\paren{\theta} \leq t \leq t_{\max}\paren{\theta}$ since $t_{\max}\paren{\theta}$ is a continuous and strictly decreasing function of $\theta$ for $\theta \in \sqparen{0, \frac{\pi}{2}}$, having minimum value $t_{\max}\paren{\frac{\pi}{2}} = \sqrt{\tau^2 + d^2}$ and $\tau \geq \sqrt{\psi^2 + 2d\psi}$. Using these observations, we can write $\PR{\relayselect\paren{\vec{U}}>t}$ as
\begin{eqnarray}
\PR{\relayselect\paren{\vec{U}}>t} &=& \frac{2}{\pi}\int_0^{a^\star} \I{t < t_{\min}\paren{\theta}} \diff \theta + \frac{2}{\pi}\int_{a^\star}^{\frac{\pi}{2}}\frac{\tau^2 - \paren{-d\cos\paren{\theta} + \sqrt{t^2 - d^2\sin^2\paren{\theta}}}^2}{\tau^2 - \psi^2} \diff \theta \nonumber \\
&=& \frac{2}{\pi} a^\star + \frac{2}{\pi}\int_{a^\star}^{\frac{\pi}{2}}\frac{\tau^2 - \paren{-d\cos\paren{\theta} + \sqrt{t^2 - d^2\sin^2\paren{\theta}}}^2}{\tau^2-\psi^2} \diff \theta \nonumber \\
&=& \frac{\tau^2-t^2}{\tau^2-\psi^2} + \frac{2a^\star\paren{t^2-\psi^2}+d^2\sin\paren{2a^\star}}{\pi(\tau^2-\psi^2)} + p_{\tau,\psi}(t, d) - p_{\tau,\psi}\paren{t,d\sin\paren{a^\star}} \nonumber
\end{eqnarray}
for $\sqrt{\psi^2 + d^2} \leq t < \psi + d$. 

Next, we consider $\psi+d \leq t < \sqrt{\tau^2 + d^2}$. In this case, we have $t_{\min}\paren{\theta} \leq t \leq t_{\max}\paren{\theta}$ for all $\theta \in \sqparen{0, \frac{\pi}{2}}$. Hence,
\begin{eqnarray}
\PR{\relayselect\paren{\vec{U}}>t} &=& \frac{2}{\pi} \int_0^{\frac{\pi}{2}} \frac{\tau^2 - \paren{-d\cos\paren{\theta} + \sqrt{t^2 - d^2\sin^2\paren{\theta}}}^2}{\tau^2-\psi^2} \diff \theta \nonumber \\
&=& \frac{\tau^2-t^2}{\tau^2-\psi^2} + p_{\tau, \psi}\paren{t, d} \nonumber
\end{eqnarray}
for $\psi + d \leq t < \sqrt{\tau^2 + d^2}$.  Finally, for $\sqrt{\tau^2 + d^2} \leq t < \tau + d$, we have $t \leq t_{\max}\paren{\theta}$ for $\theta \in \sqparen{0, b^\star}$ and $t > t_{\max}\paren{\theta}$ for $\theta \in \parenlo{b^\star, \frac{\pi}{2}}$, where $b^\star = \arccos\paren{\frac{t^2-\tau^2-d^2}{2d\tau}}$. Thus, 
\begin{eqnarray}
\PR{\relayselect\paren{\vec{U}}>t} &=& \frac{2}{\pi}\int_0^{b^\star} \frac{\tau^2 - \paren{-d\cos\paren{\theta} + \sqrt{t^2 - d^2\sin^2\paren{\theta}}}^2}{\tau^2-\psi^2} \diff \theta \nonumber \\
&=& \frac{2b^\star\paren{\tau^2-t^2} - d^2\sin\paren{2b^\star}}{\pi\paren{\tau^2-\psi^2}} + p_{\tau, \psi}\paren{t, d\sin\paren{b^\star}}, \nonumber
\end{eqnarray}
which concludes the proof. 
\end{IEEEproof}

\begin{figure*}[!t]
\begin{minipage}[t]{\textwidth}
\begin{center}
\includegraphics[scale=0.25]{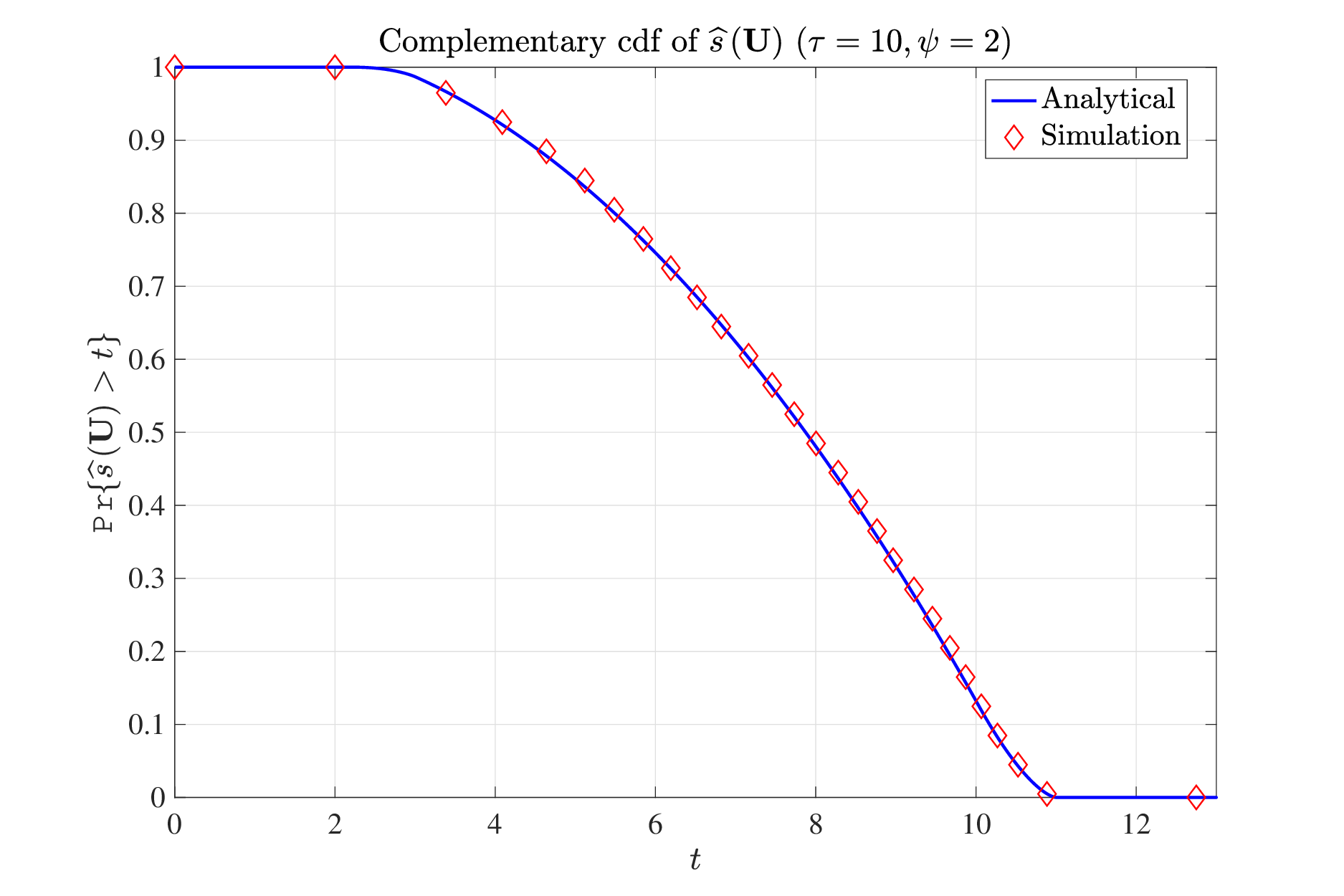}
\hspace{\fill}
\includegraphics[scale=0.25]{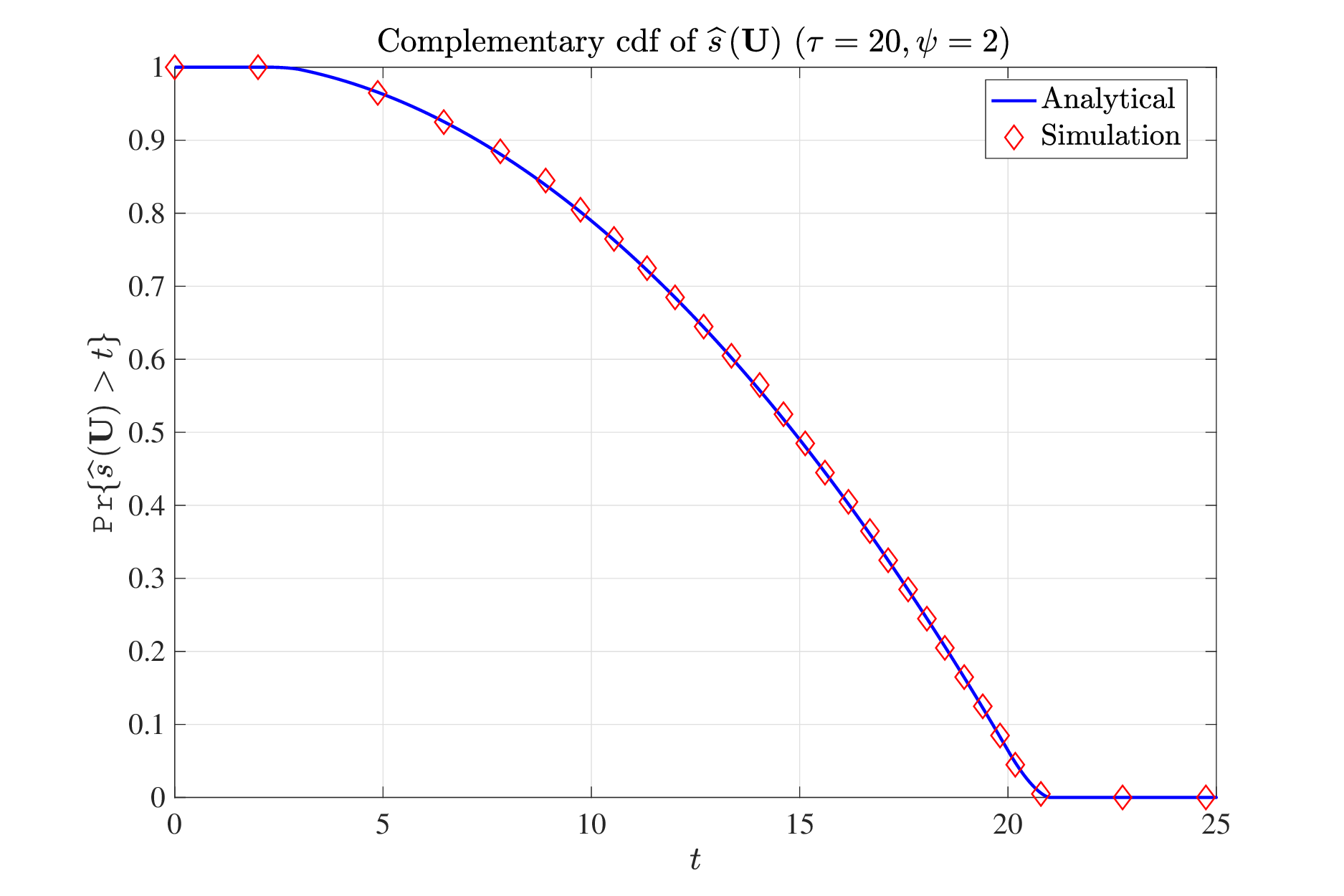}
\\
\includegraphics[scale=0.25]{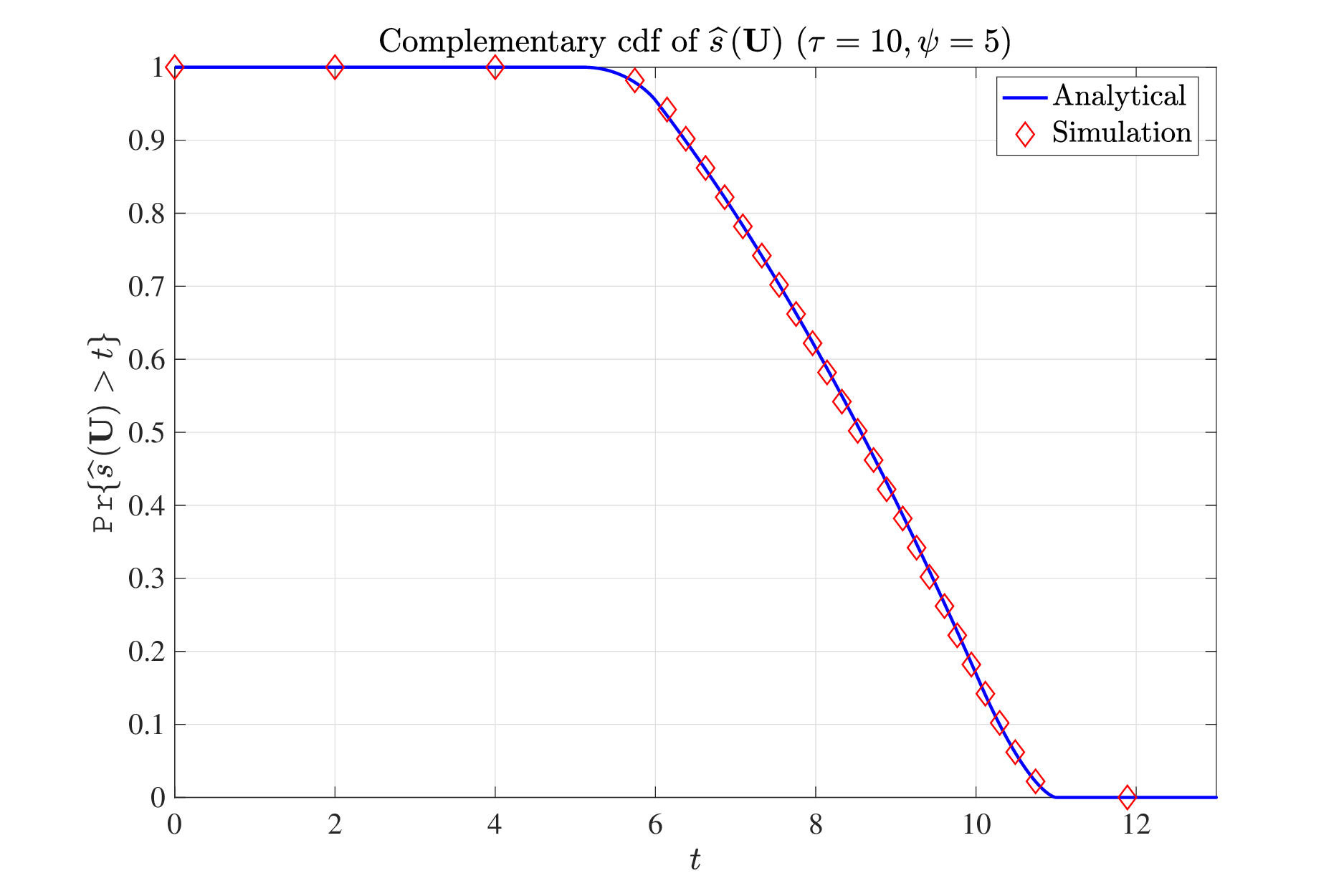}
\hspace{\fill}
\includegraphics[scale=0.25]{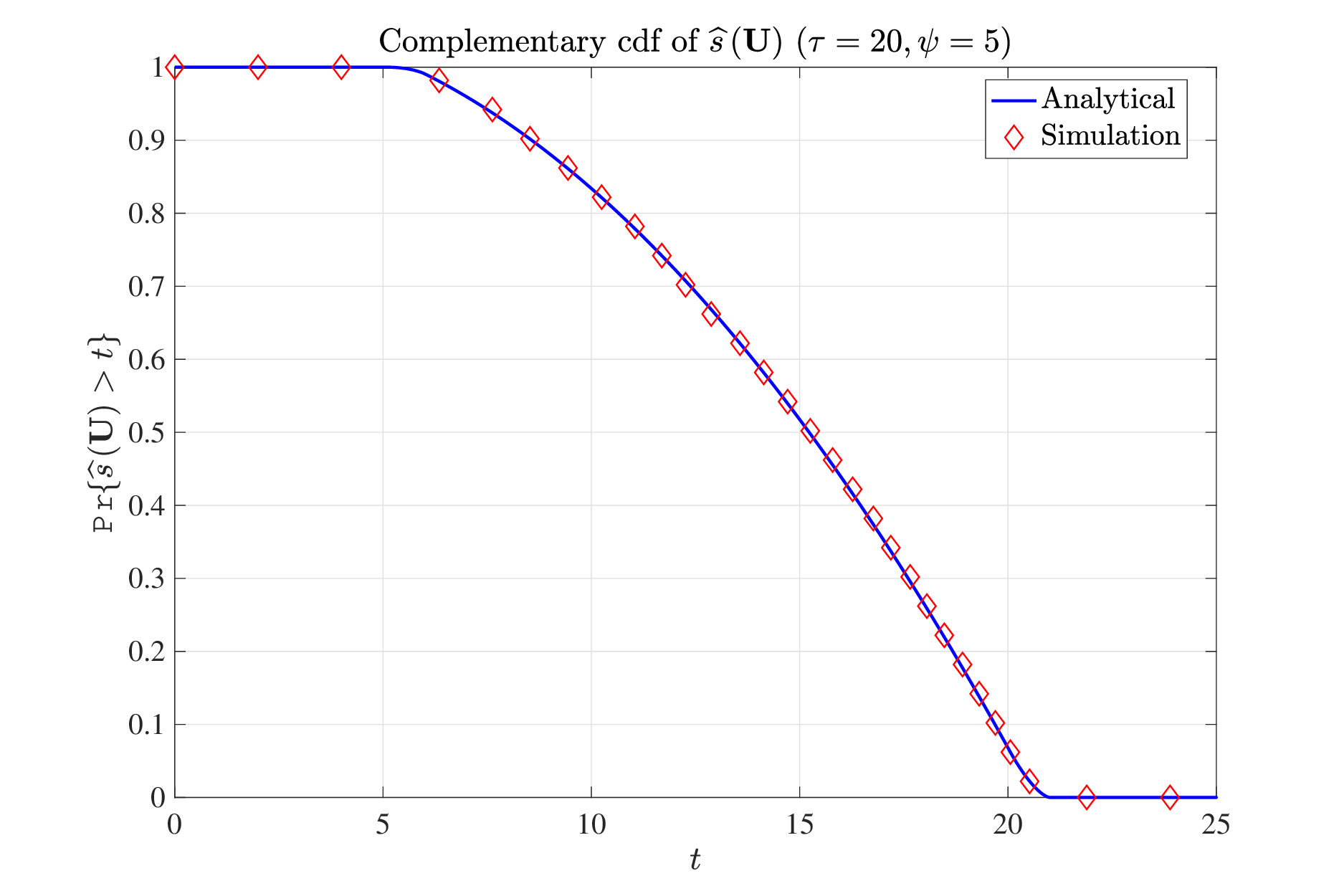}
\end{center}
\end{minipage}
\caption{Complementary cdf of $\relayselect\paren{\vec{U}}$ for $d=1$ and different values of $\tau$ and $\psi$ parameters. $\tau = 10$ and $\psi = 2$ for upper left hand-side figure; $\tau = 20$ and $\psi = 2$ for upper right hand-side figure; $\tau = 10$ and $\psi = 5$ for bottom left hand-side figure; $\tau = 20$ and $\psi = 5$ for bottom right hand-side figure.}  \label{Fig: UniformLemma}
\end{figure*}

For the sake of completeness, we illustrate the analytical expression derived for $\PR{\relayselect\paren{\vec{U}}>t}$ in Lemma \ref{Lemma: Probability Su} and our simulation results for $d=1$ and different values of $\tau$ and $\psi$ parameters in Fig. \ref{Fig: UniformLemma}. As can be seen in this figure, the analytical and simulation results perfectly match each other.  

\section{Proof of Theorem~\ref{Theorem: Optimality Probability}}\label{Appendix: Probability Xmid equal Xopt Proof}

Without loss of generality, we assume that $\xs = \paren{-d, 0}^\top$ and $\xd = \paren{d, 0}^\top$ due to stationarity  and isotropy of HPPPs.  Observing that $\relayselect\paren{\vec{X}} \neq \relayselect\paren{\vec{Y}}$ for any two different points $\vec{X}$ and $\vec{Y}$ in $\Phi$ with probability one, we can write $\PR{\Xmid = \Xopt}$ as 
\begin{eqnarray}
\PR{\Xmid = \Xopt} &=& \PR{\relayselect\paren{\Xmid} = \relayselect\paren{\Xopt}} \nonumber \\
&=& \EW_{\Xmid}\sqparen{\PR{\relayselect\paren{\Xmid} < \min_{\vec{X} \in \Phi \setminus \brparen{\Xmid}}\relayselect\paren{\vec{X}} \Big| \Xmid }}. \nonumber
\end{eqnarray}

Consider the function $g\paren{\vec{x}} = \PR{\relayselect\paren{\vec{x}} < \min_{\vec{X} \in \Phi \setminus \brparen{\vec{x}}}\relayselect\paren{\vec{X}} \Big| \Xmid = \vec{x}}$. Let $\psi = \norm{\vec{x}}$ and $\mathcal{B}\paren{\vec{0}, \tau}$ be the closed disc centered around the origin and having radius $\tau = \sqrt{\psi^2 + 2d\psi}$. For any $\vec{y} \in \R^2\setminus \mathcal{B}\paren{\vec{0}, \tau}$, we have $\relayselect\paren{\vec{y}} > \psi + d$ and $\relayselect\paren{\vec{x}} \leq \psi + d$.  Hence, we can express $g\paren{\vec{x}}$ as
\begin{eqnarray}
g\paren{\vec{x}} &=& \PR{\relayselect\paren{\vec{x}} < \min_{\vec{X} \in \Phi \cap \mathcal{B}\paren{\vec{0}, \tau} \setminus \brparen{\vec{x}}}\relayselect\paren{\vec{X}} \Big| \Xmid = \vec{x}} \nonumber \\
&=& \PR{\relayselect\paren{\vec{x}} < \min_{\vec{X} \in \Phi \cap \mathcal{D}\paren{\psi, \tau}}\relayselect\paren{\vec{X}} \Big| \Xmid = \vec{x}}, \nonumber \\
&=& \PR{\relayselect\paren{\vec{x}} < \min_{\vec{X} \in \Phi \cap \mathcal{D}\paren{\psi, \tau}}\relayselect\paren{\vec{X}}}
\end{eqnarray}
where $\mathcal{D}\paren{\psi, \tau} = \mathcal{B}\paren{\vec{0}, \tau} \setminus \mathcal{B}\paren{\vec{0}, \psi}$, the second equality follows from the fact $\Xmid$ is at the boundary of $\mathcal{B}\paren{\vec{0}, \psi}$ given $\Xmid = \vec{x}$ and there is no relay in the interior of $\mathcal{B}\paren{\vec{0}, \psi}$, and the last equality follows from the total independence property of HPPPs. Below, we will obtain the functional form of $g\paren{\vec{x}}$ to complete the proof.

To this end, let $N$ be the number of relays in $\Phi \cap \mathcal{D}\paren{\psi, \tau}$. Given the event $\brparen{N=n}$ for $n \geq 1$, all the relays in $\Phi \cap \mathcal{D}\paren{\psi, \tau}$ are uniformly distributed over $\mathcal{D}\paren{\psi, \tau}$. Therefore, Lemma \ref{Lemma: Probability Su} can be used directly to obtain the functional form for $g\paren{\vec{x}}$ by conditioning on $N$.  In particular, we have 
\begin{eqnarray}
g\paren{\vec{x}} &=& \sum_{n=0}^\infty \PR{N=n} \PR{\min_{\vec{X}\in \Phi \cap \mathcal{D}\paren{\psi, \tau}} \relayselect\paren{\vec{X}} > \relayselect\paren{\vec{x}} \big| N=n} \nonumber \\
&=& \sum_{n=0}^\infty \frac{\paren{\lambda \pi \paren{\tau^2 - \psi^2}}^n \e{-\lambda\pi\paren{\tau^2 - \psi^2}}}{n!} \PR{\relayselect\paren{\vec{U}} > \relayselect\paren{\vec{x}}}^n \nonumber \\
&=& \exp\paren{-\lambda 2\pi d \psi \PR{\relayselect\paren{\vec{U}} \leq \relayselect\paren{\vec{x}}}}, \label{Eqn: Uniform Optimality 3} 
\end{eqnarray}
where $\vec{U}$ is uniformly distributed over $\mathcal{D}\paren{\psi, \tau}$.  We observe that the second case in Lemma \ref{Lemma: Probability Su} applies to calculate $\PR{\relayselect\paren{\vec{U}} \leq \relayselect\paren{\vec{x}}}$ since $\sqrt{\psi^2 + d^2} \leq \relayselect\paren{\vec{x}} \leq \psi + d$. Defining the function $P\paren{\vec{x}} = 2\pi d \psi \PR{\relayselect\paren{\vec{U}} \leq \relayselect\paren{\vec{x}}}$ and using Lemma \ref{Lemma: Probability Su}, we obtain
\begin{eqnarray}
P\paren{\vec{x}} = \paren{\paren{\relayselect\paren{\vec{x}}}^2 - \psi^2}\paren{\pi - 2\theta} - d^2\sin\paren{2\theta} - V\paren{\vec{x}, d} + V\paren{\vec{x}, d\sin\paren{\theta}}
\end{eqnarray}
after some manipulations, where $V\paren{\vec{x}, y} = 2y\sqrt{\paren{\relayselect\paren{\vec{x}}}^2 - y^2} + 2\paren{\relayselect\paren{\vec{x}}}^2\arctan\paren{\frac{y}{\sqrt{\paren{\relayselect\paren{\vec{x}}}^2-y^2}}}$ and $\theta$ is the angle of $\vec{x}$ restricted to $\sqparen{0, \frac{\pi}{2}}$.  Let $\Psi_{\rm mid} = \norm{\Xmid}$, $\Theta_{\rm mid}$ be the angle of $\Xmid$ and $f_{\rm NN}\paren{\psi}$ be the pdf of $\Psi_{\rm mid}$. Switching to polar coordinates, using the symmetry in the problem and writing $g\paren{\vec{x}}$ and $P\paren{\vec{x}}$ in polar coordinates as $g\paren{\psi, \theta}$ and $P\paren{\psi, \theta}$ with a slight abuse of notation, it can be seen that 
\begin{eqnarray}
\PR{\Xmid = \Xopt} &=& \frac{1}{2\pi}\int_0^{2\pi}\int_0^\infty g\paren{\psi, \theta}f_{\rm NN}\paren{\psi} \diff \psi \diff \theta \nonumber \\
&=& \frac{2}{\pi} \int_0^{\frac{\pi}{2}}\int_0^\infty \exp\paren{-\lambda P\paren{\psi, \theta}} f_{\rm NN}\paren{\psi} \diff \psi \diff \theta \nonumber \\
&=& 4 \ES{\exp\paren{-\lambda P\paren{\Xmid}}\I{0 \leq \Theta_{\rm mid} \leq \frac{\pi}{2}}}, \nonumber
\end{eqnarray}
which concludes the proof.

\section{Upper Bounds on $\Rave\paren{\pol_{\rm opt}}$} \label{Appendix: Optimum Rave Upper Bounds}
We will only obtain the bound on $\Rave\paren{\pol_{\rm opt}}$ for the Rayleigh fading scenario since the derivation for the no-fading case is similar. In the Rayleigh fading case, we can express $\rateave\paren{\pol}$ for any given relay selection policy $\pol$ as
\begin{eqnarray}
\rateave\paren{\pol} &=& \frac12 \EW\sqparen{\log_2\paren{1 + \snr \abs{H}^2G\paren{\relayselect\paren{\vec{X}_\pol}}} \Big| \Phi} \nonumber \\
&=& \frac{1}{2 \ln 2} \int_0^\infty \ln\paren{1 + \snr \cdot G\paren{\relayselect\paren{\vec{X}_\pol}} x} \e{-x} \diff x \nonumber \\
&\stackrel{\rm (a)}{=}& \frac{1}{2 \ln 2} \int_0^\infty \frac{\snr \cdot G\paren{\relayselect\paren{\vec{X}_\pol}}}{1+\snr \cdot G\paren{\relayselect\paren{\vec{X}_\pol}} x} \e{-x} \diff x \nonumber \\
&\stackrel{\rm (b)}{=}& \frac{1}{2 \ln 2} \e{\frac{1}{\snr \cdot G\paren{\relayselect\paren{\vec{X}_\pol}}}} \int_1^\infty \frac{\e{-\frac{t}{\snr \cdot G\paren{\relayselect\paren{\vec{X}_\pol}}}}}{t} \diff t \nonumber \\
&=& \frac{1}{2 \ln 2} f\paren{\frac{1}{\snr \cdot G\paren{\relayselect\paren{\vec{X}_\pol}}}},
\end{eqnarray}
where (a) follows from integration-by-parts and (b) follows from the change of variables with $t = 1+\snr \cdot G\paren{\relayselect\paren{\vec{X}_\pol}} x$. Hence, the difference between $\rateave\paren{\pol_{\rm opt}}$ and $\rateave\paren{\pol_{\rm mid}}$ can be written as 
\begin{eqnarray}
\rateave\paren{\pol_{\rm opt}} - \rateave\paren{\pol_{\rm mid}} = \frac{1}{2 \ln 2} \Inb{\mathcal{E}_{\rm opt}^c} \paren{f\paren{\frac{1}{\snr \cdot G\paren{\relayselect\paren{\Xopt}}}} - f\paren{\frac{1}{\snr \cdot G\paren{\relayselect\paren{\Xmid}}}}}, \nonumber
\end{eqnarray}
where $\mathcal{E}_{\rm opt}^c$ is the event $\mathcal{E}_{\rm opt}^c \equiv \brparen{\Xmid \neq \Xopt}$. The function $f(x) = \e{x} \text{E}_1(x) = \e{x} \int_1^\infty \frac{\e{-tx}}{t} dt$ is monotone decreasing for $x > 0$ since the range of integration is $\parenro{1, \infty}$ and $\relayselect\paren{\Xopt} \geq \sqrt{\Psi^2_{\rm mid} + d^2}$ by using the arguments in the proof of Theorem \ref{Lemma: Mid-point Optimality}. Thus, we can upper bound the difference $\rateave\paren{\pol_{\rm opt}} - \rateave\paren{\pol_{\rm mid}}$ as 
\begin{eqnarray}
\lefteqn{\rateave\paren{\pol_{\rm opt}} - \rateave\paren{\pol_{\rm mid}}} \hspace{14cm} \nonumber \\
\lefteqn{\leq \frac{1}{2 \ln 2} \Inb{\mathcal{E}_{\rm opt}^c}\left( f\paren{\frac{1}{\snr \cdot G\paren{\sqrt{\Psi^2_{\rm mid} + d^2}}}} - f\paren{\frac{1}{\snr \cdot G\paren{\relayselect\paren{\Xmid}}}} \right).} \hspace{12.7cm} \label{Eqn: Rate Difference Upper Bound 1}
\end{eqnarray}

We let $\Delta_\Phi = \frac{1}{2 \ln 2} \Inb{\mathcal{E}_{\rm opt}^c}\left( f\paren{\frac{1}{\snr \cdot G\paren{\sqrt{\Psi^2_{\rm mid} + d^2}}}} - f\paren{\frac{1}{\snr \cdot G\paren{\relayselect\paren{\Xmid}}}} \right)$ and $\Delta_{\rm ave} = \ES{\Delta_\Phi}$. Then, by averaging both sides of \eqref{Eqn: Rate Difference Upper Bound 1},
%
%
we obtain the upper bound $\Rave\paren{\pol_{\rm opt}} \leq \Rave\paren{\pol_{\rm mid}} + \Delta_{\rm ave}$. The expression for $\Delta_{\rm ave}$ is derived by first conditioning on $\Xmid$ and then averaging over $\Xmid$ as below: 
\begin{eqnarray}
\lefteqn{\Delta_{\rm ave} = \ES{\ES{\Delta_\Phi \big| \Xmid}}} \hspace{15cm} \nonumber \\
\lefteqn{= \frac{1}{2 \ln 2}\EW\left[\left( f\paren{\frac{1}{\snr \cdot G\paren{\sqrt{\Psi^2_{\rm mid} + d^2}}}}\right.\right.} \hspace{14.1cm} \nonumber \\
\lefteqn{\left.\left.- f\paren{\frac{1}{\snr \cdot G\paren{\relayselect\paren{\Xmid}}}} \rule{0cm}{0.9cm}\right)\ES{\Inb{\mathcal{E}_{\rm opt}^c} \big| \Xmid}\right]} \hspace{9cm} \nonumber \\
\lefteqn{= \frac{2}{\ln 2}\EW\left[\paren{1-\e{-\lambda P\paren{\Xmid}}}\left(f\paren{\frac{1}{\snr \cdot G\paren{\sqrt{\Psi_{\rm mid}^2 + d^2}}}}\right.\right.} \hspace{14.1cm} \nonumber \\ \lefteqn{\left.\left.-f\paren{\frac{1}{\snr \cdot G\paren{\relayselect\paren{\Xmid}}}}\rule{0cm}{0.9cm} \right)\I{0 \leq \Theta_{\rm mid} \leq \frac{\pi}{2}} \right],} \hspace{7.5cm} \nonumber
\end{eqnarray}
where the last equality follows from the symmetry of $\relayselect\paren{\Xmid}$ over $\parenro{0, 2\pi}$ with respect to the angle of $\Xmid$ and from the proof of Theorem \ref{Theorem: Optimality Probability} where it was shown that the conditional probability $\PR{\Xmid = \Xopt \big| \Xmid = \vec{x}}$ is equal to $\PR{\Xmid = \Xopt \big| \Xmid = \vec{x}} = \e{-\lambda P\paren{\vec{x}}}$ when the angle of $\vec{x}$ is restricted to $\sqparen{0, \frac{\pi}{2}}$.

\section{Proof of Theorem~\ref{Theorem: Optimal dis distribution}}\label{Appendix: optimal dis distribution Proof}

We first divide the relay locations into two types:
\begin{eqnarray}
\Phi_{\rm right} = \Phi \bigcap \R^2_{\rm right} \,\, \text{and}\,\,\Phi_{\rm left} = \Phi \bigcap \R^2_{\rm left}, 
\end{eqnarray}
where $\R^2_{\rm right} = \brparen{\paren{x_1, x_2}^\top \in \R^2: x_1 \geq 0}$ and $\R^2_{\rm left} = \brparen{\paren{x_1, x_2}^\top \in \R^2: x_1 < 0}$. That is, the relays in $\Phi_{\rm right}$ are closer to the destination node, whereas the ones in $\Phi_{\rm left}$ are closer to the source node. Then, we define 
\begin{eqnarray}
\Gamma_{\rm opt}^{\rm right} \defeq \min_{\vec{X} \in \Phi_{\rm right}} \relayselect\paren{\vec{X}} \,\, \text{and}\,\, \Gamma_{\rm opt}^{\rm left} \defeq \min_{\vec{X} \in \Phi_{\rm left}} \relayselect\paren{\vec{X}}. 
\end{eqnarray}
Due to stationarity of HPPPs and symmetry of the problem, $\Gamma_{\rm opt}^{\rm right}$ and $\Gamma_{\rm opt}^{\rm left}$ are identically distributed random variables. Further, they are also independent due to the complete randomness property of Poisson point processes \cite{Kingman93}. Hence, it will be enough to obtain the cdf of $\Gamma_{\rm opt}^{\rm right}$ to prove Theorem \ref{Theorem: Optimal dis distribution} since $\Gamma_{\rm opt} = \min\brparen{\Gamma_{\rm opt}^{\rm right}, \Gamma_{\rm opt}^{\rm left}}$. More specifically,   
$F_{\Gamma_{\rm opt}}\paren{\gamma} = 1 - \paren{1 - F_{\Gamma_{\rm opt}^{\rm right}}\paren{\gamma}}^2$. 
To obtain $F_{\Gamma_{\rm opt}^{\rm right}}\paren{\gamma}$, we further define $\Phi_{{\rm right}, \tau} \defeq \Phi_{\rm right} \cap \mathcal{B}\paren{\vec{0}, \tau}$ and $\Gamma_{{\rm opt}, \tau}^{\rm right} \defeq \min_{\vec{X} \in \Phi_{{\rm right}, \tau}}\relayselect\paren{\vec{X}}$, where $\mathcal{B}\paren{\vec{0}, \tau}$ is the closed disc centered around the origin $\vec{0}$ and having radius $\tau$. We note that $\Gamma_{{\rm opt}, \tau}^{\rm right}$ converges almost surely to $\Gamma_{\rm opt}^{\rm right}$ as $\tau$ tends to infinity. Thus, the cdf of $\Gamma_{{\rm opt}, \tau}^{\rm right}$ will also converge to the cdf of $\Gamma_{\rm opt}^{\rm right}$ pointwise as $\tau$ tends to infinity \cite{Billingsley95}. We will derive the cdf of $\Gamma_{\rm opt}^{\rm right}$ by first obtaining the cdf of $\Gamma_{{\rm opt}, \tau}^{\rm right}$ and then taking the limit $\tau \ra \infty$.    

Let $N$ be the number of relays in $\Phi_{{\rm right}, \tau}$. Given the event $\brparen{N=n}$ for $n\geq 1$, all the relays in $\Phi_{{\rm right}, \tau}$ will be uniformly distributed over the half-disc centered at $\vec{0}$, having radius $\tau$ and containing only those points of $\R^2$ with non-negative first coordinates. Let $\vec{U}$ be such a uniformly distributed random relay location. Let also $\Gamma = \relayselect\paren{\vec{U}}$, which is equal to the distance between $\vec{U}$ and the source node. $\Gamma$ can be written as 
$\Gamma = \sqrt{\Psi^2 + 2d\Psi \cos\Theta + d^2}$
by using the law of cosines, where $\Psi = \norm{\vec{U}}$ and $\Theta$ is the angle between the positive $x$-axis and the line segment connecting $\vec{0}$ and $\vec{U}$. $\Theta$ is uniformly distributed over $\sqparen{-\frac{\pi}{2}, \frac{\pi}{2}}$, and independent of $\Psi$ due to the uniformly distributed nature of $\vec{U}$. Hence, the conditional cdf of $\Gamma$ given $\Theta = \theta$ can be expressed as 
\begin{eqnarray}
F_{\Gamma | \Theta}\paren{\gamma | \theta} 
= \PR{\Psi \leq \sqrt{\gamma^2 - d^2\sin^2\theta} - d\cos\theta} \label{Eqn: Conditional Gamma CDF}
\end{eqnarray}
for $d \leq \gamma \leq \sqrt{\tau^2 + 2d\tau\cos\theta + d^2}$, where the last equation follows from the fact that the convex quadratic function $f(x) = x^2 + 2d \cos\theta x + d^2-\gamma^2$ has only one positive root at $\sqrt{\gamma^2 - d^2\sin^2\theta} - d\cos\theta$ when $\gamma \geq d$. For $\gamma < d$, we have $F_{\Gamma | \Theta}\paren{\gamma | \theta} = 0$ since $\Gamma$ is always greater than or equal to $d$. For $\gamma > \sqrt{\tau^2 + 2d\tau\cos\theta + d^2}$, we have $F_{\Gamma | \Theta}\paren{\gamma | \theta} = 1$ since $\Gamma$ is always smaller than or equal to $\sqrt{\tau^2 + 2d\tau\cos\theta + d^2}$ when $\Theta = \theta$. As a result, using the cdf of $\Psi$, which is equal to $F_{\Psi}\paren{\psi} = \frac{\psi^2}{\tau^2}$, we have 
\begin{eqnarray}\label{Eqn: Conditional Distance CDF}
F_{\Gamma | \Theta}(\gamma | \theta)
=\left\{
\begin{array}{ll}
0 & \mbox{ if } \gamma < d\\
\frac{\left(\sqrt{\gamma^2-d^2 \sin^2\theta}-d\cos\theta\right)^2}{\tau^2} & \mbox{ if } d \leq \gamma \leq \sqrt{\tau^2 + 2d\tau\cos\theta + d^2} \\   
1 & \mbox{ if } \gamma > \sqrt{\tau^2 + 2d \tau \cos\theta + d^2}
\end{array}\right..
\end{eqnarray}
\begin{figure*}
\begin{equation}\label{Eqn: Distance Distribution}
F_{\Gamma}(\gamma)
=\left\{
\begin{array}{ll}
0& \mbox{ if } \gamma < d \\
\frac{2}{\pi \tau^2}\left(\gamma^2 \arcsec\left(\frac{\gamma}{d}\right)- d \sqrt{\gamma^2-d^2}\right)& \mbox{ if } d\leq \gamma \leq \sqrt{\tau^2+d^2} \\
\frac{2\gamma^2}{\pi \tau^2}\left(\rule{0cm}{0.55cm} \arcsec\left(-\frac{2 d \tau}{\tau^2+d^2-\gamma^2}\right)\right.& \\
\hspace{1.0cm} \left.- \arctan\left(\frac{\sqrt{4 d^2 \tau^2-\left(\tau^2+d^2-\gamma^2\right)^2}}{\tau^2-d^2+\gamma^2}\right) \right)&  \\
\hspace{1.0cm} - \frac{2 \arccsc\left(\frac{2 d \tau}{\tau^2+d^2-\gamma^2}\right)}{\pi} &\\
\hspace{1.0cm} -\frac{\sqrt{(\tau-d+\gamma)(\tau+d-\gamma)(d-\tau+\gamma)(\tau+d+\gamma)}}{\pi \tau^2}& \mbox{ if } \sqrt{\tau^2+d^2} < \gamma \leq \tau+d \\ 
1& \mbox{ if } \gamma > \tau+d
\end{array}.\right.
\end{equation}
\hrule
\end{figure*}
We will obtain $F_{\Gamma}(\gamma)$ by averaging \eqref{Eqn: Conditional Distance CDF} over $\Theta$.
To this end, we need to consider four cases separately. If $\gamma < d$, then $F_{\Gamma | \Theta}(\gamma | \theta) = 0$ for all $\theta \in \sqparen{-\frac{\pi}{2}, \frac{\pi}{2}}$. Hence, $F_{\Gamma}(\gamma) = 0$ when $\gamma < d$. If $d \leq \gamma \leq \sqrt{\tau^2 + d^2}$, the second condition in \eqref{Eqn: Conditional Distance CDF} is always satisfied, and we have 
\begin{eqnarray*}
F_{\Gamma}(\gamma) = \frac{2}{\pi \tau^2}\int_0^{\frac{\pi}{2}} \paren{\sqrt{\gamma^2-d^2\sin^2\theta} - d\cos\theta}^2 \diff \theta
\end{eqnarray*}
for this range of $\gamma$. If $\sqrt{\tau^2 + d^2} < \gamma \leq \tau+d$, the second and third conditions in \eqref{Eqn: Conditional Distance CDF} are satisfied for $\theta \in \sqparen{-\theta^\star, \theta^\star}$ and $\theta \in \parenro{-\frac{\pi}{2}, -\theta^\star} \cup \parenlo{\theta^\star, \frac{\pi}{2}}$, respectively, where $\theta^\star = \arccos\paren{\frac{\gamma^2 - \tau^2 - d^2}{2d\tau}}$. Thus, we have  
\begin{eqnarray*}
F_{\Gamma}(\gamma) = \frac{2}{\pi \tau^2}\hspace{-0.1cm}\int_0^{\theta^\star} \hspace{-0.2cm}\paren{\sqrt{\gamma^2-d^2\sin^2\theta} - d\cos\theta}^2\hspace{-0.1cm}\diff \theta + 1 -\frac{2\theta^\star}{\pi}  
\end{eqnarray*}
for this range of $\gamma$. Finally, if $\gamma > \tau+d$, then $F_{\Gamma | \Theta}(\gamma | \theta) = 1$ for all $\theta \in \sqparen{-\frac{\pi}{2}, \frac{\pi}{2}}$, and therefore $F_{\Gamma}(\gamma) = 1$ if $\gamma > \tau + d$. Combining all four cases and evaluating the integrals, we obtain $F_{\Gamma}(\gamma)$ as in \eqref{Eqn: Distance Distribution}. 
Using $F_{\Gamma}(\gamma)$, we obtain 
\begin{eqnarray}
F_{\Gamma_{{\rm opt}, \tau}^{\rm right}}(\gamma) &=& \sum_{n=0}^\infty \paren{1-\paren{1-F_{\Gamma}\paren{\gamma}}^n} \PR{N=n} \nonumber \\
&=& 1 - \sum_{n=0}^\infty \paren{1-F_{\Gamma}\paren{\gamma}}^n \frac{\paren{\frac{\lambda\pi \tau^2}{2}}^n \e{-\frac{\lambda \pi \tau^2}{2}}}{n!} \nonumber \\
&=& 1 - \e{-\frac{\lambda \pi \tau^2}{2}F_{\Gamma}\paren{\gamma}}. \label{Eqn: Distance Distribution 2} 
\end{eqnarray}
As stated earlier, $F_{\Gamma_{{\rm opt}}^{\rm right}}(\gamma) = \lim_{\tau \ra \infty} F_{\Gamma_{{\rm opt}, \tau}^{\rm right}}(\gamma)$ since $\Gamma_{{\rm opt}, \tau}^{\rm right}$ converges to $\Gamma_{{\rm opt}}^{\rm right}$ almost surely. Thus, by rearranging the terms in \eqref{Eqn: Distance Distribution} and using \eqref{Eqn: Distance Distribution 2}, we have  
\begin{eqnarray} 
F_{\Gamma_{{\rm opt}}^{\rm right}}(\gamma) &=& \lim_{\tau \ra \infty} F_{\Gamma_{{\rm opt}, \tau}^{\rm right}}(\gamma) \nonumber \\  
&=& 1 - \lim_{\tau \ra \infty} \e{-\frac{\lambda \pi \tau^2}{2}F_{\Gamma}\paren{\gamma}} \nonumber \\
&=& \left\{
\begin{array}{ll}
0& \mbox{ if } 
\gamma < d \\
1-\e{-\lambda d^2 \paren{\paren{\frac{\gamma}{d}}^2 \arcsec\left(\frac{\gamma}{d}\right) - \sqrt{\paren{\frac{\gamma}{d}}^2-1} }}& \mbox{ if } 
\gamma \geq d 
\end{array}.\right. \label{Eqn: Distance Distribution 3}
\end{eqnarray}
Finally, using \eqref{Eqn: Distance Distribution 3} 
and the identity $F_{\Gamma_{\rm opt}}\paren{\gamma} = 1 - \paren{1 - F_{\Gamma_{\rm opt}^{\rm right}}\paren{\gamma}}^2$, we conclude the proof. 
\section{Proof of Theorem \ref{Theorem: Feedback}}\label{Appendix: Feedback}
To prove this theorem, we first focus on the relay nodes located inside the closed disc $\mathcal{B}\paren{\vec{0}, \tau}$.  
%
%
Let $\mu\paren{T, \tau}$ be the average number of relays located in $\mathcal{B}\paren{\vec{0}, \tau}$ and feeding their channel quality indicators back to the source node. $\mu\paren{T, \tau}$ is given by 
\begin{eqnarray*}
\mu\paren{T, \tau} = \ES{\sum_{\vec{X} \in \Phi \cap \mathcal{B}\paren{\vec{0}, \tau}} \I{\relayselect\paren{\vec{X}} \leq T}}.
\end{eqnarray*}

Using the monotone convergence theorem, it can be seen that $\mu\paren{T} = \lim_{\tau \ra \infty} \mu\paren{T, \tau}$. Let $N$ be the number of relays in $\Phi \cap \mathcal{B}\paren{\vec{0}, \tau}$. Given the event $\brparen{N = n}$, all the relays are independently and uniformly distributed over $\mathcal{B}\paren{\vec{0}, \tau}$, and hence $\ES{\sum_{\vec{X} \in \Phi \cap \mathcal{B}\paren{\vec{0}, \tau}} \I{\relayselect\paren{\vec{X}} \leq T} \Big| N = n} = n \PR{\relayselect\paren{\vec{U} \leq T}}$, where $\vec{U}$ is a {\em generic} random variable uniformly distributed over $\mathcal{B}\paren{\vec{0}, \tau}$. Using this observation, we can write $\mu\paren{T, \tau}$ as
\begin{eqnarray}
\mu\paren{T, \tau} = \lambda \pi \tau^2 \PR{\relayselect\paren{\vec{U} \leq T}}. \label{Eqn: mu(T, tau) Expression} 
\end{eqnarray}

We will use Lemma \ref{Lemma: Probability Su} to conclude the proof, and it is enough to focus only on the case where $d \leq T \leq \sqrt{d^2 + \tau^2}$. In particular, it can be seen by using this lemma that $\PR{\relayselect\paren{\vec{U} \leq T}} = 0$ for all values of $T$ smaller than $d$. Therefore, $\mu\paren{T} = \lim_{\tau \ra \infty} \mu\paren{T, \tau} = 0$ for $T < d$.  For other cases of this lemma where $T \geq \sqrt{d^2 + \tau^2}$, the threshold value grows without any bound when $\tau$ tends to infinity, which is equivalent to the all-feedback case investigated in Section \ref{Section: Optimum Relay Selection}. 

For $d \leq T \leq \sqrt{d^2 + \tau^2}$, we have
\begin{eqnarray*}
\PR{\relayselect\paren{\vec{U} \leq T}} = \frac{T^2}{\tau^2} - \frac{2d\sqrt{T^2 - d^2}}{\pi \tau^2} - \frac{2 T^2}{\pi \tau^2}\arctan\paren{\frac{d}{\sqrt{T^2 - d^2}}}
\end{eqnarray*}
by using Lemma \ref{Lemma: Probability Su}. As a result, 
\begin{eqnarray*}
\mu\paren{T, \tau} = \lambda \pi T^2 - 2d\lambda\sqrt{T^2 - d^2} - 2 T^2 \lambda \arctan\paren{\frac{d}{\sqrt{T^2 - d^2}}}.
\end{eqnarray*}
Taking the limit as $\tau$ tends to infinity, we obtain \eqref{Eqn: Average Feedback Load 2}. To obtain the distribution of $N_{\rm FB}$, we first observe that the sum $\sum_{\vec{X} \in \Phi \cap \mathcal{B}\paren{\vec{0}, \tau}} \I{\relayselect\paren{\vec{X}} \leq T}$ has the characteristic function $\varphi_\tau(t) = \exp\paren{\mu\paren{\tau, T}\paren{\e{\jmath t} - 1}}$, where $\jmath = \sqrt{-1}$. Since $N_{\rm FB} = \lim_{\tau \ra \infty} \sum_{\vec{X} \in \Phi \cap \mathcal{B}\paren{\vec{0}, \tau}} \I{\relayselect\paren{\vec{X}} \leq T}$ almost surely, we conclude that $N_{\rm FB}$ has a Poisson distribution with mean $\mu\paren{T}$ \cite{Billingsley95}.  

\section{Derivation of Distribution Functions for $\Gamma_{\rm mid}$ and $\Gamma_{\rm C2\star}$ with $\star \in \brparen{\rm D, S}$} \label{Appendix: Distributions for Benchmark Strategies}
 
In this appendix, we will derive distribution functions for $\Gamma_{\rm mid}$ and $\Gamma_{\rm C2\star}$, $\star \in \brparen{\rm D, S}$, given in \eqref{Eqn: Mid Point CDF} - \eqref{Eqn: Closest-to-Destination PDF}. We will start with $F_{\Gamma_{\rm mid}}\paren{\gamma}$ and $f_{\Gamma_{\rm mid}}\paren{\gamma}$.

\subsection{Probability Distribution for $\Gamma_{\rm mid}$} 

Let $\Xmid$ be the closest point of $\Phi$ to the origin, and $\Psi_{\rm mid} = \norm{\Xmid}$ and $\Theta_{\rm mid}$ be the angle of $\Xmid$. $\Psi_{\rm mid}$ and $\Theta_{\rm mid}$ are independent of each other, with $\Psi_{\rm mid}$ having the nearest-neighbor pdf $f_{\rm NN}\paren{\psi}$ for HPPPs as its pdf, and $\Theta_{\rm mid}$ being uniformly distributed over $[0, 2\pi)$. Let $\mathcal{E}$ be the event $\mathcal{E} = \brparen{\Theta \in \sqparen{0, \frac\pi2}}$. Due to symmetry in the problem, we have $F_{\Gamma_{\rm mid}}\paren{\gamma} = F_{\Gamma_{\rm mid} | \mathcal{E}}\paren{\gamma}$. Hence, we will focus on $F_{\Gamma_{\rm mid} | \mathcal{E}}\paren{\gamma}$ for the rest of the derivation below.

We first observe that $\Gamma_{\rm mid}$ is always greater than $d$, and hence $F_{\Gamma_{\rm mid} | \mathcal{E}}\paren{\gamma} = 0$ for $\gamma < d$. For $\gamma \geq d$, we define the function $g\paren{\gamma, \psi}$ as
\begin{eqnarray}
g\paren{\gamma, \psi} = F_{\Gamma_{\rm mid} | \mathcal{E}, \Psi_{\rm mid}}\paren{\gamma | \psi}. \nonumber
\end{eqnarray}
Given $\Psi_{\rm mid} = \psi$, the smallest value $\Gamma_{\rm mid}$ can take is $\sqrt{\psi^2 + d^2}$, which is attained when $\Theta_{\rm mid} = \frac\pi2$. The largest value, on the other hand, is $\psi+d$, which is attained at $\Theta_{\rm mid} = 0$. Hence, we can write $g\paren{\gamma, \psi} = 0$ if $\gamma < \sqrt{\psi^2 + d^2}$, and $g\paren{\gamma, \psi} = 1$ if $\gamma > \psi+ d$. For $\sqrt{\psi^2 + d^2} \leq \gamma \leq \psi + d$, we have
\begin{eqnarray*}
g\paren{\gamma, \psi} &=& \PR{\psi^2 + 2d\psi\cos\paren{\Theta} + d^2 \leq \gamma^2 \ \big| \ \mathcal{E}} \\
&=& \PR{\cos\paren{\Theta} \leq \frac{\gamma^2 - \psi^2 - d^2}{2d\psi} \ \Big| \ \mathcal{E}} \\
&=& 1 - \frac{2}{\pi} \arccos\paren{\frac{\gamma^2 - \psi^2 - d^2}{2d\psi}}. 
\end{eqnarray*}

Using independence between $\Psi_{\rm mid}$ and $\Theta_{\rm mid}$, and integrating $g\paren{\gamma, \psi}$ over $\psi$ with respect to $f_{\rm NN}\paren{\psi}$, we obtain $F_{\Gamma_{\rm mid}}\paren{\gamma}$ for $\gamma \geq d$ according to 
\begin{eqnarray}
F_{\Gamma_{\rm mid}}\paren{\gamma} &=& \int_0^\infty g\paren{\gamma, \psi} f_{\rm NN}\paren{\psi} \diff\psi \nonumber \\
&=& \int_0^{\gamma-d} f_{\rm NN}\paren{\psi}d\psi + \int_{\gamma - d}^{\sqrt{\gamma^2 - d^2}} f_{\rm NN}\paren{\psi}\paren{1 - \frac{2}{\pi} \arccos\paren{\frac{\gamma^2 - \psi^2 - d^2}{2d\psi}}} \diff\psi \nonumber \\
&=& F_{\rm NN}\paren{\sqrt{\gamma^2 - d^2}} - \frac2\pi\int_{\gamma-d}^{\sqrt{\gamma^2 - d^2}}f_{\rm NN}\paren{\psi} \arccos\paren{\frac{\gamma^2 - \psi^2 - d^2}{2d\psi}} \diff \psi \nonumber \\
&=& F_{\rm NN}\paren{\sqrt{\gamma^2 - d^2}} - 2\lambda \int_{\gamma-d}^{\sqrt{\gamma^2 - d^2}}2\psi\e{-\lambda\pi\psi^2} \arccos\paren{\frac{\gamma^2 - \psi^2 - d^2}{2d\psi}} \diff \psi \label{Eqn: Gamma_mid CDF}
\end{eqnarray}

Using change of variable $x = \psi^2$ in \eqref{Eqn: Gamma_mid CDF} and the fact that $F_{\Gamma_{\rm mid}}\paren{\gamma} = 0$ for $\gamma < d$, we arrive at
\begin{eqnarray}
F_{\Gamma_{\rm mid}}\paren{\gamma} = \paren{F_{\rm NN}\paren{\sqrt{\gamma^2 - d^2}} - 2\lambda \int_{\paren{\gamma-d}^2}^{\gamma^2 - d^2}\e{-\lambda \pi x} \arccos\paren{\frac{\gamma^2 - x - d^2}{2d\sqrt{x}}} \diff x}\I{\gamma \geq d}. \label{Eqn: Gamma_mid CDF 2}
\end{eqnarray}


Using $F_{\Gamma_{\rm mid}}\paren{\gamma}$ and Leibniz rule for differentiation under integral sign, we obtain
\begin{eqnarray}
f_{\Gamma_{\rm mid}}\paren{\gamma} = \paren{4\lambda\gamma\int_{\paren{\gamma-d}^2}^{\gamma^2 - d^2}\frac{\e{-\lambda \pi x}}{\sqrt{4d^2x - \paren{\gamma^2 - x - d^2}^2}}\diff x}\I{\gamma \geq d}. 
\end{eqnarray}
We note that $f_{\Gamma_{\rm mid}}\paren{\gamma}$ is well defined at $\gamma = d$ since the left and right derivatives are equal to zero. 

\subsection{Probability Distribution for $\Gamma_{\rm C2\star}$ with $\star \in \brparen{\rm D, S}$}
In this part, we will derive the distribution for CQI for the closest-to-destination $\pol_{\rm C2D}$ and closest-to-source $\pol_{\rm C2S}$ relay selection policies. Due to symmetry in the problem, it is enough to focus on only one of these policies. Below, we will provide the derivation for $\pol_{\rm C2D}$.  Let $\vec{X}_{\rm C2D}$ be the closest point of $\Phi$ to $\xd$, and $\Psi_{\rm C2D} = \norm{\vec{X}_{\rm C2D} - \xd}$ and $\Theta_{\rm C2D}$ be the angle between the positive $x$-axis and the line segment connecting $\xd$ with $\vec{X}_{\rm C2D}$. Let $\mathcal{A} = \brparen{\Theta_{\rm C2D} \in \paren{\frac\pi2, \frac{3\pi}{2}}}$ and $\mathcal{E} = \brparen{\Theta_{\rm C2D} \in \sqparen{0, \frac\pi2} \cup \left[\frac{3\pi}{2}, 2\pi\right)}$. Using law of total probability, we can write
\begin{eqnarray} \label{Eqn: Gamma C2D CDF}
F_{\Gamma_{\rm C2D}}\paren{\gamma} = \frac12 F_{\Gamma_{\rm C2D}|\mathcal{A}}\paren{\gamma} + \frac12 F_{\Gamma_{\rm C2D}|\mathcal{E}}\paren{\gamma}. 
\end{eqnarray}

The derivation of $F_{\Gamma_{\rm C2D}|\mathcal{E}}\paren{\gamma}$ follows from the above arguments we used to obtain $F_{\Gamma_{\rm mid}}\paren{\gamma}$ since the geometry of the problem is exactly the same except with the change of $d$ with $2d$ in this case. Hence, we have 
\begin{eqnarray}
F_{\Gamma_{\rm C2D}|\mathcal{E}}\paren{\gamma} = \paren{F_{\rm NN}\paren{\sqrt{\gamma^2 - 4d^2}} - 2\lambda \int_{\paren{\gamma-2d}^2}^{\gamma^2 - 4d^2}\e{-\lambda \pi x} \arccos\paren{\frac{\gamma^2 - x - 4d^2}{4d\sqrt{x}}} \diff x}\I{\gamma \geq 2d}. \label{Eqn: Gamma C2D CDF 1}
\end{eqnarray}

For derivation of $F_{\Gamma_{\rm C2D}|\mathcal{A}}\paren{\gamma}$, we define the function $g\paren{\gamma, \psi}$ as 
\begin{eqnarray}
g\paren{\gamma, \psi} = F_{\Gamma_{\rm C2D}|\mathcal{A}\ \Psi_{\rm C2D}}\paren{\gamma | \psi}. 
\end{eqnarray}
We will analyze the cases $\psi \in [0, d]$ and $\psi \in (d, \infty)$ separately to obtain a functional form for $g\paren{\gamma, \psi}$. We will first consider the case $\psi \in [0, d]$. Given $\Psi_{\rm C2D} = \psi \in [0, d]$, we have $\Gamma_{\rm C2D} = \sqrt{\psi^2 - 4d\psi\abs{\cos\paren{\Theta_{\rm C2D}}} + 4d^2}$, which lies in $\left[2d - \psi, \sqrt{\psi^2 + 4d^2}\right)$ when $\Theta_{\rm C2D} \in \paren{\frac\pi2, \frac{3\pi}{2}}$. Hence, we can write $g\paren{\gamma, \psi}$ as
\begin{eqnarray} \label{Eqn: Gamma C2D CDF 2}
g\paren{\gamma, \psi} = \left\{ \begin{array}{ll}
0& \mbox{ if }  \gamma < d \\
\frac{2}{\pi}\arccos\paren{\frac{\psi^2 + 4d^2 - \gamma^2}{4d\psi}}\I{2d-\gamma \leq \psi \leq d} & \mbox{ if } d \leq \gamma < 2d \\
\I{0 \leq \psi \leq \sqrt{\gamma^2 - 4d^2}} + \frac2\pi \arccos\paren{\frac{\psi^2 + 4d^2 - \gamma^2}{4d\psi}} \I{\sqrt{\gamma^2 - 4d^2} < \psi \leq d}& \mbox{ if } 2d \leq \gamma < \sqrt{5} d \\
\I{\psi \in \sqparen{0, d}} & \mbox{ if } \gamma \geq \sqrt{5} d
\end{array}
\right. 
\end{eqnarray}
for four different ranges of $\gamma$ in this case.

Now, we consider the case $\Psi_{\rm C2D} = \psi \in \paren{d, \infty}$. In this case, $\Gamma_{\rm C2D} = \psi$ when $\Theta_{\rm C2D} \in \sqparen{\pi - \arccos\paren{\frac{d}{\psi}}, \pi + \arccos\paren{\frac{d}{\psi}}}$. On the other hand,  $\Gamma_{\rm C2D} = \sqrt{\psi^2 - 4d\psi\abs{\cos\paren{\Theta_{\rm C2D}}} + 4d^2}$ when $\Theta_{\rm C2D} \in \paren{\frac\pi2, \pi - \arccos\paren{\frac{d}{\psi}}} \bigcup \paren{\pi + \arccos\paren{\frac{d}{\psi}}, \frac{3\pi}{2}}$. Hence, we can write $g\paren{\gamma, \psi}$ as
\begin{eqnarray} \label{Eqn: Gamma C2D CDF 3}
g\paren{\gamma, \psi} = \left\{ \begin{array}{ll}
0& \mbox{ if }  \gamma < d \\
\frac{2}{\pi}\arccos\paren{\frac{\psi^2 + 4d^2 - \gamma^2}{4d\psi}}\I{d < \psi \leq \gamma} & \mbox{ if } d \leq \gamma < \sqrt{5}d \\
\I{d < \psi \leq \sqrt{\gamma^2 - 4d^2}} + \frac2\pi \arccos\paren{\frac{\psi^2 + 4d^2 - \gamma^2}{4d\psi}} \I{\sqrt{\gamma^2 - 4d^2} < \psi \leq \gamma}& \mbox{ if } \gamma > \sqrt{5} d
\end{array}
\right. 
\end{eqnarray}
for three different ranges of $\gamma$.  Combining \eqref{Eqn: Gamma C2D CDF 2} and \eqref{Eqn: Gamma C2D CDF 3}, we arrive at
\begin{eqnarray} \label{Eqn: Gamma C2D CDF 4}
g\paren{\gamma, \psi} = \left\{ \begin{array}{ll}
0& \mbox{ if }  \gamma < d \\
\frac{2}{\pi}\arccos\paren{\frac{\psi^2 + 4d^2 - \gamma^2}{4d\psi}}\I{2d - \gamma < \psi \leq \gamma} & \mbox{ if } d \leq \gamma < 2 d \\
\I{0 \leq \psi \leq \sqrt{\gamma^2 - 4d^2}} + \frac2\pi \arccos\paren{\frac{\psi^2 + 4d^2 - \gamma^2}{4d\psi}} \I{\sqrt{\gamma^2 - 4d^2} < \psi \leq \gamma}& \mbox{ if } \gamma \geq 2 d
\end{array}.
\right. 
\end{eqnarray}

The identity in \eqref{Eqn: Gamma C2D CDF 4} gives us the complete functional representation for the function $g\paren{\gamma, \psi}$. Using \eqref{Eqn: Gamma C2D CDF 4}, we can write $F_{\Gamma_{\rm C2D}|\mathcal{A}}\paren{\gamma}$ as
\begin{eqnarray}
\lefteqn{F_{\Gamma_{\rm C2D}|\mathcal{A}}\paren{\gamma}} \hspace{15.5cm} \nonumber \\
\lefteqn{= \int_{0}^\infty g\paren{\gamma, \psi} f_{\rm NN}\paren{\psi} \diff \psi} \hspace{15.3cm} \nonumber \\
\lefteqn{= \left\{ \begin{array}{ll}
0& \mbox{ if }  \gamma < d \\
\frac{2}{\pi}\int_{2d - \gamma}^\gamma \arccos\paren{\frac{\psi^2 + 4d^2 - \gamma^2}{4d\psi}}f_{\rm NN}\paren{\psi} \diff \psi & \mbox{ if } d \leq \gamma < 2 d \\
F_{\rm NN}\paren{\sqrt{\gamma^2 - 4d^2}} + \frac2\pi\int_{\sqrt{\gamma^2 - 4d^2}}^\gamma \arccos\paren{\frac{\psi^2 + 4d^2 - \gamma^2}{4d\psi}} f_{\rm NN}\paren{\psi}\diff\psi& \mbox{ if } \gamma \geq 2 d
\end{array}.
\right.} \hspace{15.3cm} \label{Eqn: Gamma C2D CDF 5} 
\end{eqnarray}

Using the identities $\arccos\paren{-x} = \pi - \arccos{x}$, $F_{\rm NN}\paren{x} = 0$ for $x <0$, and \eqref{Eqn: Gamma C2D CDF}, \eqref{Eqn: Gamma C2D CDF 1} and \eqref{Eqn: Gamma C2D CDF 5}, we arrive at  
\begin{eqnarray}
F_{\Gamma_{\rm C2D}}\paren{\gamma} =  \paren{F_{\rm NN}\paren{\gamma - 2d} + \lambda \int_{\paren{\gamma - 2d}^2}^{\gamma^2}\e{-\lambda\pi x}\arccos\paren{\frac{x+4d^2-\gamma^2}{4d\sqrt{x}}}\diff x}\I{\gamma \geq d}.  
\end{eqnarray}

Using $F_{\Gamma_{\rm C2D}}\paren{\gamma}$ and Leibniz rule for differentiation under integral sign, we obtain
\begin{eqnarray}
f_{\Gamma_{\rm C2D}}\paren{\gamma} = 2\lambda\gamma\paren{\arccos\paren{\frac{d}{\gamma}}\e{-\lambda\pi\gamma^2} + \int_{\paren{\gamma - 2d}^2}^{\gamma^2} \frac{\e{-\lambda\pi x}}{\sqrt{16d^2x - \paren{x + 4d^2 - \gamma^2}^2}}\diff x}\I{\gamma \geq d}. 
\end{eqnarray}
As for $\Gamma_{\rm mid}$, $f_{\Gamma_{\rm C2D}}\paren{\gamma}$ is well defined at $\gamma = d$ since the left and right derivatives are equal to zero. 

\section{Proof of Theorem \ref{Theorem: Isotropic Extension}} \label{Appendix: Isotropic Extension}
The proof of this theorem follows from the proof of Theorem \ref{Theorem: Optimal dis distribution} until \eqref{Eqn: Conditional Gamma CDF}, where we obtained the conditional cdf of $\Gamma = \relayselect\paren{\vec{U}}$ given the angle $\Theta$ of any random relay location $\vec{U}$ in $\mathcal{B}\paren{\vec{0}, \tau} \bigcap \R^2_{\rm right}$ and given that there are $n$, $n \geq 1$, relays in $\Phi_{{\rm right}, \tau}$. The following lemma establishes the independence property for the angle and magnitude of $\vec{U}$ to proceed with the rest of the proof for isotropic PPPs.   
\begin{lemma} \label{Lemma: Independence for Isotropic PPPs}
Let $N$ be the number of relays in $\Phi_{{\rm right}, \tau}$ and $\vec{U}$ be any random relay location in $\Phi_{{\rm right}, \tau}$ given $\brparen{N=n}$ for $n \geq 1$. Then, its angle $\Theta$ is uniformly distributed over $\sqparen{-\frac{\pi}{2}, \frac{\pi}{2}}$ and is independent from its magnitude $\Psi = \norm{\vec{U}}$. 
\end{lemma}
\begin{IEEEproof}
Using the Poisson property and isotropy \cite{Kingman93}, we can express $\PR{\vec{U} \in \mathcal{S}}$ as 
\begin{eqnarray}
\PR{\vec{U} \in \mathcal{S}} = \frac{2\Lambda\paren{\mathcal{S}}}{\Lambda\paren{\mathcal{B}\paren{\vec{0}, \tau}}} \nonumber
\end{eqnarray}
for all Borel subsets $\mathcal{S}$ of $\mathcal{B}\paren{\vec{0}, \tau} \bigcap \R^2_{\rm right}$. We now consider an auxiliary random variable $\widetilde{\vec{U}}$ with distribution given according to 
\begin{eqnarray}
\PR{\widetilde{\vec{U}} \in \mathcal{S}} = \frac{\Lambda\paren{\mathcal{S}}}{\Lambda\paren{\mathcal{B}\paren{\vec{0}, \tau}}} \nonumber
\end{eqnarray}
for all Borel subsets $\mathcal{S}$ of $\mathcal{B}\paren{\vec{0}, \tau}$. $\widetilde{\vec{U}}$ is a spherically symmetric random variable because 
\begin{eqnarray}
\PR{\Pi\paren{\widetilde{\vec{U}}} \in \mathcal{S}} &=& \PR{\widetilde{\vec{U}} \in \Pi^{-1}\paren{\mathcal{S}}} \nonumber \\
&=& \frac{\Lambda\paren{\Pi^{-1}\paren{\mathcal{S}}}}{\Lambda\paren{\mathcal{B}\paren{\vec{0}, \tau}}} \nonumber \\
&=& \frac{\Lambda\paren{\mathcal{S}}}{\Lambda\paren{\mathcal{B}\paren{\vec{0}, \tau}}}
\end{eqnarray}
for all rotations $\Pi$ around the origin, where the last equality follows from the isotropy property. Hence, the magnitude $\widetilde{\Psi}$ of $\widetilde{\vec{U}}$ is independent of its angle $\widetilde{\Theta}$, which is uniformly distributed over $\parenro{0, 2\pi}$, i.e., see \cite[Theorem 2.3]{Fang90}. Now, we consider conditional distribution of $\widetilde{\vec{U}}$ given $\widetilde{\Theta} \in \sqparen{-\frac{\pi}{2}, \frac{\pi}{2}}$. For any Borel subset $\mathcal{S}$ of $\mathcal{B}\paren{\vec{0}, \tau} \bigcap \R^2_{\rm right}$, it is equal to 
\begin{eqnarray}
\PR{\widetilde{\vec{U}} \in \mathcal{S} \Big| \widetilde{\Theta} \in \sqparen{-\frac{\pi}{2}, \frac{\pi}{2}}} &=& \frac{\PRP{\brparen{\widetilde{\vec{U}} \in \mathcal{S}} \bigcap \brparen{\widetilde{\Theta} \in \sqparen{-\frac{\pi}{2}, \frac{\pi}{2}}}}}{\PR{\widetilde{\Theta} \in \sqparen{-\frac{\pi}{2}, \frac{\pi}{2}}}} \nonumber \\
&=& 2\PR{\widetilde{\vec{U}} \in \mathcal{S}} \nonumber \\
&=& \frac{2\Lambda\paren{\mathcal{S}}}{\Lambda\paren{\mathcal{B}\paren{\vec{0}, \tau}}},
\end{eqnarray}
which is the same distribution with that of $\vec{U}$. Using this identity and taking $\mathcal{S}$ to be $\mathcal{S}_1 = \brparen{\paren{\psi\cos\theta, \psi\sin\theta}: \theta \in \sqparen{\theta_1, \theta_2}, \psi \in \sqparen{0, \tau}}$, $\mathcal{S}_2 = \brparen{\paren{\psi\cos\theta, \psi\sin\theta}: \theta \in \sqparen{-\frac{\pi}{2}, \frac{\pi}{2}}, \psi \in \sqparen{\psi_1, \psi_2}}$ and $\mathcal{S}_3 = \brparen{\paren{\psi\cos\theta, \psi\sin\theta}: \theta \in \sqparen{\theta_1, \theta_2}, \psi \in \sqparen{\psi_1, \psi_2}}$ for $-\frac{\pi}{2} \leq \theta_1 \leq \theta_2 \leq \frac{\pi}{2}$ and $0 \leq \psi_1 \leq \psi_2 \leq \tau$, it can be seen that $\Psi$ and $\Theta$ are independent and $\Theta$ is uniformly distributed over $\sqparen{-\frac{\pi}{2}, \frac{\pi}{2}}$.    \end{IEEEproof}

Using Lemma \ref{Lemma: Independence for Isotropic PPPs}, we can express $F_{\Gamma | \Theta}(\gamma | \theta)$ for isotropic PPPs according to
\begin{eqnarray}\label{Eqn: Conditional Gamma CDF for Isotropic PPPs}
F_{\Gamma | \Theta}(\gamma | \theta)
=\left\{
\begin{array}{ll}
0 & \mbox{ if } \gamma < d\\
\frac{\Lambda\left(\mathcal{B}\paren{\vec{0}, \sqrt{\gamma^2 - d^2\sin^2\theta} - d\cos\theta}\right)}{\Lambda\paren{\mathcal{B}\paren{\vec{0}, \tau}}} & \mbox{ if } d \leq \gamma \leq \sqrt{\tau^2 + 2d\tau\cos\theta + d^2} \\   
1 & \mbox{ if } \gamma > \sqrt{\tau^2 + 2d \tau \cos\theta + d^2}
\end{array}\right..
\end{eqnarray}
We will obtain $F_{\Gamma}\paren{\gamma}$ by averaging \eqref{Eqn: Conditional Gamma CDF for Isotropic PPPs} over the distribution of $\Theta$ by considering four different cases. Two of them are trivial. For $\gamma < d$, $F_{\Gamma}\paren{\gamma} = 0$, and $F_{\Gamma}\paren{\gamma} = 1$ for $\gamma > \tau + d$. For $d \leq \gamma \leq \sqrt{\tau^2 + d^2}$, we have 
\begin{eqnarray}
F_{\Gamma}\paren{\gamma} = \frac{2}{\pi \Lambda\paren{\mathcal{B}\paren{\vec{0}, \tau}}} \int_0^\frac{\pi}{2} \Lambda\left(\mathcal{B}\paren{\vec{0}, \sqrt{\gamma^2 - d^2\sin^2\theta} - d\cos\theta}\right) \diff \theta, \nonumber
\end{eqnarray}
while we have 
\begin{eqnarray}
F_{\Gamma}\paren{\gamma} = 1 + \frac{2}{\pi \Lambda\paren{\mathcal{B}\paren{\vec{0}, \tau}}} \int_0^{\theta^\star} \Lambda\left(\mathcal{B}\paren{\vec{0}, \sqrt{\gamma^2 - d^2\sin^2\theta} - d\cos\theta}\right) \diff \theta - \frac{2\theta^\star}{\pi} \nonumber
\end{eqnarray}
for $\sqrt{\tau^2 + d^2} < \gamma \leq \tau + d$, where $\theta^\star = \arccos\paren{\frac{\gamma^2 - \tau^2 - d^2}{2d\tau}}$. Using the same definitions in Appendix \ref{Appendix: optimal dis distribution Proof}, averaging over $N$ and taking the limit as $\tau$ goes to infinity, we obtain 
\begin{eqnarray}
F_{\Gamma_{{\rm opt}}^{\rm right}}(\gamma) = \left\{
\begin{array}{ll}
0& 
\gamma < d \\
1-\e{-\frac{1}{\pi} \int_0^\frac{\pi}{2} \Lambda\left(\mathcal{B}\paren{\vec{0}, \sqrt{\gamma^2 - d^2\sin^2\theta} - d\cos\theta}\right)\diff \theta}& 
\gamma \geq d 
\end{array}.\right. \nonumber
\end{eqnarray}
for isotropic PPPs. The identity \eqref{Eqn: Gamma Opt CDF for Isotropic PPPs} holds since $F_{\Gamma_{\rm opt}}\paren{\gamma} = 1 - \paren{1 - F_{\Gamma_{\rm opt}^{\rm right}}\paren{\gamma}}^2$. 

Now, we assume that $\Lambda$ is absolutely continuous with respect to the Lebesgue measure and has the unique Radon-Nikodym derivative $\lambda: \R^2 \mapsto \Rp$, i.e., $\Lambda\paren{\mathcal{S}} = \int_\mathcal{S} \lambda\paren{\vec{x}} d\vec{x}$ for all Borel subsets $\mathcal{S}$ of $\R^2$. It can be seen that $\lambda$ is a spherically symmetric function due to isotropy. Hence, by switching to polar coordinates, we can express $F_{\Gamma_{{\rm opt}}^{\rm right}}(\gamma)$ in this case as 
\begin{eqnarray}
F_{\Gamma_{{\rm opt}}^{\rm right}}(\gamma) = \left\{
\begin{array}{ll}
0& \mbox{ if } 
\gamma < d \\
1-\e{-4\int_0^\frac{\pi}{2} g\paren{\gamma, \theta} \diff \theta}& \mbox{ if } 
\gamma \geq d 
\end{array}\right., \nonumber
\end{eqnarray}
where $g\paren{\gamma, \theta} = \int_0^{\sqrt{\gamma^2-d^2\sin^2\theta} - d\cos\theta} \lambda\paren{\psi}\psi \diff \psi$. The pdf of $\Gamma_{\rm opt}$ is then equal to \begin{eqnarray}
f_{\Gamma_{\rm opt}}\paren{\gamma} &=& \frac{\diff}{\diff\gamma} F_{\Gamma_{\rm opt}}\paren{\gamma} \nonumber \\
&=& 4 \paren{\frac{\diff}{\diff \gamma}\int_0^\frac{\pi}{2}g\paren{\gamma, \theta} \diff \theta} \e{-4\int_0^\frac{\pi}{2} g\paren{\gamma, \theta} \diff \theta} \I{\gamma \geq d}. \label{Eqn: Gamma Opt PDF for Isotropic PPPs}
\end{eqnarray}
$g\paren{\gamma, \theta}$ is a continuous function of $\gamma$ and $\theta$, and its partial derivative with respect to $\gamma$ 
\begin{eqnarray}
\frac{\partial}{\partial \gamma} g\paren{\gamma, \theta} = \frac{\gamma \paren{\sqrt{\gamma^2 - d^2\sin^2\theta} - d\cos\theta}}{\sqrt{\gamma^2 - d^2\sin^2\theta}} \lambda\paren{\sqrt{\gamma^2 - d^2\sin^2\theta} - d\cos\theta} \nonumber
\end{eqnarray}
is also a continuous function of $\gamma$ due to continuity of $\lambda$, which can be obtained by applying Leibniz rule for differentiation under integral sign. Hence, applying Leibniz rule one more time to differentiate $\int_0^\frac{\pi}{2}g\paren{\gamma, \theta} \diff \theta$ with respect to $\gamma$, we obtain
\begin{eqnarray}
\frac{\diff }{\diff \gamma}\int_0^\frac{\pi}{2}g\paren{\gamma, \theta} \diff \theta &=& \int_0^\frac{\pi}{2} \frac{\partial}{\partial \gamma}g\paren{\gamma, \theta} \diff \theta \nonumber \\
&=& \int_0^\frac{\pi}{2} \frac{\gamma \paren{\sqrt{\gamma^2 - d^2\sin^2\theta} - d\cos\theta}}{\sqrt{\gamma^2 - d^2\sin^2\theta}} \lambda\paren{\sqrt{\gamma^2 - d^2\sin^2\theta} - d\cos\theta} \diff \theta. \nonumber
\end{eqnarray}

Plugging the above expression into \eqref{Eqn: Gamma Opt PDF for Isotropic PPPs} and using the definition of the function $g\paren{\gamma, \theta}$, we obtain the pdf of $\Gamma_{\rm opt}$ for isotropic PPPs as stated in Theorem \ref{Theorem: Isotropic Extension}.   

\section{Proof of Theorem \ref{Theorem: Diff SNR Distribution}} \label{Appendix: Different SNR Extension}

The main proof idea is similar to the one given for Theorem~\ref{Theorem: Optimal dis distribution}. We first argue why $\Gamma_{\rm opt, diff}$ cannot be smaller than $2d\left(\frac{\widetilde{\SNR}_1\widetilde{\SNR}_2}{\widetilde{\SNR}_1+\widetilde{\SNR}_2}\right)$. The optimum relay location $\vec{x}^\star$ minimizing $\relayselectdiffsnr\paren{\vec{x}}$ over $\R^2$ must be located on the line segment $\mathcal{L}$ connecting $\xs$ and $\xd$. Otherwise, we can always consider the projection of $\vec{x}^\star$ on $\mathcal{L}$, and obtain a strictly smaller value for $\relayselectdiffsnr\paren{\vec{x}}$. At $\vec{x}^\star$, we must also have $\widetilde{\snr}_1 \norm{\xs - \vec{x}^\star} = \widetilde{\snr}_2 \norm{\vec{x}^\star - \xd}$. Otherwise, we can obtain a strictly smaller value for $\relayselectdiffsnr\paren{\vec{x}}$ in a neighborhood of $\vec{x}^\star$ by moving towards $\xs$ or $\xd$. Combining these observations and using $\norm{\xs - \vec{x}^\star} + \norm{\vec{x}^\star - \xd} = 2d $, we obtain $\relayselectdiffsnr\paren{\vec{x}^\star} = 2d\frac{\widetilde{\SNR}_1\widetilde{\SNR}_2}{\widetilde{\SNR}_1+\widetilde{\SNR}_2}$.    

For the rest of the proof, we will assume $\widetilde{\snr}_2 > \widetilde{\snr}_1$ without loss of generality, . The analysis for $\widetilde{\snr}_1 > \widetilde{\snr}_2$ is the same due to symmetry in the problem. The case $\widetilde{\snr}_2 = \widetilde{\snr}_1$ reduces to the equal $\snr$ case analyzed in Theorem \ref{Theorem: Optimal dis distribution}. We categorize the relay locations into two classes according to  
\begin{eqnarray}
\Phi_{\rm right} = \Phi \bigcap \R^2_{\rm right} \,\, \text{and}\,\,\Phi_{\rm left} = \Phi \bigcap \R^2_{\rm left}, 
\end{eqnarray}
where $\R^2_{\rm right} = \brparen{\paren{x_1, x_2}^\top \in \R^2: x_1 \geq 0}$ and $\R^2_{\rm left} = \brparen{\paren{x_1, x_2}^\top \in \R^2: x_1 < 0}$. Let $\Gamma_{\rm opt, diff}^{\rm right}$ and $\Gamma_{\rm opt, diff}^{\rm left}$ be defined as $\Gamma_{\rm opt, diff}^{\rm right} \defeq \min_{\vec{X} \in \Phi_{\rm right}} \relayselectdiffsnr\paren{\vec{X}}$ and $\Gamma_{\rm opt, diff}^{\rm left} \defeq \min_{\vec{X} \in \Phi_{\rm left}} \relayselectdiffsnr\paren{\vec{X}}$. With these definitions, we have $\Gamma_{\rm opt, diff} = \min\brparen{\Gamma_{\rm opt, diff}^{\rm right}, \Gamma_{\rm opt, diff}^{\rm left}}$. $\Gamma_{\rm opt, diff}^{\rm right}$ and $\Gamma_{\rm opt, diff}^{\rm left}$ are independent but not identically distributed random variables due to scaling with different $\snr$ values. Hence, the cdf of $\Gamma_{\rm opt, diff}$ is given by 
\begin{eqnarray}
F_{\Gamma_{\rm opt, diff}}\paren{\gamma} = 1 - \paren{1 - F_{\Gamma_{\rm opt, diff}^{\rm right}}\paren{\gamma}} \cdot \paren{1 - F_{\Gamma_{\rm opt, diff}^{\rm left}}\paren{\gamma}}. \label{Eqn: Diff SNR CDF Fucntional Form} 
\end{eqnarray}

We observe that if $\vec{X} \in \Phi_{\rm left}$, then $\relayselectdiffsnr\paren{\vec{X}} = \widetilde{\snr}_2\norm{\vec{X} - \xd}$ (i.e., $\widetilde{\snr}_2 > \widetilde{\snr}_1$), and the analysis in the proof of Theorem \ref{Theorem: Optimal dis distribution} directly applies to obtain $F_{\Gamma_{\rm opt, diff}^{\rm left}}\paren{\gamma}$. For the cdf of $\Gamma_{\rm opt, diff}^{\rm left}$, we also proceed as in the proof of Theorem \ref{Theorem: Optimal dis distribution}. Consider the restriction of $\Phi_{\rm right}$ to a disc $\mathcal{B}\paren{\vec{0}, \tau}$ centered at the origin with radius $\tau$, which we call $\Phi_{{\rm right}, \tau}$. Let $N$ be the number of relays in $\Phi_{{\rm right}, \tau}$. Given the event $\brparen{N = n}$, all relays will be uniformly distributed over the half disc $\mathcal{B}\paren{\vec{0}, \tau} \bigcap \R^2_{\rm right}$. Let $\vec{U}$ be such a uniformly distributed relay location and define $\Gamma = \relayselectdiffsnr\paren{\vec{U}}$. Depending on the angle $\Theta$ and magnitude $\Psi$ of $\vec{U}$, we have $\Gamma$ is either equal to $\widetilde{\snr}_2\norm{\vec{U} - \xd}$ or $\widetilde{\snr}_1\norm{\vec{U} - \xs}$. In particular, when $\Theta \in \sqparen{\theta^\star, \frac{\pi}{2}} \bigcup \sqparen{-\frac{\pi}{2}, -\frac{\pi}{2} + \theta^\star}$, we always have $\Gamma = \widetilde{\snr}_2\norm{\vec{U} - \xd}$, where $\theta^\star = \arccos\paren{\frac{c^2+1}{c^2-1}}$ and $c \defeq \frac{\widetilde{\snr}_2}{\widetilde{\snr}_1}$. For $\Theta \in \paren{-\theta^\star, \theta^\star}$, we have $\Gamma = \widetilde{\snr}_2\norm{\vec{U} - \xd}$ when $\Psi \in \sqparen{0, r_1\paren{\Theta}} \bigcup \sqparen{r_2\paren{\Theta}, \tau}$ and $\Gamma = \widetilde{\snr}_1\norm{\vec{U} - \xs}$ when $\Psi \in \paren{r_1\paren{\Theta}, r_2\paren{\Theta}}$, where the functions $r_1\paren{\theta}$ and $r_2\paren{\theta}$ are defined as $r_1\paren{\theta} = \frac{c^2+1}{c^2-1}d\cos\paren{\theta} - d \sqrt{\paren{\frac{c^2+1}{c^2-1}}^2\cos^2\paren{\theta}-1}$ and $r_2\paren{\theta} = \frac{c^2+1}{c^2-1}d\cos\paren{\theta} + d \sqrt{\paren{\frac{c^2+1}{c^2-1}}^2\cos^2\paren{\theta}-1}$.\footnote{Here, we assume that $\tau \geq \frac{c^2+1}{c^2-1}d + d\sqrt{\paren{\frac{c^2+1}{c^2-1}^2}-1}$.} Analyzing these different cases, averaging over $N$, taking the limit $\tau \ra \infty$ and using \eqref{Eqn: Diff SNR CDF Fucntional Form}, we arrive at \eqref{e_cdfmaxminDdiffSNR}.

\bibliographystyle{IEEEtran}

\bibliography{1references}
\end{document}